 \documentclass[ aps,  amsfont, amsmath,amssymb]{revtex4}
\usepackage{amsmath,amsthm}
\usepackage{amssymb}
 \usepackage{mathrsfs}

 \usepackage{epsfig}
 
 \usepackage{blindtext}
\usepackage{amsmath, amssymb, graphicx, mathtools, multirow,makecell,subfigure,color,float,stmaryrd,mathrsfs}

\usepackage{enumitem}

\usepackage[utf8]{inputenc}

\usepackage{ulem}
\usepackage{natbib}

 \usepackage{mathrsfs}
 \usepackage{subfigure}
 \usepackage{graphicx}
 \usepackage{graphics}
\DeclareGraphicsExtensions{.png,.jpg,.pdf,.gif,.eps}
\begin{document}

\title{Inertia-induced  power-law scaling in martensites}
\author{O. U. Salman$^{1,2}$, A. Finel$^{2}$, and L. Truskinovsky$^{3}$}

\affiliation{$^{1}$Laboratoire des Sciences des Procédés et des Matériaux (LSPM), CNRS, Université Sorbonne Paris Nord, UPR 3407, F-93430, Villetaneuse, France}

\affiliation{$^{2}$Laboratoire d'Etudes des Microstructures, ONERA, 29 Avenue de la Division Leclerc, 92322 Chatillon, France}

\affiliation{$^{3}$Physique et Mécanique des Milieux Hétérogènes (PMMH), ESPCI Paris, PSL University, Sorbonne Université, Université de Paris, CNRS, F-75005 Paris, France}

\date{\today}

\begin{abstract}
 Martensites  subjected to quasistatic  deformation  are known to  exhibit power law distributed acoustic emission in a broad range of scales,  however,  the origin of the observed scaling behavior and the mechanism of self-organization towards  apparent criticality remains obscure.  Here  we argue that the power law  structure of intermittent   fluctuations    can be  at least partially interpreted    as an effect of inertia.   We build on the insight   that  inertial dynamics, evidenced by acoustic emission,  can  become  an important   factor  if the  underlying mechanical system is only marginally stable. We first illustrate the possibility of inertia-induced  \textit{heavy-tailed}  avalanche size distributions  using   a prototypical example of a discrete chain with bi-stable springs. We then  explore the effects of inertia in fully realistic two- and three-dimensional continuum models  of  elastic phase transitions. In particular, we demonstrate that a three-dimensional model can produce  not only qualitative but also quantitative agreement with experiment. 
\end{abstract}

 \maketitle
 
\quad \quad \quad \quad \quad \quad\quad\quad       The paper is dedicated to  Marcelo Epstein on the occasion of his 80th birthday.
\section{Introduction}

Elastic crystals undergoing  diffusionless (martensitic)  phase  transformations are known to belong to the class of   systems exhibiting avalanche-type dynamics under quasi-static driving \cite{Planes2017, Porta2019, Planes2013-ag,PerezReche2017, Balandraud2015}. They are then examples of inherently nonlinear complex systems with intermittent response  resulting from cascades of internal instabilities  \cite{Sethna2017, Alava2006, Alava2014, Miguel2001, Zaiser2006, daRocha2022, Salman2011-ij, Zhang2020-ax, Salman2012}.  Typical examples of such systems,  usually characterized by threshold type nonlinearity, long range interactions and self-induced disorder,  include  terrestrial earthquakes, Barkhausen noise   and  amorphous plasticity, to mention just a few ~\cite{Fisher1985, Rosso2022, LeBlanc2014, Denisov2016, Liu2016, Chan2010, Friedman2012}. A salient feature of all these systems is that the  observable quantities,   representing the underlying  collective behavior,  exhibit heavy-tailed   distributions. If no fine tuning   is    needed to reach  the    scaling regime,   the emerging \textit{quasi-criticality} is  usually   interpreted  either  in the framework of  the concept of self-organized criticality (SOC)  ~\cite{PhysRevLett.59.381,  Dickman2000, Jensen1998, Dhar2006, Pruessner2012, Sornette2006, Gros2024, Tadic2021} or  within  a  closely related paradigm  of  marginal stability \cite{Muller2015, Shang2020, Ovaska2017, Franz2017}. 

In this paper we address the origin of intermittent response  in  quasi-statically driven martensites. We recall that martensitic  transformations  are structural,   athermal, shear-dominated displacive  phase  transitions \cite{nishiyama2012martensitic,roitburd1978martensitic}.
The presence of   lattice mismatch between the low symmetry  martensite and the high symmetry austenite  phases  leads to   long-range elastic interactions which conspire with elastic energy nonconvexity  in producing  complex multiscale  microstructures \cite{james2000martensitic,khachaturyan2013theory, Bhattacharya2003}. 

It has been long known that  mechanically (or thermally) driven  martensitic phase transitions are    accompanied by  broadly distributed bursts (avalanches) which are audible. More recently, it has become clear that   the intermittent transformation advance generates detectable acoustic emission (AE) which carries an important information about the mechanism of martensitic  transformations.  Thus, it reveals that between consecutive avalanches the transformation is suppressed while  the  system deforms purely elastically. Such elastic branches terminate   when  the system reaches   consecutive instability thresholds.  It was established empirically that  when the driving is quasi-static, which  prevents the overlap of individual avalanches, the amplitudes and durations of AE hits  (induced by individual elastic instability events)    exhibit power law behavior.  The value of the corresponding   exponents were shown to depend only  on  crystal symmetry which points towards  the universality of the underlying nonequilibrium steady states \cite{PhysRevB.69.064101, Planes2017, Porta2019, Balandraud2015, Salman2012, Chandni2009, Rosinberg2011, Vives2011, Gallardo2010, Beke2019, Illa2015,PhysRevB.80.180101,Planes2013-ag,Torrents2017-kk,Beke2018-xv,Song2013-os}. 

The exact origin of power law distributed avalanches in martensites   is still a  subject of debate.  For instance, it   has been  argued   that   the main factor   behind the observed scaling behavior  is  quenched  disorder    \cite{Sethna2001}. To corroborate this idea,  a driven zero-temperature Random Field Ising Model (RFIM)  with short range interactions of ferromagnetic type was studied extensively  \cite{PhysRevLett.70.3347,PhysRevLett.75.4528,PhysRevB.53.14872}; same idea is behind  the parallel studies of  the  Random Bond Ising Model \cite{PhysRevB.50.3839}, the Diluted Ising Model \cite{PhysRevLett.77.3843}  and the  Random Anisotropy Ising Model \cite{PhysRevB.63.134431}.  A  shortcoming of   these  models is that  a   power law  emerges there only at a critical level of quenched  disorder  \cite{PhysRevLett.70.3347,PhysRevLett.75.4528,PhysRevB.53.14872,Sethna2001,Perez-Reche2008}, which    points towards  'tuned' rather than 'generic' scaling behavior.  It has been argued, however,   that the near-critical domain  in the corresponding  parameter space  may be  so  large  that  it emerges in experiment as an extended  criticality  \cite{PhysRevLett.70.3347, Sethna2001, Sethna2006}.

A different, but closely related, interpretation of the generic   nature of scaling in martensites was proposed in \cite{Perez-Reche:2007et}, where it  was linked to the presence of a limited dislocational activity taking place concurrently with the  martensitic transformation. In this interpretation  an external  tuning of disorder is   replaced by self-tuning of the annealed inhomogeneity which takes the form of a  co-evolving dislocation distribution.  Based on the analysis of the corresponding  stylized model  it was shown  that the implied   self-induced  disorder   is   correlated  which facilitates  the self-organization of the system towards criticality \cite{Perez-Reche:2007et,PerezReche2016, PerezReche2009, PerezReche2017}.  This idea was corroborated to some extent by  the observations that  the   power-law behavior in martensites emerges  only after  cyclic loading (known as 'training') which  ensures that the necessary level of self-organization  has been reached \cite{Carrillo1998-jq, PhysRevB.69.064101}. Despite its plausibility, the universality of such an explanation has been  also challenged with the reference to  some experiments where  power law distributed avalanches were observed without training, already during the first loading  cycle and within samples   exhibiting  no clear signs of plastic slip \cite{Bonnot2008-qg}.

Yet another proposed interpretation of scale invariance in martensites, still relying on  the umbrella  RFIM framework, is linking it to the  antiferromagnetic nature of elastic long range interactions. It suggests that  the  crucial ingredient  missing in the standard RFIM model, is the anisotropy of   elastic interactions  which makes the  corresponding interaction kernels sign indefinite \cite{Eshelby1957, Picard2004}. It was shown  that  the presence  of  such kernels changes   the nature of  interactions  fundamentally and can  drastically modify the  spatial structure of   avalanches.  In particular, it was shown that  in a model  setting  of this type a robust self-organization towards  scaling regime  becomes possible \cite{Perez-Reche2008}. This idea was  recently reinforced  by an extensive numerical study  of  a comprehensive  long range version of the RFIM \cite{Rossi2023}. However, given that in the proposed models  the role of disorder remains opaque while  the robust nature of the scaling is not sufficiently manifested, the conclusion about the 'generic' nature of the associated criticality remains still debatable. 

%Therefore, in order to model adequately  the quadrupolar plastic events associated with elementary martensitic transformation transformations   one should  introduce anti-ferromagnetic components into the corresponding kernels \cite{Eshelby1957, Picard2004}.
%
%It was also argued  that  the corresponding change in the nature of the
%interactions   drastically modifies
%the geometry and the spatial structure of the large avalanches and  then  disorder emerges as  an irrelevant parameter suggesting  self organization towards  scaling regime\cite{Perez-Reche2008, Eshelby1957, Picard2004, Rossi2023}. 
%To this end one can use for instance  the Eshelby-RFIM  where the interaction kernel is not positive definite which makes the corresponding automaton system no longer Abelian  \cite{???}. ( \textbf{add references})
%L. Kurbah and P. Shukla, Hysteresis in the antiferromagnetic
%random-field Ising model at zero temperature, Phys. Rev. E 83,
%061136 (2011).
% J. N. Nampoothiri, K. Ramola, S. Sabhapandit, and B.
%Chakraborty, Gaps between avalanches in one-dimensional
%random-field Ising models, Phys. Rev. E 96, 032107 (2017).

%
% We nbote that fine tuning  nature of the  scaling regime is also characteristic   depinning transitions  \cite{???}. 
%% D. S. Fisher, Collective transport in random media: From super-
%%conductors to earthquakes, Phys. Rep. 301, 113 (1998).
%
%Models were proposed where scaling during  martensitic transformations was linked to the concurrent  development of plastic deformation, however the associated self organization to criticality  requires an extended training.

In the present paper we complement those   studies  by focusing on the possibility  that \textit{inertia} may also be  one of the    factors  responsible for  generic (extended)  scaling  behavior during martensitic transformations.  
 We elaborate  on the   evidence that  none of the above models  takes   into account the fact that   martensitic phase boundaries  move close to sound velocity, not only  producing lattice scale   waves but also  generating elastic shocks which may affect the transformation \cite{clapp1995would,miao2020explosive,schwabe2022speed,lin1998pulse,truskinovsky2002nucleation,PhysRevB.67.024114,Truskinovsky2008-fn,steinbach2011phase,cho2012finite,liu2024modeling}. The   implied scattering and focusing of elastic waves suggests    rich  dynamics  which is  revealed  by the detected  intense  acoustic emission  \cite{PhysRevLett.72.1694}. The underlying   dynamic activity is not dissimilar to the one   generated during  terrestrial earthquakes \cite{Manosa1989-fn,baro2013statistical}. In this respect, of particular importance to our study is the ability of seismic waves  to nucleate secondary slips, which is  the phenomenon  known as \textit{dynamic triggering} of  earthquakes \cite{gomberg2005dynamic,johnson2005nonlinear,kocharyan2021nucleation}. This analogy is a   reminder  that  during martensitic transformations the   mechanical system  should be considered as  fundamentally  \textit{underdamped}. 

%
%While the issue of inertia  has   been then underplayed previously as a factor essential for criticality in martensites,  it is clear that a driven system of this type  should exhibit  rich intermittent dynamics as it is  revealed   by intense  acoustic emission activity \cite{PhysRevLett.72.1694}.
%
% Indeed, as martensitic  transformation advances, phase boundaries propagate with close to sonic spoeed and   elastic waves are emitted in the ultrasonic range  in front of  such dynamic interfaces.   To summarize, similar to the case of   earthquakes,  the emergence  of  elastic waves during martensitic transformations 

An  important   role of inertia in the process of self-organization towards criticality and scaling  has been  long  realized  in the studies   of avalanches in sandpiles. For instance,  effects of inertia were engaged  to  explain the failure to achieve full  criticality in  physical experiments with actual sand  and to justify skewed nature of the associated avalanche shapes \cite{Khfifi2008, Maimon2004, Papanikolaou2016, Carlson1989, Prado1992, Held1990, Jaeger1989}.  However,  in  these studies   the underdamped nature of the system was  modeled  only \textit{indirectly}, by accounting   for inertia-induced threshold weakening.  Inertial effects were also  implied in the closely related   models of critical behavior caused by  kinetic softening where the underdamped nature of the model was represented  by the non-monotonicity of the flow curve \cite{ Marchetti2006-oa, Denisov2017, Nicolas2016, Karimi2016, deGeus2024}.  A well known  example of such studies  is the Burridge-Knopoff  model  of earthquakes where inertia  is hidden under the phenomenological assumption that dynamic friction is lower than  static friction \cite{Clancy2005}. Behind all these indirect representations of inertia  is the idea that   dynamically generated   acoustic waves allow the system  to jump over  barriers.  The implied   positive feedback   was shown to be   responsible for stick-slip  mechanical response which is ultimately 
behind the observed  intermittency \cite{Salerno2012, Salerno2013, Karimi2016, Karimi2017, Nicolas2016}. An interesting salient  feature  of the   velocity-weakening-type models of  inertial effects   is  the prediction of the    prevalence of large avalanches  known as supercriticality.   
 
A crucial importance of the  direct account for inertia in the modeling of martensitic transformations has been also long realized \cite{truskinovsky1994transition,clapp1995would,reid1997pattern,lin1998pulse}.
Thus, already in  an early study of a one dimensional continuum model of a  prototypical  martensitic transformation, it was  shown that  the formation of twin microstructure  can  be interpreted as  a purely \textit{inertial} effect which disappears in an overdamped setting \cite{PhysRevLett.67.3412}. Inertial effects were then  shown to be behind the development of intrmittency in a closely related  fully dynamic Frenkel-Kontorova model where power law scaling was  shown to emerge  without any quenched or annealed disorder \cite{Elmer1994}. The possibility  of intermittency and scaling in a molecular dynamics model of a martensitic transformation  was   demonstrated   in \cite{ding2013dynamically},  however, the implicitly present  inertial effects were not specifically identified  as a factor contributing to self-organization towards criticality.  

The  first \textit{continuum} model dedicated   directly  to the study of inertia-induced criticality  in martensites  was proposed   in \cite{Ahluwalia2001-cy,PhysRevB.70.224105}, see  also  its subsequent developments in \cite{paul2008non}. The authors considered  a square-to-rectangle transition in the   setting of   Ginzburg-Landau model with kinetic energy  taken into account.  Numerical experiments with such a model,  imitating  cooling and heating runs,  showed   the intermittent  nature of   time series describing the  volume fraction of  the  emerging phase. It was also shown that  energy is dissipating in the form of bursts (avalanches)   whose  statistical distribution    follows a power law. To overcome some technical difficulties the authors  had to use   a phenomenological  truncated elastic kernel so the elastic interactions in this  model were effectively of a short range type. However,   even in such a reduced model, the presence of inertia was shown to ensure  that the robust scaling regime  can be reached  without  any involvement of quenched disorder.  Notwithstanding its  pioneering nature, this work  left  unanswered the questions about  the  role  of the  degree of underdamping,   the  effect of  antiferromagnetic  elastic interactions  and,   the potential importance of  elastic  incompatibility   \cite{ball2015incompatible}. 

In view of its role played  in what follows,  we recall that  the implied \textit{incompatibilit}y concerns the structure of the stretch tensor $\bold U$  that maps the undeformed 
austenite lattice to one of the equivalent variants of the undeformed martensite lattice. Specifically martensite ad austenite  phases are considered to be geometrically compatible if there exists a rotation $\bold R$ such that   $\bold R \bf U-\bf I$ is a rank-one tensor. Here  $\bf I$ is a unit tensor. As it is well known \cite{Bhattacharya2003}, in the case of compatibility there exist   stress-free coherent interfaces between  austenite and  martensite phase that  are  very mobile allowing the resulting 'thermoelastic'  transformation to proceed almost reversibly with only a minimal hysteresis \cite{bouville2007effect, zhang2009energy,atli2013influence, yang2023nonthermoelastic}. We note that such an algebraic compatibility of the energy wells should not be confused with the differential  elastic compatibility constraint imposed  on  the strain tensor \cite{Shenoy:1999ri,Ahluwalia2006-uz,paul2008non}.
 
The goal of the present paper, where we limit our attention to   athermal systems and, to avoid spurious time scales,  focus  exclusively on the case of quasi-static   driving, is twofold. First, we develop an intentionally oversimplified   model of an inertial  martensitic transformation that can be used to demonstrate  the crucial role of the  underdamping for reaching the scaling regime. This is achieved by  the systematic study of the \textit{crossover} between the limiting  overdamped and undamped regimes,  neither of which  by itself is compatible with power law scaling. Second, to complement the insights obtained from  the study of a prototypical  model, we  perform large scale numerical experiments  with   realistic continuum 2D and 3D models simulating some   basic martensitic transformations.  In particular, this allows us to  juxtapose  the results for models with  compatible and incompatible energy wells.  To compare numerical  predictions with experimental results, we use  in our 3D modeling   the parameters that match the available data for  single crystals of Fe$_{68.8}$Pd$_{31.2}$ undergoing  a martensitic cubic-to-tetragonal transition.  We show that  even a very  basic continuum model of this transformation accounting for both inertia and incompartibility, delivers   an excellent  numerical  reproduction  of the power law statistics of avalanches  with exponent  values  very close to the ones  measured in   acoustic emission (AE) experiments previously conducted on the corresponding crystals \cite{Bonnot2008-qg}. 

Our  important finding on the theoretical side, is that a simple 1D discrete snap-spring model is already capable of generating   heavy-tailed avalanche size distributions given that the value  of a non-dimensional parameter, quantifying the relative strength of inertia vs damping, is chosen in an appropriate   range.  Interestingly, the ability of such  a model to generate  intermittent fluctuations  is lost if one takes   a straightforward  continuum  limit.  The problem persists if  we move from  1D to 2D continuum setting, while maintaining elastic compatibility of the energy wells.   The heavy-tailed distribution  is  recovered  in the continuum setting only if we  advance  further towards  a 3D model  while also exacerbating the effect of incompatibility between the energy wells by accounting for a volumetric effect of the transformation. These observations  suggests that the  abundance of metastable energy wells,  emerging in  our  1D model due to its \textit{discreteness},   can be viewed as an oversimplified reproduction of the actual complexity of the energy landscape in a 3D continuum model with elastic \textit{incompatibility}.  We argue that both of these models effectively describe the same mechanism ensuring  'generic' scale invariance.
  
The rest of the paper is organized as follows. Our  prototypical 1D lattice model is introduced in Section II.  We analyze separately  damped,  undamped and  intermediate underdamped regimes and  show that only in the underdamped  regime the system exhibits intermittency and scaling.  We   compute numerically the corresponding power law exponents and confirm the validity of the  known  scaling relations. Our Section III contains the analysis of a  1D continuum analog of our discrete model obtained by taking a formal continuum (thermodynamic) limit. This model is shown to exhibit neither    intermittency nor  scaling.   In Section IV we consider a more realistic 2D  continuum model and in this way limit our attention  to the case of  compatible energy wells. Here again it turns out  that the desired self-organization to criticality is not achieved. Still we are able to present in the 2D case a revealing comparison  of  the transformation mechanisms in  overdamped  and  underdamped regimes showing  a remarkable difference in the complexity of the microstructures reachable by these two different types of dynamics.  Our Section  V is dedicated to the numerical study of a fully realistic 3D  model of the  cubic-to-tetragonal transition with incompatible energy wells.  We  show that in this case the system  manages to reach  the  scaling regime.   We  verify that the  computed  power law  exponents both, respect the known  scaling relations  and match the available experimental data. Our conclusions are summarized in the final Section VI where we also formulate some open questions.

\section
{One dimensional discrete model} 
Consider a one dimensional chain composed of mass particles linked by elastic springs. Suppose that each particle interacts with nearest (NN) and next to nearest neighbors (NNN) on each side. Denote by $u_i$ the horizontal displacement of a  particle   with index  $i=0,..., N$. We can then     write  the  total energy of the   chain  in the form 
 %\begin{equation} \label{E1}
\begin{equation}\label{E1}
 \mathscr E = \mathscr K+ \mathscr F,
 \end{equation}
%\end{equation}
 where the first term  is the kinetic energy 
\begin{equation}
\mathscr K= \frac{1}{2}\rho \epsilon \sum_{i=2}^{N-3}\dot u_{i}^2,
\end{equation}
and  the superimposed dot denotes partial time derivative, $\epsilon$ is the reference inter-particle distance, $\rho$ is the reference mass density.  The second term in \eqref{E1} is the 
elastic   energy 
\begin{equation}
 \mathscr F = \epsilon \sum_{i=2}^{N-3}  \left(\phi(e_i)+  \phi_1(e_i, e_{i-1}) \right).
\end{equation}
Here we introduced the elastic strain
 \begin{equation}
 e_i = \frac{u_{i+1}-u_{i}}{\epsilon}. 
  \end{equation}
   The functions  $\phi $ and  $\phi_1$  introduce nearest neighbor  (NN)  and  next to nearest neighbor  (NNN)  elastic interactions, respectively. 
   
    To model   the simplest  elastic  (martensitic) phase transition  we   assume that the NN  potential is bi-stable so that the function $\phi (e)$  has a double well structure with two energy wells  representing different elastic phases. For analytical simplicity   we use  in what follows   the simplest  piece wise quadratic function 
 \begin{equation}
  \phi (e_i)=\frac{\kappa}{2}(e_i - d)^2, 
 \end{equation}
   where $d=0$ for $e_i<e^c$ and $d=a$ for $e_i>e^c$. Here $\kappa$ is the  elastic modulus,   $a$  is the  transformation strain and  $e^c$ is the critical strain. 

To  capture  the ferromagnetic nature of the RFIM-type short range interactions in the simplest form we assume  that the  NNN   potential $\phi_1$ is  of the form   
 \begin{equation}
 \phi_1(e_i, e_{i-1})  =  \frac{\mu}{2\epsilon^2}(e_i-e_{i-1})^2, 
  \end{equation}
where $\mu>0$ is the   second order   elastic modulus.  

Finally, to account for \textit{environmental friction}  we introduce the standard  quadratic Rayleigh function 
 \begin{equation}
 \mathscr R   =\frac{\gamma}{2} \sum_{i=2}^{N-3} \dot u_i^2, 
 \end{equation}
where $\gamma$ is the effective viscosity coefficient.

 The dynamic equations describing  the resulting finite dimensional discrete mechanical system  can be writen the form   \begin{equation}
\frac{\delta \mathscr L}{\delta u_i}= \frac{\delta \mathscr R}{\delta  \dot u_i}
\end{equation} 
where \begin{equation}
\mathscr L=\mathscr K-\mathscr F
\end{equation}
is the Lagrangian and 
\begin{equation}
 \delta \mathscr L / \delta u  = \partial \mathscr L / \partial u- \partial / \partial t(\partial \mathscr L / \partial \dot u )
 \end{equation}
  is the  variational derivative.
We   assume  that  the system is loaded  in a hard device and therefore set the boundary conditions in the form
\begin{equation} \label{extrah}
u_0=u_1=0, \dot u_{N-1}= \dot u_{N-2} = v,
\end{equation} 
where $v$ is the   parameter characterizing the loading rate. 

%\subsection{Non-dimensionalization}

%Due to  weak non-locality of the ensuing model, the  boundary conditions imitating a hard loading device  must involve    specification of the displacements for two particles at each end. Accordingly, 

It will be convenient to  write the resulting system of equations in a dimensionless form. If we  normalize lengths using the  system size   $L=\epsilon N$ and  times  using  the    viscous time scale  $\tau = \gamma/\kappa$,  we obtain the system    
\begin{equation} 
\begin{array}{ll}
C \delta \ddot {\tilde u}_i =(\tilde\phi '((\tilde u_{i+1}- \tilde u_{i})/\delta) -\tilde\phi '((\tilde u_{i}-\tilde u_{i-1})/\delta) 
-   \dot  {\tilde u}_i\\
 +(G/\delta^3) (\tilde u_{i+2}+ \tilde u_{i-2}-4\tilde u_{i+1}-4\tilde u_{i-1}+ 6 \tilde u_i))).
\end{array}
 \label{3}
\end{equation} 
%\begin{equation}  \label{2}
%\begin{array}{ll}
%M\delta \ddot {\tilde u}_i =(\tilde\phi_1'((\tilde u_{i+1}- \tilde u_{i})/\delta) -\tilde\phi_1'((\tilde u_{i}-\tilde u_{i-1})/\delta)\\
%+(W/\delta)(\dot {\tilde u}_{i+1}+\dot {\tilde u}_{i-1}-2\dot  {\tilde u}_i) \\
%+(E/\delta) (\tilde u_{i+2}+ \tilde u_{i-2}-4\tilde u_{i+1}-4\tilde u_{i-1}+ 6 \tilde u_i)),
%\end{array}
%\end{equation} 
where $\tilde u_{i }=  u_{i }/L$, $\tilde t=t/\tau$  and  $\tilde\phi = \phi/\kappa$. The   dimensionless parameter characterizing the  degree of discreetness is  
 \begin{equation}
 \delta=  \frac{1}{N}. 
 \end{equation}
The relative role  of inertia vs damping is measured by the non-dimensional parameter 
 \begin{equation}
 C=\frac{\rho \kappa L^2}{\gamma^2}.
 \end{equation}
The   effect of NNN interactions scales with dimensionless parameter 
  \begin{equation}
 G = \frac{\mu}{\kappa}. 
 \end{equation}
Yet another dimensionless ratio entering the problem through the boundary conditions characterizes  the rate of loading
  \begin{equation}
 V=\frac{v \gamma}{\kappa L}. 
 \end{equation}
%$M= v^2\rho/\kappa$  and  $W= \gamma v/(L\kappa)$, respectively.
% Finally, 
%Note that  both $M$ and $W$  approach  zero  in the quasi-static  limit which is of our main interest in this paper. therefore, it will be convenient to work with  the ratio of these  parameters $C=M/W^2$ is independent of the driving rate $v$  rewrite the   system \eqref{2} in the form
%$$
%\begin{array}{ll}
%CW^2\delta \ddot {\tilde u}_i =(\tilde\phi_1'((\tilde u_{i+1}- \tilde u_{i})/\delta) -\tilde\phi_1'((\tilde u_{i}-\tilde u_{i-1})/\delta)\\
%+(W/\delta)(\dot {\tilde u}_{i+1}+\dot {\tilde u}_{i-1}-2\dot  {\tilde u}_i)\\
% +(E/\delta) (\tilde u_{i+2}+ \tilde u_{i-2}-4\tilde u_{i+1}-4\tilde u_{i-1}+ 6 \tilde u_i))).
%\end{array}
%$$
%OLDPICTURES
%\begin{figure}
%\subfigure[][]{\label{ex1}\includegraphics[scale=0.2]{./fig2a.pdf}}
%\subfigure[][]{\label{se1}\includegraphics[scale=0.2]{./fig2b.pdf}}
%%\subfigure[][]{\label{tex1}\includegraphics[scale=0.15]{./abscissa_2/durations.pdf}}\\
%%\subfigure[][]{\label{tex2}\includegraphics[scale=0.15]{./abscissa_2/surfaces.pdf}}
%%\subfigure[][]{\label{tex3}\includegraphics[scale=0.15]{./abscissa_2/evst3.pdf}}
%\caption{\label{p1ex}\small}
%\end{figure}

In what follows we consider   the behavior of  the   system \eqref{3}   under the assumption that the two non-dimensional parameters $V$ and $\delta$ are small but finite.  The limit $V \rightarrow 0$ corresponds  to quasi-static driving  and in our numerical experiments  we use the value  $V \sim 10^{-5}$.  In the limit $\delta \rightarrow 0$  we obtain continuum  model, however,   in all our numerical experiments  we use  a small but finite value  $\delta \sim 10^{-3}$. 
The  parameter  $G$,  characterizing the strength of ferromagnetic short range interactions, is  kept   sufficiently small to ensure that  phase boundaries are   localized; specifically   throughout the paper we use the value $G=10^{-4}$.  

In contrast, the parameter $C$, playing the role analogous to a Reynolds number in the theory of turbulence  as it characterizes  the ratio of inertia to dissipation, is  varied in a broad  range $C \sim 10-10^5$. In this way we  are  able to cover  different regimes,  from fully damped to fully  undamped. The goal is to obtain a phase (regime) diagram and  search for   the range  where the system may exhibit power law scaling.  To increase statistics we studied  cyclic loading and in each of the regimes performed hundreds of cycles.
%More systematic study, involving the analysis of the whole parameter space,  will be presented separately.
 
To solve the equations of motion \eqref{3}  numerically we carried out  a time-stepping approach  using a fourth-order  explicit predictor-corrector algorithm. For    spatial discretization  we  used  the FFT method.   The numerical algorithm is discussed in more detail in  Appendix \ref{app:discretisation}.
% see also \cite{Salman2009-qv}.
 
% \subsection{ Observables }
 In an attempt to reproduce the results of AE experiments  \cite{Bonnot2008-qg}, we recorded  at each avalanche the associated  \textit{dimensionless}  energy dissipation.  We started with discretization of the   time domain  $[0, T]$  introducing  the  points  $ t_a = a \Delta t,$ where $ a = 0, 1, 2, \ldots, N$  and  $N =  T/\Delta t$.  As  the system is being loaded,  an avalanche begins when  at least one NN spring  starts changing its energy well and we suppose that this takes place   at  the  time  $t=t_i$. During the avalanche the  energy dissipation $\mathscr R(t_a)$   is recorded at each time  step $ t=t_a $ between the time $t=t_i$  and the  avalanche  termination time  $t=t_f$ when the energy wells are not switched any more.  We   then compute for such an avalanche  the total  dissipation 
\begin{equation} \label{e2} 
E = \sum_{t=t_i}^{t_f}  \mathscr R(t_a)\Delta t,
\end{equation}
where  the summation is over the corresponding discrete time points  \textbf{$t_a$}. In what follows  we associated   dissipation only  with the fluctuating part of the displacement field 
 \begin{equation} 
 w_i(t)=  u_i(t)-\frac{i+1}{N}  u_{N-1}(t).
 \end{equation}

While it may be  natural to adopt as a measure of  avalanche size   the total transformation strain, 
%$
%  \sum_{i=2}^{N-3} |\dot e_i|\Delta t,
%$
one can  show  \cite{Bonnot2008-qg} that  an almost equivalent result  is  obtained if we use instead a closely related  quantity with the same dimensionality 
   \begin{equation}
S = \sum_{t=t_i}^{t_f} V(t_a)\Delta t,
\label{defs1}
\end{equation}
 where 
\begin{equation}
  V(t) = \mathscr R(t)^{1/2}.
  \label{defs2}
  \end{equation}
  In addition to the two macroscopic variables $E$ and  $S$,  which are computed for  each  avalanche,  we  also record  the third macroscopic observable $T$ defined as  the duration   of  same avalanche 
   \begin{equation}
   T= t_f-t_i.
    \end{equation}
In this way we can check the validity of various general scaling relations  linking   fluctuations in the observables $E$, $S$ and $T$.

 To complement  the  avalanche-related information, we also studied  a different type of observable which   is not tied to  intermittent structure of the AE signal. Specifically,   we recorded the power spectrum  of the time series $V(t)$ defined by the formula  \cite{kertesz1990noise,PhysRevB.62.11699,laurson2005power,laurson20061,nandi2022scaling,ruseckas2014scaling}:  
% Kertész, J., & Kiss, L. (1990). The noise spectrum in the model of self-organised criticality. Journal of Physics A: Mathematical and General, 23(9), L433.
%Laurson, L., Alava, M. J., & Zapperi, S. (2005). Power spectra of self-organized critical sandpiles. Journal of Statistical Mechanics: Theory and Experiment, 2005(11), L11001. 
%Kuntz, M. C., & Sethna, J. P. (2000). Noise in disordered systems: The power spectrum and dynamic exponents in avalanche models. Physical Review B, 62(17), 11699. 
%Laurson, L., & Alava, M. J. (2006). 1∕ f noise and avalanche scaling in plastic deformation. Physical Review E—Statistical, Nonlinear, and Soft Matter Physics, 74(6), 066106.
%Nandi, M. K., Sarracino, A., Herrmann, H. J., & de Arcangelis, L. (2022). Scaling of avalanche shape and activity power spectrum in neuronal networks. Physical Review E, 106(2), 024304.
%Ruseckas, J., & Kaulakys, B. (2014). Scaling properties of signals as origin of 1/f noise. Journal of Statistical Mechanics: Theory and Experiment, 2014(6), P06005.
  \begin{equation} \label{PS}
PS(f) =\biggl\lvert\sum_{k=1}^{n} V_ke^{-i2\pi f k } \biggr\rvert^2,
\end{equation}
where  $V_k=V(t_k)$ is an abstract dimensionless  discrete signal  with   $t_k=k/n$ and $i=1,...,n$.   

It is clear that by  analyzing   the function \eqref{PS}  one can   identify  the presence   of  correlations in the time series even in the absence of intermittency and avalanches. Indeed, consider an  auto-correlation function of the  discrete signal $V_k$     
\begin{equation}
 C(s) = \frac{1}{(N-s)\langle  V_k^2\rangle}\sum_{k=1}^{n-s}  V_k  V_{k+s},
 \end{equation} 
 where
  \begin{equation}
 \langle  V_k \rangle= \frac{1}{n}\sum_{k=1}^{n}V_k. 
 \end{equation} 
While  in the case of a fully  uncorrelated  white  noise   $C(s)=0$,   in the presence of short-range correlations one can expect  an exponentially decaying behavior  $C(s)\sim \exp(-s/\tau)$ (Lorentzian  spectrum).   Of particular interest to us will be the time series exhibiting long-range correlations.  In this case one can   expect   a  slower, power law decay behavior  for  both, the auto-correlation function
 \begin{equation}
 C(s)\sim s^{-\gamma}.
  \end{equation} 
We'll be using the fact that  the corresponding power spectrum  is
    \begin{equation}
 PS(f) \sim 1/f^{\alpha}.
  \end{equation}    
with  $\alpha=1-\gamma$; in the limiting  case   $\alpha=1$ the auto-correlation function  is known to exhibit   a slower  logarithmic decay \cite{kay1993fundamentals}.

\subsection{Overdamped regimes}
The regime where dissipation completely  overcomes inertia corresponds to the damped limit $C \to 0$. This regime in our 1D discrete setting has been studied analytically before, see for instance \cite{Truskinovsky2004-xp,Puglisi2005-lg,Mielke2011-ck,Salman2012a}. It was shown that under  cyclic quasi-static loading the system  exhibits rate independent hysteresis with individual springs changing phase (energy well) sequentially, one after another. Dissipation can be then represented as a sequence of periodically spaced identical events (trivial avalanches). Here we present for completeness the results of   numerical experiments  in the overdamped regime which  corroborate  these  analytical findings.
\begin{figure} [h!] 
\includegraphics[scale=.2]{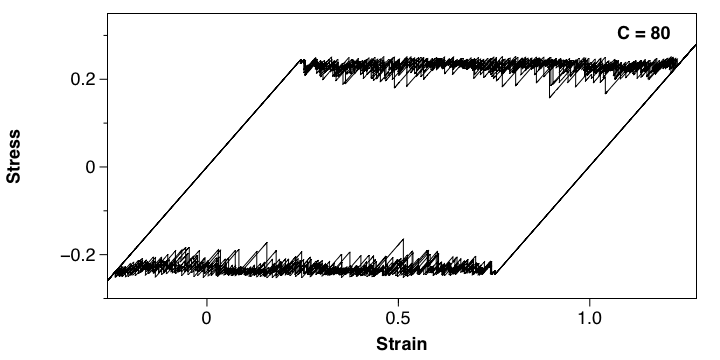}
\caption{Hysteretic stress-strain response of an overdamped chain subjected to quasi-static cyclic loading in a hard device.  Parameters:  $N=1000$ and $C=80$. }
\label{S1}
\end{figure}
 
% Due to the oversymmetric nature of our system  one can expect in this regime certain dynamic degeneracy when the perfectly elastic stage ends and all springs reach  the elastic stability (spinodal) limit.    In such point the  system  can be expected to exhibit    extreme sensitivity    \cite{10.1007/1-4020-3559-4_16,Mielke2011-ck}, however the problem disappears  we  take the value of $C$ to be small but finite. 
 
% which in numerical simulations  would  be identical for each spring  only with some finite level of accuracy. Therefore the deterministic approach  in the presence of  such massive instability will no longer be adequate
 
% and also replace   in \eqref{3}  the  visco-elastic dissipation  by  the   environmental  viscous  dissipation which does not   affect any of our main conclusions (\textbf{ why this replacement is needed, what is going on here? need to clarify!}).  Specifically, we assumed that in this subsection only the Rayleigh function is chosen in a different form  
%   \begin{equation}
% \mathscr R   =\frac{\gamma}{2}\sum_{i=2}^{N-3} \dot u_i^2, 
% \end{equation}
%where $\gamma$ is again the corresponding viscosity coefficient. 
%
%We therefore  instead of  \eqref{3} consider  a closely related but different equation 
% \begin{equation}  \label{4}
%\begin{array}{ll}
%C \delta \ddot {\tilde u}_i =(\tilde\phi '((\tilde u_{i+1}- \tilde u_{i})/\delta) -\tilde\phi '((\tilde u_{i}-\tilde u_{i-1})/\delta) 
%-  \dot {\tilde u}_{i} \\
% +(G/\delta^3) (\tilde u_{i+2}+ \tilde u_{i-2}-4\tilde u_{i+1}-4\tilde u_{i-1}+ 6 \tilde u_i))).
%\end{array}
%\end{equation} 
 In our numerical experiments we  loaded   the system   from a homogeneous state where   springs were in the first   energy well until all   springs have switched   into the second energy  well and the system reached again  a  homogeneous configuration. We then unloaded  the chain following the same protocol with both loading and unloading   performed quasistatically. The details of our numerical implementation  can be found in Appendix \ref{app:discretisation}.
\begin{figure} [h!] 
\includegraphics[scale=.33 ]{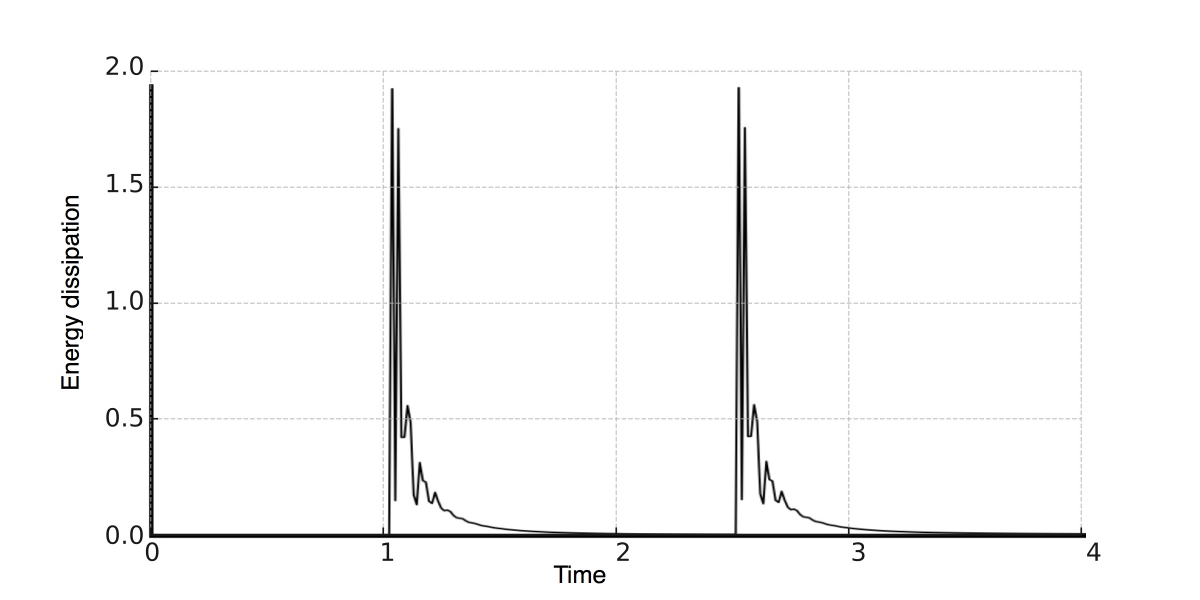}
\caption{Two consecutive individual avalanches in the overdamped regime which is shown in Fig. \ref{S1}.  Parameters:  $N=1000$ and $C=80$.   }
\label{S2}
\end{figure}
 
We found  that in the range  $C \sim 20-80$  the  dissipation is still sufficiently strong to deliver the expected over-damped hysteretic response.  The computed  strain-stress relation is shown in Fig. \ref{S1}. In the horizontal yielding segments of the hysteresis cycle   the system exhibits a succession of  almost equal size stress drops.  The typical individual  avalanches, responsible for such drops, are shown in  Fig. \ref{S2}. The avalanches  are separated by   the  quiescent intervals and  each avalanche has  roughly the same shape with  an   exponentially decaying tail. While in  the studied  overdamped regimes with $C \sim 20-80$  only a very small number of springs  switch   from one energy  well to another during each  avalanche. 

An intriguing feature of  Fig. \ref{S1} is the  absence of the nucleation peak predicted theoretically for NNN system in \cite{Truskinovsky2004-xp}. Such a peak would correspond  to a collective nucleation event when many springs change phase simultaneously.  It allows subsequent  transformation to proceed through  a growth of a nucleus and  implies  spring-by-spring advance of a   phase boundary. There are several reasons why the situation here may be   different. First of all,  we use  "extra hard device" boundary conditions \eqref{extrah}, which prevent  the formation of a nucleus at the boundary of the domain. Then, instead of viscoelastic dissipation which  allows one to close  the problem in terms of strain variables only, we use environmental viscous friction formulated  in terms of displacements. Note that the latter  represent  nonlocal functions of strains. Finally, we performed our numerical experiments at small but finite inertia which apparently  also contributed to the  inhibition  collective nucleation. As we show in what follows, the  reduction of  friction leads to   re-emergence  of  the nucleation peak, however, it will then  represent   a different physical phenomenon vis-a-vis the nucleation peak in a purely viscoelastic overdamped system.

 \begin{figure} [h!] 
\includegraphics[scale=.2]{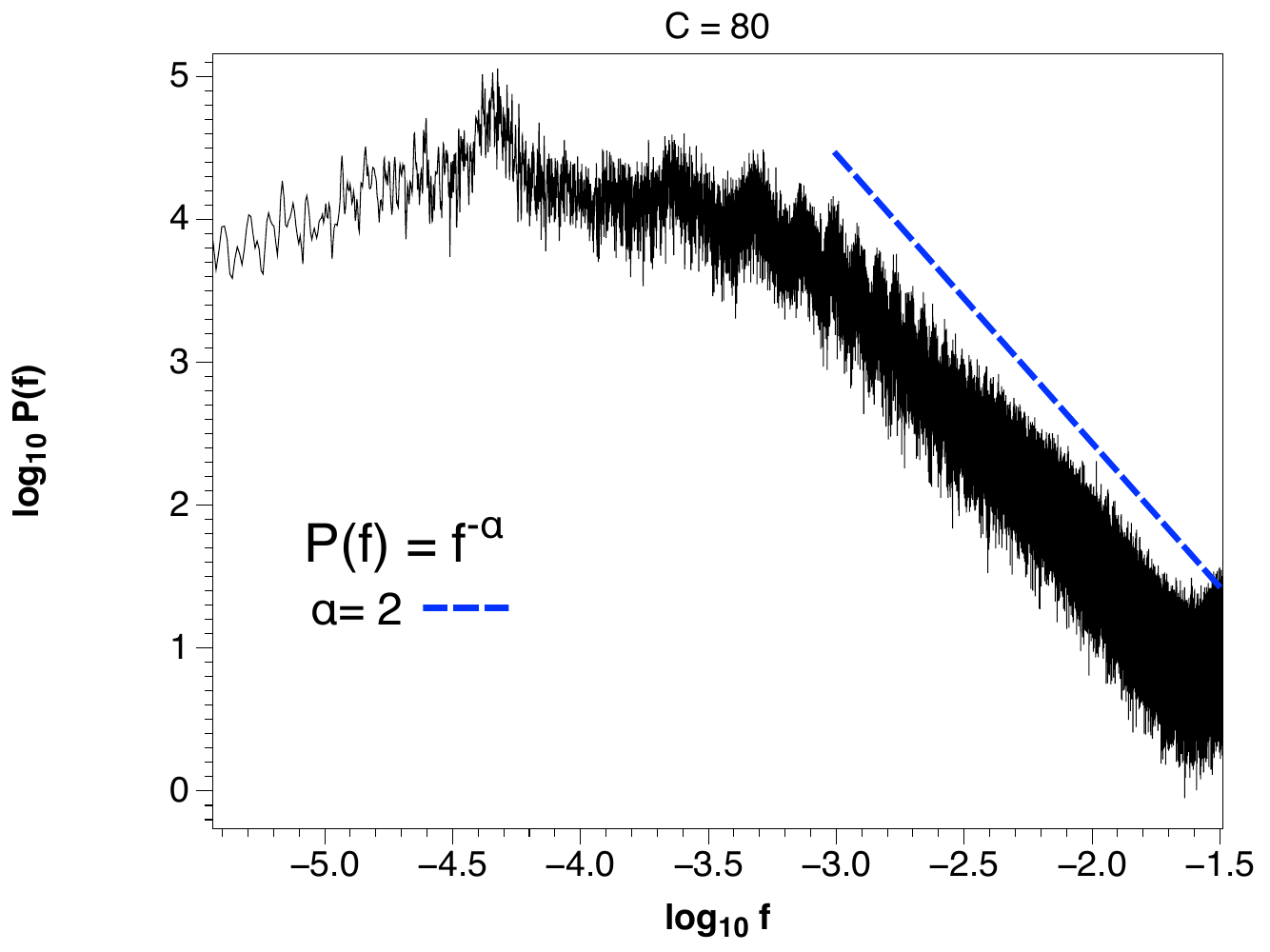}
\caption{Power spectrum  of the signal $V(t)$  in the overdamped regime which is shown in Fig. \ref{S1}.  Here $C=80$ and $N=1000$.}
\label{S3}
\end{figure}

One can see that   transformational yield in  Fig. \ref{S1}  takes the form of   a sequence   of similar instability events,   each one  representing a  transition from   a \textit{marginally} stable state  to the  nearest  meta-stable state.  Then, in the limit $\delta \to 0$ the yielding emerges  as  a collection of infinitely many events that are  all infinitely small.  One can say that  in this limit  all  of these events merge together and   the system can be viewed as  remaining all the time in a \textit{marginally} stable state.  The emerging coarse grained  macroscopic    dissipative  potential   is known to be a (non-Onsagerian) homogeneous function of degree one which replaces  the  classical  Onsagerian   dissipative potential operative at the   level of individual mass points \cite{Puglisi2005-lg,Truskinovsky2005-tq}.

The power spectrum \eqref{PS} computed in this regime is  shown in Fig. \ref{S3}.  At high frequencies  it exhibits  a range of  the power law behavior  
 \begin{equation} 
PS(f) \sim 1/f^2. 
\end{equation} 
There is  an almost  flat segment  at  small frequencies with perhaps a single  characteristic frequency expressed in a more pronounced way. 
\begin{figure} [h!] 
\includegraphics[scale=.2]{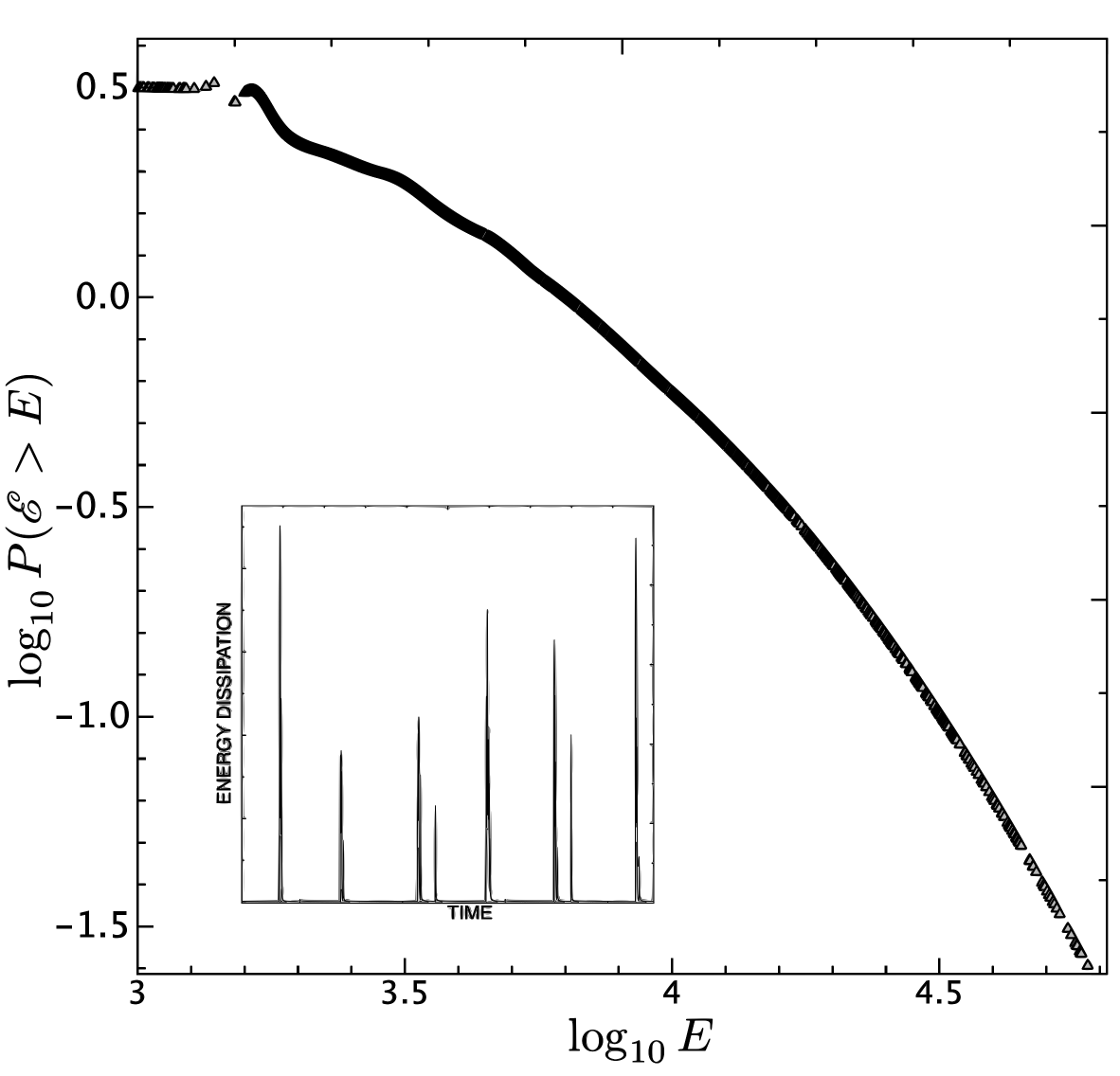}
\caption{Cumulative probability distribution representing statistics of avalanches in the overdamped   regime which is shown in Fig. \ref{S1}.  Here $C=80$ and $N=1000$.  Inset:  several consecutive  avalanches  occurred in a short time interval. }
\label{S4}
\end{figure}

 To explain these observations we turn to  Fig. \ref{S2}   showing the structure of  two consecutive  avalanches.  
Observe that each avalanche  can be approximated  by an exponentially decaying function 
 \begin{equation} \label{R1}
 R(t) = R_0e^{-\lambda t}, 
  \end{equation} 
   where $\lambda$ is the decay rate which appear to be the same for both avalanches. The   Fourier transform of \eqref{R1} is given by the formula
    \begin{equation} 
    \hat R(f)  = \frac{R_0}{\lambda +if} 
     \end{equation} 
     and  the corresponding   power spectrum takes the form  
      \begin{equation} 
      PS(f) = \frac{R_0^2}{\lambda ^2+f^2}.
      \end{equation} 
One can see that both    regimes visible  in Fig. \ref{S3}   are captured by this formula, which suggests that    the computed   power spectrum    is a   representation of  a sum of almost identical individual avalanches of exponential shape.  Note also that  the transition between high and low frequency regimes  is positioned around the frequency characterizing the   decay  rate of the exponential tail. The presence of a characteristic frequency in the low frequency range may be the  sign of the superimposed  'ringing' in the system due to   small but nonzero inertia in the system.  

 Finally, in Fig. \ref{S4} we show the cumulative probability distribution of avalanche sizes represented by the variable $E$. One can see that the distribution is localized   on a small interval of avalanche sizes with an exponential cut off tail.  This   suggests  that in the overdamped regime avalanches   are   over-correlated  vis-a-vis  a  potential  power law scaling regime  with    individual avalanches  being  both almost equidistant and having almost the same size. We  can conclude that  in  the ovedamped regime   small inertia is almost invisible,  playing   the  role   of a  quenched quasi-Gaussian  disorder, see also ~\cite{Salman2012a}.

\subsection{Underdamped regimes }

The fully undamped limit $C \rightarrow \infty$ with neglected NNN interactions ( $G=0$) was studied analytically and numerically in \cite{Efendiev2010-kq}. In this case there is no hysteresis and in cyclic quasi-static  loading the system  exhibits, after a one-cycle  transient, a reversible behavior reminiscent of  entropic elasticity. 
This means that already during the first cycle the system fully thermalizes after the system departs from  the homogeneous elastic branch. In other words, a state of equipartition is reached in the course of the development of  the associated spinodal instability.  After reaching the state of thermal equilibrium  the system   behaves  under quasi-static driving as an adiabatically loaded  thermoelastic body. Our numerical experiments fully corroborate these analytical and numerical findings. 

\begin{figure} [h!] 
\includegraphics[scale=.35]{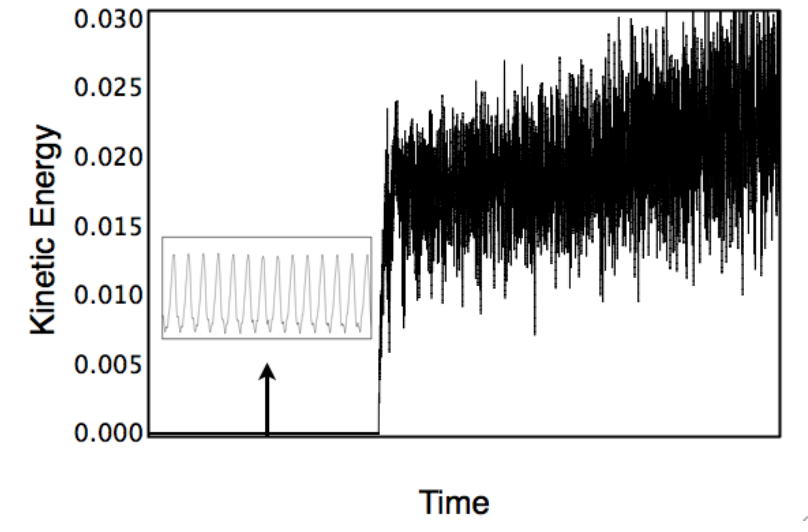}
\caption{Evolution of  an undamped discrete chain subjected to quasiststic loading in a hard device:  the first loading cycle. Here $N=1000$ and $C=100000$.  Inset: a zoom on the kinetic energy evolution in the pre-transition time interval. }
\label{Q2}
\end{figure}

The response shown in  Fig.~\ref{Q2}, where $C=100000$, is very close to the behavior of the system with  $C =\infty$. The latter   is described by the  equation 
\begin{equation} \label{eq5}
 \begin{array}{ll}
 \delta \ddot {\tilde u}_i   =    (\tilde\phi '((\tilde u_{i+1}- \tilde u_{i})/\delta) -\tilde\phi '((\tilde u_{i}-\tilde u_{i-1})/\delta) + \\
 (G/\delta^3) (\tilde u_{i+2}+ \tilde u_{i-2}-4\tilde u_{i+1}-4\tilde u_{i-1}+ 6 \tilde u_i),  
 \end{array}
 \end{equation}
where the time scale  is chosen differently than in the case $C\neq 0$. Specifically,    we  assumed in \eqref{eq5} that  $\tau=\sqrt{\rho/\kappa}L$ and $V=v \sqrt{\rho/\kappa}$. We again started with   a homogeneous  state where   all springs were in a single well and then loaded the  system quasi-statically (and effectively isentropically) in a hard device.   
According to Fig. \ref{Q2} the originally 'cold' system  first deforms homogeneously  while maintaining an   affine configuration. One can see, however, the  presence of superimposed small elastic 'ringing' because our loading is still characterized  by small but finite rate. When the system reaches  the elastic instability limit, the  elastic  homogeneous state breaks down giving rise to  a complex dynamical regime. 
\begin{figure} [h!]
 \includegraphics[scale=.08]{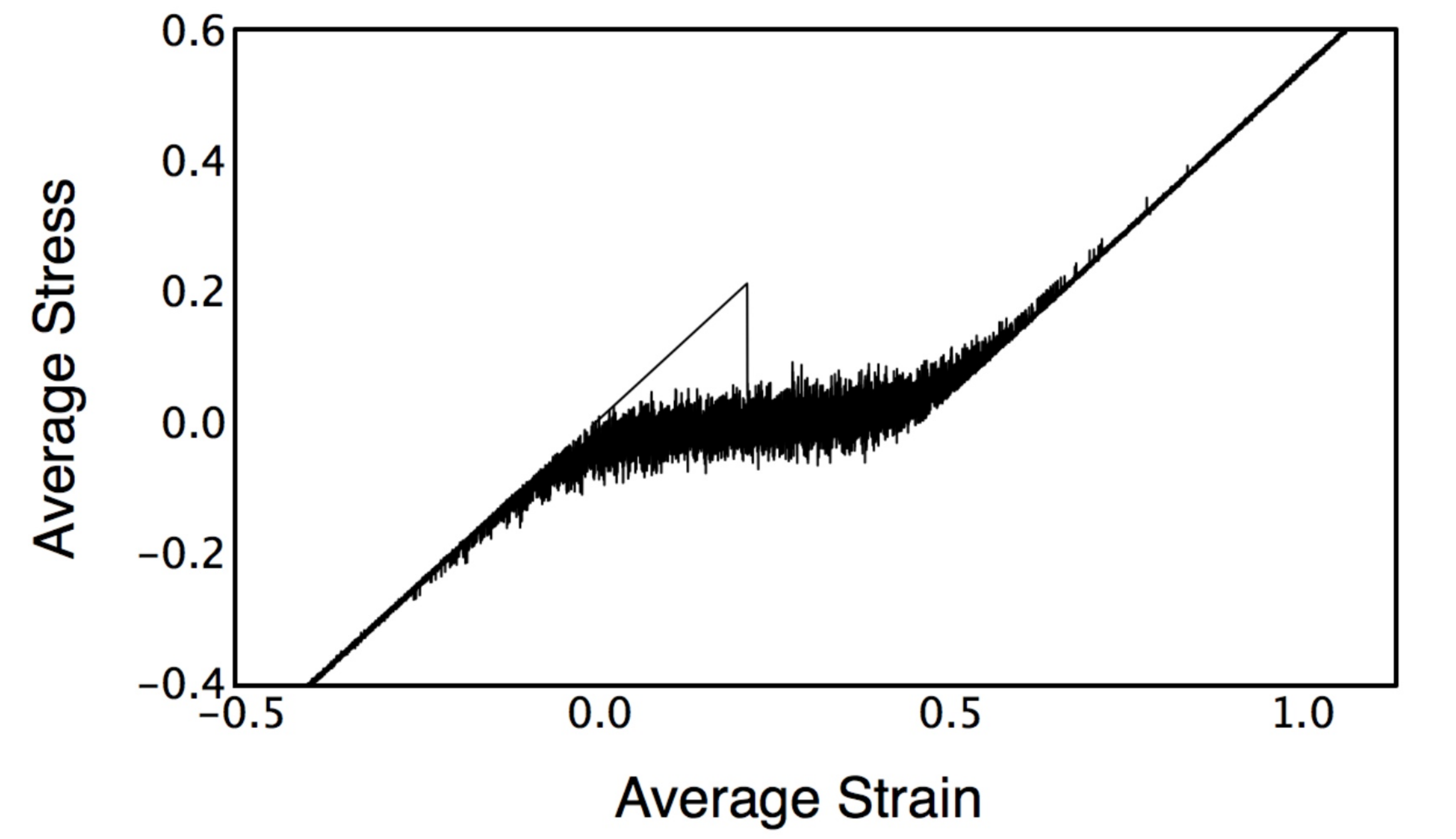} 
%\subfigure[][]{\label{ex3}\includegraphics[scale=.25]{C80_copy}}
%\subfigure[][]{\label{ex2}\includegraphics[scale=0.22]{C8000_ss copy}}
\caption{Stress-strain response  of  an undamped discrete chain subjected to quasiststic cyclic loading in a hard device. Here $N=1000$ and $C=100000$.}
\label{Q1}
\end{figure}

 As loading continues beyond the ensuing massive instability, we observe a quasi-deterministic  average stress-strain response with  superimposed chaotic fluctuations, see Fig. \ref{Q1}.  In accordance with \cite{Efendiev2010-kq}   all  modes acquire exactly the same energy which allows one to introduce the idea of temperature. The subsequent  quasi-thermoelastic, hysteresis-free averaged behavior proceeds in full agreement with the formulas of classical thermodynamics describing the corresponding  adiabatic processes. See   \cite{Efendiev2010-kq} for  a fully explicit analysis  in the  case of bi-quadratic NN potential and $G=0$ when both  free energy and entropy can be computed analytically.

%More specifically, all springs start to oscillate simultaneously and vigorously giving raise to the overall macroscopic thermoelastic behavior  with all modes acquiring the same  temperature.
%This phenomenon is illustrated in Fig. \ref{Q2} where we show how, during the first proportional loading segment  the original small 'ringing' associated with a finite rate of loading  of  linear elastic system  turns into a chaotic  dynamic regime.
%
%
%Subsequent  quasi-static cyclic deformation proceeds reversibly with the averaged behavior  in full agreement with the formulas of classical thermodynamics of adiabatic processes, see Fig. \ref{Q1} We note that in the case of bi-linear nonlinearity this behavior is amendable to a fully explicit analysis \cite{Efendiev2010-kq}. 
%

According to  Fig. \ref{Q1}, at  $C=100000$ the overall behavior of the system is practically  indistinguishable from the limiting case $C=\infty$.  While   in the  limiting  Hamiltonian  regime the actual dissipation  vanishes,  in the analysis of the underdamped regime we can still  resort to the direct study of the  quantity
 \begin{equation}
 V(t) = \sqrt{\sum_i(\dot u_i)^2 }. 
 \end{equation} 
 where summation is over all the mass points. Therefore, the statistical study of environmental dissipation in the underdamped system is somewhat similar to the study  of temperature fluctuations  in the limiting undamped limit.  
  
\begin{figure} [h!] 
\includegraphics[scale=.2 ]{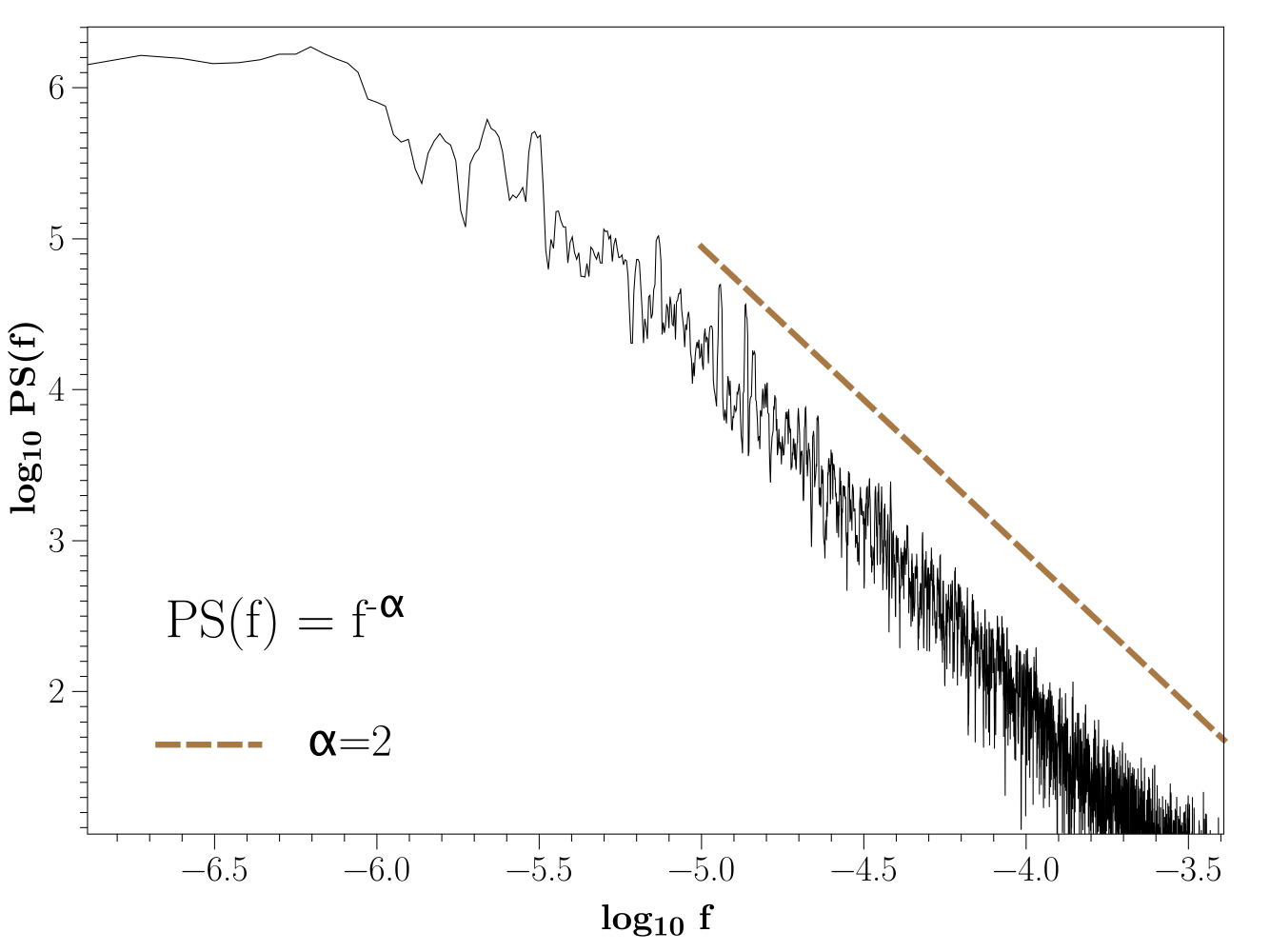}
\caption{Power spectrum  of the signal $V(t)$  for the undamped discrete chain subjected to quasiststic cyclic loading in a hard device, which is shown in Fig. \ref{Q1}.  Here $C=100000$ and $N=1000$.}
\label{Q3}
\end{figure}

 The corresponding power spectrum \eqref{PS}, which is not accessible by purely thermodynamic analysis, is shown in  Fig. \ref{Q3}. We  observe fully  uncorrelated equilibrium fluctuations   at  small frequencies  producing   a flat part of the spectrum  $PS(f) \sim 1/f^{\alpha}$ with $\alpha =0$. This indicates that long wave collective modes are effectively subjected to white noise. At larger  frequencies the power law behavior persists with exponent stabilizing at the value  $\alpha=2$. Such behavior is indicative  of the underlying Brownian motion of individual mass points which can be interpreted as a  Brown noise. It suggests that  these points  are subjected to fully uncorrelated random forces  which should be  expected in the state of thermal equilibrium. More generally, the Lorentz-like broad structure of the computed power spectrum is indicative of  an exponential decay of correlations in the time domain which in turn points to  fast relaxation times characteristic of strongly chaotic systems involving a wide range of frequencies. 

%\begin{figure} [h!] 
%\includegraphics[scale=.2]{differentCs.pdf}
%\caption{}
%\label{Q3new}
%\end{figure} 

\begin{figure} [h!] 
\includegraphics[scale=0.9]{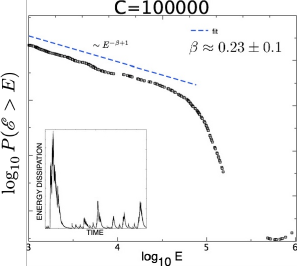}
\caption{Cumulative probability distribution for the magnitude of the effective energy dissipation during individual abalanches in  the undamped discrete chain subjected to quasiststic cyclic loading in a hard device, which is shown in Fig. \ref{Q1}. Inset shows a fragment of the actual time series.  Here  $C=100000$ and $N=1000$.}
\label{Q4}
\end{figure}

Even though we largely lose correlations and intermittency in the  underdamped regime, some  analog of  avalanche distribution   can be still constructed from the obtained time series by appropriate thresholding. In  Fig. \ref{Q4}  we show  the  distribution of the implied (quasi) avalanches where the quantity   $V^2(t)$  was used  as the  analog  of  dissipation.   According to Fig. \ref{Q4}, the resulting distribution is close to Gaussian with an almost  flat behavior at small event sizes. We only show here the distribution beyond the lower bound $x_{\text{min}}$ determined by minimizing the Kolmogorov-Smirnov distance.    At intermediate range we observe a small value of the   exponent  $\beta \approx 0.23$ which suggests  that inertia leads to  over-correlation among   quasi-avalanches.  At  larger event sizes we observe  a  characteristic  exponential cut off reflecting the finite size of the  system. 

To summarize, while we have seen that in the overdamped regimes  the dynamics is close to being fully deterministic and regular,  which suggests very limited  complexity,  in the underdamped regimes  the dynamics is close to being fully chaotic,   which  again  means that  complexity is  minimal. As we show below, more interesting structure of correlations emerges in the intermediate regimes where the system in neither overdamped no underdamped.

\subsection{Intermediate regimes}

At finite values of the parameter $C$  the direct integration of the system  \eqref{3} subjected  to quasi-static driving produces intermittent dynamics with avalanches of widely different  sizes.  Moreover, as we show below, in a well defined interval of the values of $C$,  the distribution of avalanche  sizes  exhibits a range of  power law behavior indicating   scale free internal organization of dynamic microstructure.

%\begin{figure}
%\subfigure[][]{\label{tex1}\includegraphics[scale=0.15]{./abscissa_2/durations.pdf}}
%\subfigure[][]{\label{tex2}\includegraphics[scale=0.15]{./abscissa_2/surfaces.pdf}}
%\subfigure[][]{\label{tex3}\includegraphics[scale=0.15]{./abscissa_2/evst.pdf}}
%\caption{\label{p1ex}\small}
%\end{figure}

%In this subsection, we present results for a finite value of $C$.  In this regime, we observe that avalanches may have different sizes depending on the value of $C$. To characterize avalanches, we define (an irrelevant) threshold and we measure the total dissipated energy above  and  we calculate the resulting probability distribution of events.
 
 \begin{figure} [h!] 
\includegraphics[scale=0.22 ]{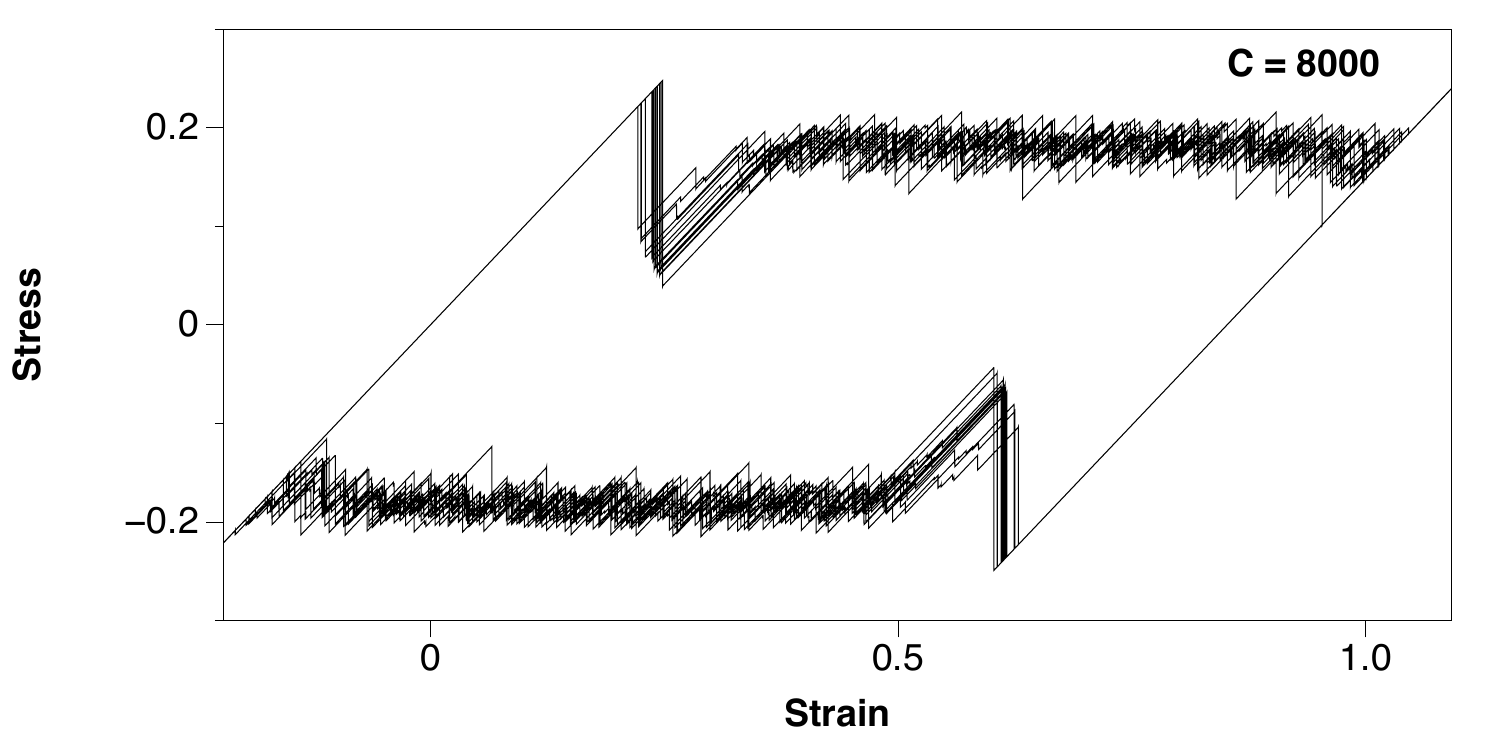}
\caption{Stress-strain response of a quasistatically driven chain  with inertia and dissipation  balanced at $C=8000$. Here  $N=1000$.}
\label{P3}
\end{figure}

A  characteristic stress-strain response  of the system at $C=8000$  under cyclic  loading over 20 cycles   is shown  in Fig. \ref{P3}.  In each cycle   we observe recurrent    system size characteristic event  when initially homogeneous (affine) configuration is breaking down due to a massive elastic instability. Using the language of plasticity theory, one can say that the system undergoes brittle yielding \cite{Zhang2020-ax, Perez-Reche2008}.

According to  Fig. \ref{P3}, after the first   system size avalanche   the subsequent avalanches are all of smaller size  and  in each cycle  the system apparently reaches  a  steady yield regime. While the average stress in this regime is maintained at an almost constant level, we observe  a broad, heavy-tailed distribution of stress drops.

\begin{figure} [h!] 
\includegraphics[scale=0.22]{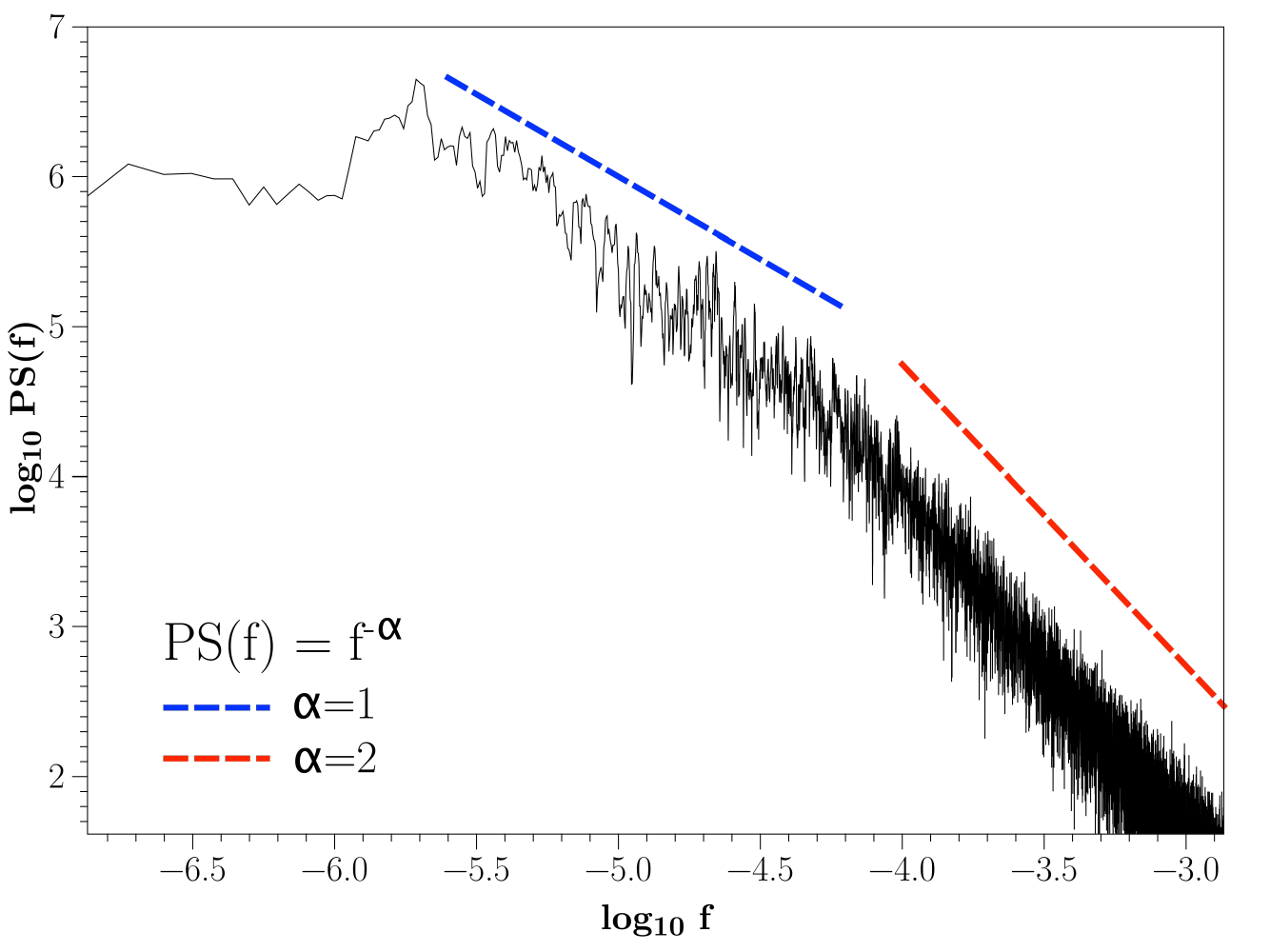}
\caption{Power spectrum of  the underdamed system shown in Fig. \ref{P3}. Here   $C=8000$ and $N=1000$.}
\label{P2}
\end{figure}  

The computed  power spectrum  is shown  in Fig. \ref{P2}, where one can  identify  three   regimes corresponding to   small, intermediate and large  frequencies. In the range of small frequencies the spectrum  is  flat  indicating the absence of persistent system size collective  correlations. The characteristic times associated with  such  frequencies, which  would have characterized the durations of the corresponding correlated events, are much bigger than the durations of the largest avalanches. It suggests that this range has nothing to do with intermittency and the fluctuational response is reminiscent of what we have seen in the underdamped regime. In the range of large frequencies, we  again see  the Lorenzian behavior with  $1/f^2$ type decay which is similar to what we have seen  in the overdamped regime. This is  apparently due to   the dominant role played in this range by the time scale  of   viscous dissipation and it  reflects  the relaxational behavior  of individual masses at the end of each avalanche.

The most interesting behavior is observed in the range of intermediate frequencies, where we  see a  power law decay  with a nontrivial exponent $\alpha=1$  characteristic of a so-called 1/f noise \cite{West1989-gr,Milotti2002-pt,kertesz1990noise,PhysRevB.62.11699,laurson2005power,ruseckas2014scaling}. 
The associated  frequencies correspond roughly to the   durations of avalanches  whose magnitudes are inside  the  'inertial' (power law) range, see below. Note   that a similar 1/f noise has been also recorded in the closely related discrete models of crystal plasticity  \cite{Salman2012a,laurson20061}.  If small  avalanches   die off in the intermediate  regime,  in the  range of frequencies where   $PS(f) \sim 1/f$, such a decay is practically invisible because at these scales dissipation is negligible and the structure of avalanches is shaped by dynamics as much as by dissipation.
\begin{figure} [h!] 
\includegraphics[scale=0.3]{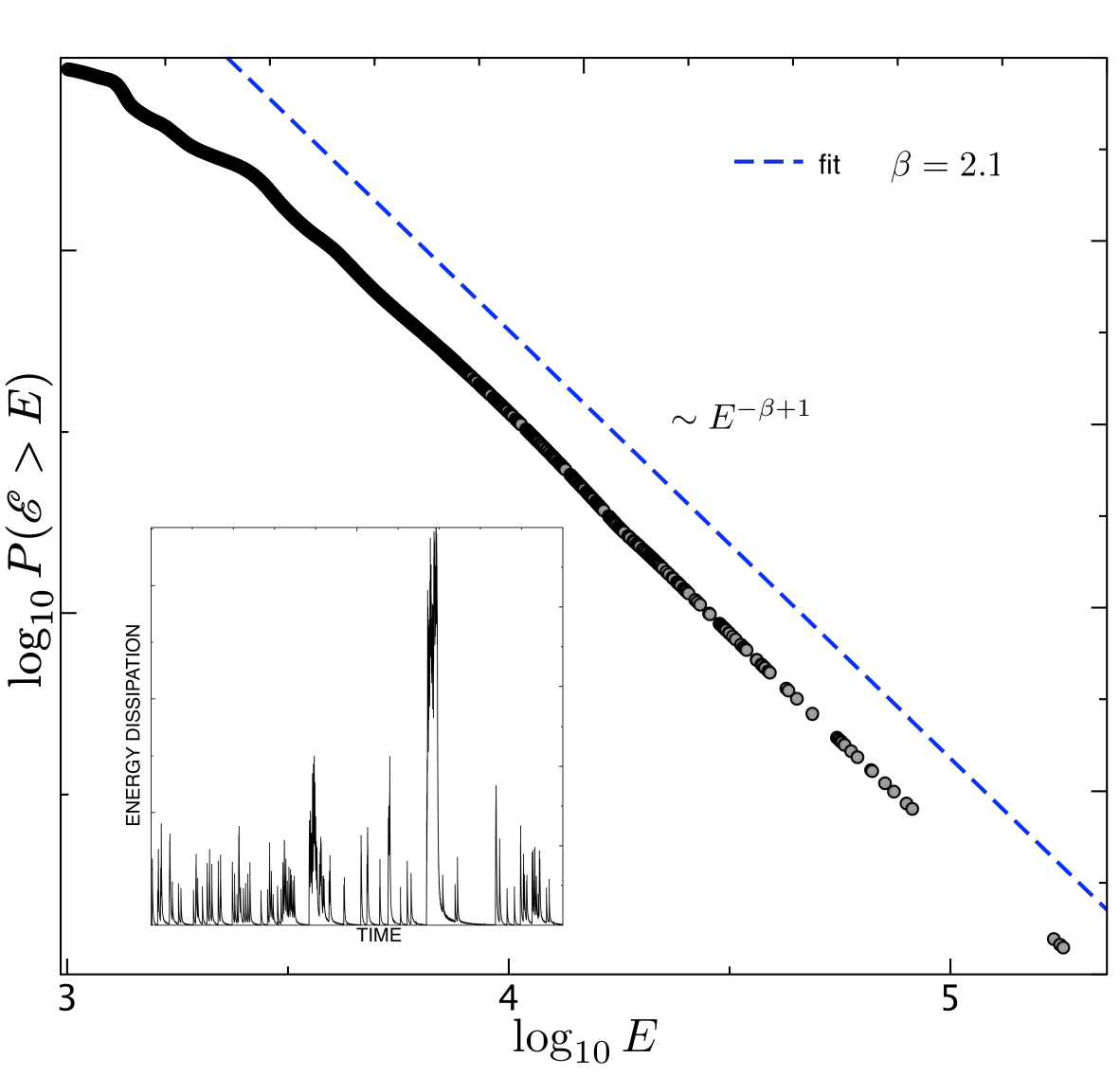}
\caption{Cumulative probability distribution  in log-log scale of the  energy dissipated by avalanches in  the underdamed system loaded quasistatically in a hard device and shown in Fig. \ref{P3}. Here  $C=8000$ and $N=1000$. Inset:  a fragment  of  the actual  time series.
}
\label{P4}
\end{figure}   
%\textbf{ For bigger values of $C$, we observe that correlation times are longer, the flat parts disappear (or it appears for much longer times). The exponents of PSs  become bigger than 2 that may  show   that inertia leads to   an over-correlation between avalanches. In  purely dissipative systems, the exponent is always smaller than $2$ and there are only temporal correlations within each avalanche. It may explain why the exponent of PS  does not satisfy the   relation found by Kuntz and  Sethna.
% }
% 

\begin{figure}[h!] 
\subfigure[][]{\label{tex1}\includegraphics[scale=0.3]{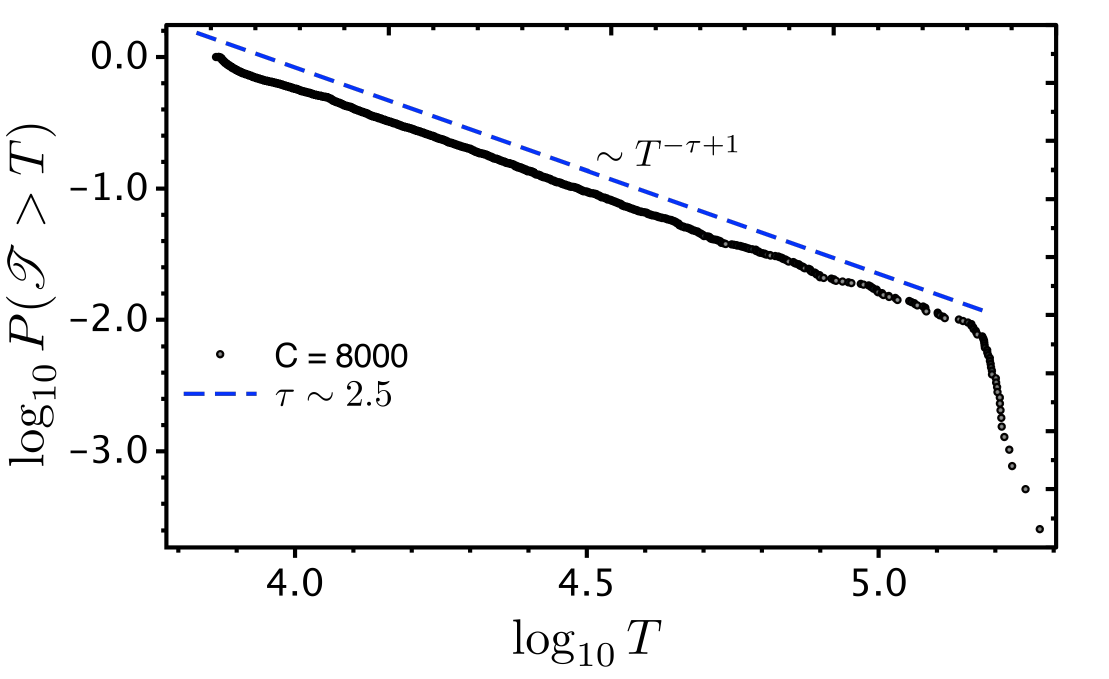}}
\subfigure[][]{\label{tex2}\includegraphics[scale=0.3]{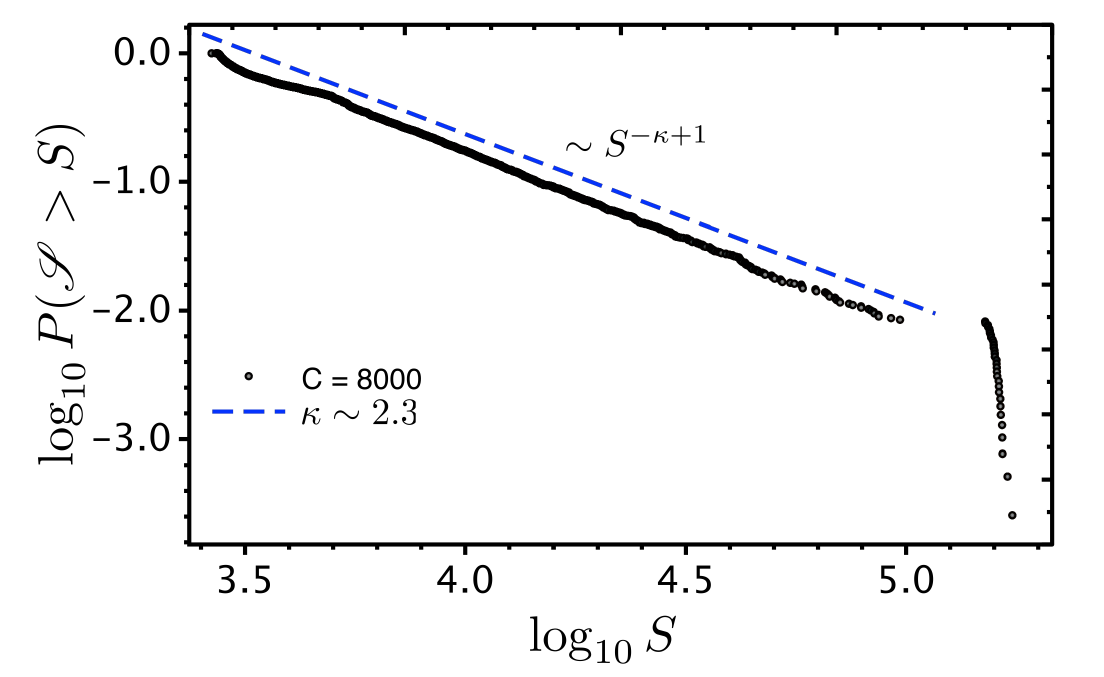}}
\subfigure[][]{\label{tex3}\includegraphics[scale=0.3]{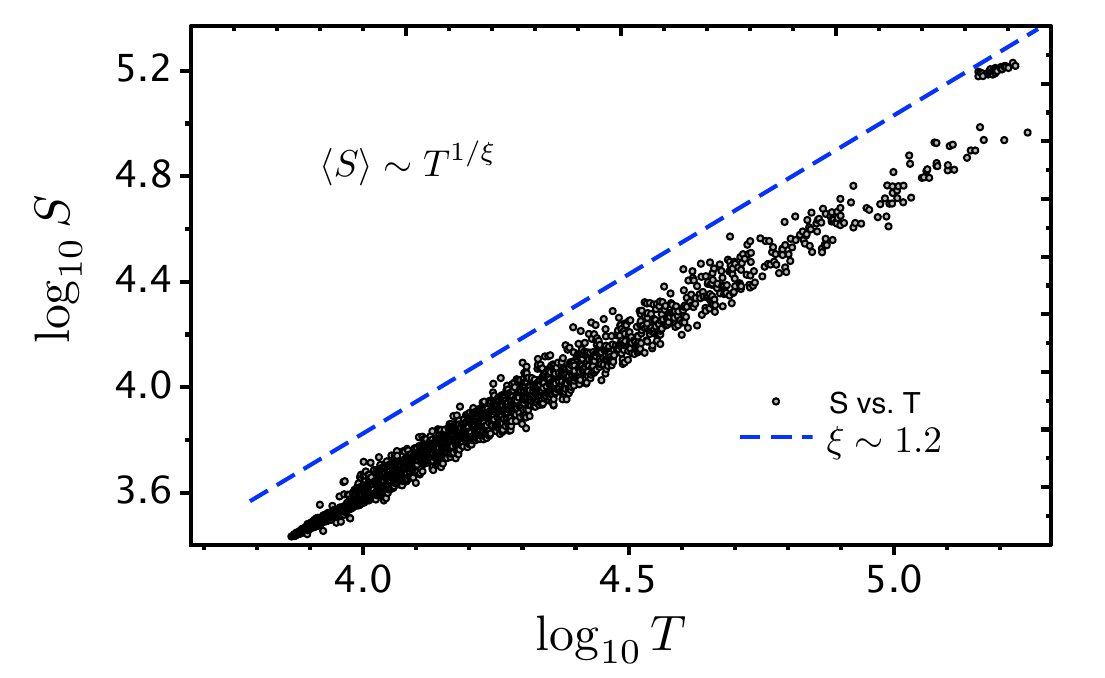}}
\caption{ Statistical signatures  of avalanche behavior for the system shown in Fig. \ref{P3} : (a) Cumulative probability distribution of avalanche durations. (b) Cumulative probability distribution of avalanche sizes. (c) Cloud plot showing joint distribution of avalanche sizes and durations and revealing strong  correlation between these two quantities.}
\label{P1}
\end{figure}

Finally, we show  in Fig. \ref{P4} the associated statistics of avalanche sizes  represented  through  distribution of the dissipated energies $E$.  Here we zoomed into  a robust range of  power law behavior 
 \begin{equation}
P(E)\sim E^{-\beta}
\end{equation}
stretching over two decades  with exponent  $\beta \approx 2.1$. In the inset we show the structure of the corresponding intermittent signal.  Our Fig.  \ref{P1} compares this distribution with the   distribution of the  avalanche  magnitudes $S$ which has again a power law form 
\begin{equation}
P(S)\sim S^{-\kappa},
\end{equation}
now with the exponent $\kappa\approx 2.2$, see Fig.  \ref{P1}(a), and the distribution of avalanche durations also exhibiting a power law behavior
 \begin{equation}
P(T)\sim T^{-\tau},
\end{equation}
now with the exponent  $\tau \approx 2.5$, see Fig.  \ref{P1}(b).  The revealed ubiquity  of  power law behaviors corroborates the idea that the system develops in this range   robust scale free   correlations. 

It has been shown in  \cite{PhysRevB.62.11699}  that  there exists  a relation between the exponents representing the  power   spectrum and the exponents characterizing the power law tail of the probability distribution for avalanche sizes of the form  
\begin{equation} \label{P5}
PS(f) =f^{-(3-\kappa)/\xi}. 
\end{equation} 
Here the parameter $\xi$ can be found from another  scaling relation 
 \begin{equation}
\langle S \rangle\sim {\langle T\rangle}^{1/\xi}, 
\end{equation} 
and our cloud plot juxtaposing distributions of   sizes and  durations,  shown in Fig.  \ref{P1}(c), suggests that  
in our case  $1/\xi\approx1.2$. Given that we also obtained the values 
$\kappa\approx 2.2$ and 
$\alpha \approx 1$,  the implied universal  relation 
 \begin{equation}
\alpha= \frac{3-\kappa}{\xi}
\end{equation} 
is   respected. 
We also note  that the exponents of  the power laws  for  durations $ \tau$  and  energies $ \beta$   can be related to $\kappa$ and $\xi$ through another two scaling relations discussed in \cite{PhysRevB.62.11699,kertesz1990noise,nandi2022scaling}:   
\begin{equation}
\tau =1+\frac{\kappa-1}{\xi}
\end{equation}  
and
\begin{equation}
\beta =1+\frac{\kappa-1}{2-\xi}.
\end{equation}   
Using the computed values of the exponents we again find an agreement which  points to  consistency of our numerical results. 

We highlight  the emergence in Fig.~\ref{P1}  of a feature which can be interpreted as a supercritical  behavior in the range of large events: it may not look as a conventional 'bump' because instead of PDF we present here a cumulative probability distribution (CPD).  This feature reveals the presence of recurrent large `brittle' nucleation events that occur in each cycle, which are clearly visible in the stress-strain curves shown in Fig.~\ref{P3}. We reiterate  that supercritical `bumps' have been found to be characteristic of underdamped behavior in many other systems~\cite{Salerno2012, Salerno2013, Karimi2016, Karimi2017, Nicolas2016}. In our representation, these large events present themselves  as almost a single data point because  in the cumulative probability distributions shown in Fig.~\ref{tex1}(a--c)  we do not perform any  binning.

Finally, as a word of caution,  we  stress    that   our three  numerically observed  fluctuation patterns  still need to be subjected to careful statistical analysis as  it is usually difficult to distinguish between different   types of heavy-tailed distributions on finite intervals and from  noisy data.  In particular,  power-law and log-normal distributions are often confused with the latter sometimes interpreted erroneously as  power laws with exponential decay \cite{polizzi2021power}.  We plan to address  this technical question in a separate study.

\subsection{Regime diagram} 

The goal of this section is  to consolidate and summarize the results obtained so far. Our  Fig. \ref{p1ex} presents   the  regime diagram illustrating qualitatively different responses of the system in the space of dimensionless parameters 
  \begin{equation}
M= v^2\rho/\kappa,\,\,\ V= \gamma v/(L\kappa).
 \end{equation} 
  Note that we have chosen both parameters $M$ and $V$  to be  dependent on the rate of loading  $v$. In this way  we  can highlight  that our  assumption of   quasi-static loading  implies  a   double asymptotics   
   \begin{equation}\label{MV}
M \to 0,\,\,\,     V\to 0 .
 \end{equation} 
Therefore in  Fig. \ref{p1ex}  we are, first of all,  interested in the behavior of the system around  the origin of the parameter space.  In view of non-commutative nature of the limits \eqref{MV}    the behavior of the system  turns out to be dependent on the   asymptotic path  controlled by the value  of  the   parameter
  \begin{equation}
C=M/V^2. 
 \end{equation} 
In other words, qualitatively different fluctuational  patterns   can be expected   to take place as one  changes  the relative strength of   inertia vs. the strength of dissipation. But this is exactly what we have seen in our numerical experiments which were all conducted in  a broad range of values of  $C$ even though we reported above the results only for selected values of this parameter.

\begin{figure}[h!]
\label{ex1}\includegraphics[scale=.03]{./phase_diagram.pdf}
\caption{Regime diagram in the space of parameters showing the stress-strain response of the quasistatically driven chain in underdamed,  overdamped and  intermediate  regimes. Here $M= v^2\rho/\kappa$  and  $V= \gamma v/(L\kappa)$.  The boundaries between different regimes   are delineated approximately in accordance with multiple numerical tests  conducted at  different values  of the parameter $C=M/V^2$.  The particular regimes a,b and c are discussed in the above in more detail.
\label{p1ex}}
\end{figure}

According to Fig. \ref{p1ex},  there are three main regimes with qualitatively different behavior.  As we have seen, in the two of them,  an  overdamped/damped  regime (small $C$) and an  underdamped/undamped  regime (large $C$), the  behavior of the system is relatively simple and well understood. It is either uncorrelated (disorder) or overcorrelated (order) behavior and therefore  in none of these regimes the system  shows  \textit{complexity} exemplified by   intermittency and  scaling.  In  the  intermediate regime, which is identified  in Fig.  \ref{p1ex} by the word  'scaling',  the   inertia and  the dissipation are  balanced in the sense that  none of them dominates. Our numerical experiments   showed that distributions with a robust  scaling range over two decades  are  observed over a finite range of values of the parameter $C$. The characteristic feature of the scaling regime is   a delicate  interplay between \textit{inertia}, \textit{dissipation} and \textit{discretness} allowing the  system to self-organize towards  scale free   non-equilibrium steady state.

%\textbf{Paco writes: "results for each of the studied regimes are presented in the previous sections for specific values of $C$. In this  section  these results are apparently   generalized  which allows the authors  to provide an intuitively appealing framework summarized in Fig. 13. However, it is not explained to what extent has the behavior of the model been explored for several values of $C$ within each regime?" Question: did you have  to cover  a continuous range of values of $C$ to obtain the boundaries between the regimes indicated in Fig. 13?} \textcolor{blue}{it wasn't computationally  possible at the time; I tried several simulations by translating the points a,b,c to the right and left and there was a good match with the claims of the figure }

\subsection{Extended criticality}  

While the  interval of parameters $C$  where scaling was observed is relatively narrow and while the power law response associated with this interval extends only along a finite range of event sizes, here we argue that one can  still interpret the observed   behavior as self-organization towards 'generic'   quasi-critical regime. 

%
% \textbf{Paco continues: "This is particularly important to discern whether there is a scaling region in the (M,V) space or there is instead a critical line indicating that behind it there is a real critical point.  These possibilities should   be discussed and an 'extended'  scaling region not given for granted unless there are solid reasons for it." This is very important: do we have a 'generic' scaling (interval of C's ) or only a 'tuned' scaling (one particular value of C)? } \textcolor{blue}{It is clearly generic criticality. We have many values of C for which we got criticality.}
%
One  reason to think that we deal here with extended criticality  is  the  near  \textit{marginality} of the  underlying mechanical system. We have seen that this is definitely the case for  the overdamped regime  where the system in a large $N$ limit  exhibits a quasi-plastic yield on a stress-strain plateau.  It is clear that the   yielding  takes place because in  the  overdamped regime the system explores only a  part of the energy landscape where the  barriers, separating  different metastable states, are extremely small.  
Then, in the scaling regime the system can be thought as being  on one side  close to barrierless marginal stability limit and, on the other side,  being able to easily cross  the remaining energy barriers due to underdamping. Apparently, it is  the implied  delicate balance  between   near marginality with its abundant  elastic instabilities  and  the 'fluidity' of navigating among  small energy  barriers   which  allows the system to self-organize toward a dynamical critical state.  While the resulting yielding takes place around a fixed average stress, the 'tuning' is only apparent because stress is not a control parameter  in the  hard device loading experiments.

%Instead, in the underdamped regime the system operates away from such marginally stable states as the system acquires  an ability  to easily cross energy barriers.

In this perspective the emergence of 'generic' scale invariance can be also associated with what has been coined in \cite{sornette1994sweeping,sornette1994power,vives1995distribution, Sethna2001} as 'sweeping' of an instability. If we interpret transformational yield as such an instability,  we see that that the  threshold is indeed continuously   swept  due to the presence of a  feedback provided  by long range elastic interactions. In particular,    a hard device loading ensures that the threshold is overshot with subsequent stress drop bringing the system back into a stable state.  Then the system is  destabilized again so on. Various  prototypical  models of the implied  feedback allowing the system  to  mainten near marginal stability, and safeguarding the 'generic' nature of the scaling behavior, were studied  in \cite{Salman2012a, Perez-Reche2008}. 
  
  % separating absorbing (pinned phase boundaries) and active (mobile phase boundaries) regimes. 
 
The fact that the observed  dynamic activity with a broad distribution of scales  takes place generically, without any special tuning of parameters,   can be also interpreted as self-organized criticality  (SOC).   In this respect it may be appropriate to mention    that our  account for  NNN  interactions  is  a poor man's attempt to  imitate  the  effect of demagnetizing forces in the theory of Barkhausen effect,  which is crucial for reaching  a robust critical regime \cite{Urbach1995-ui,Durin1995-cg,PhysRevLett.75.4528}. Our   NNN interactions can be also viewed as imitating strain incompatibility between austenite and martensite which is  another  important factor of reaching robust criticality \cite{Song2013-os,Balandraud2015}. Specifically, the presnce of  NNN interactions  contributes to the creation of metastability by  preventing most of the avalanches  from sweeping over the whole system. The classical analogs of the implied  metastability are the  stable critical nuclei in isochoric thermodynamic systems   \cite{schmelzer1987thermodynamics}, see also \cite{salman2019origin}.
 
Yet another factor  which may explain the 'generic' nature of scaling in our model is the presence of  inertia-induced dynamic disorder. While  the quenched disorder in RFIM had to be tuned to reach the critical state, in our model the effectively  annealed dynamic disorder is both self-induced and self-tuning. One can then argue that  the  implied  feedback loops  regulate the level of such disorder   driving the system to criticality  without  external fine tuning.

% with the latter ensuring in both cases a limited  stability of the effective spinodal region. 
%
%
%We show that the emergence of power laws in this system  depends crucially on both discretness and the presence of ferromagnetic  interactions. 

%
%Accordingly, they  disappear in   continuum (thermodynamic) limit which  suggests that the power law scaling in the inertial 1D model may be just  a finite size effect.
 
Finally we recall  that    bringing  inertia into the  model   leads to   softening of the  kinetic relation  which regulates propagation of phase boundaries \cite{Truskinovsky2005-tq,Truskinovsky2008-fn}. As we have already mentioned, such softening is  an important source of intermittent stick-slip behavior and a factor often used to explain the robust emergence of  'generic' scaling  \cite{Carlson1991-jh,https://doi.org/10.1029/96JB00653,Elmer1996-mm}. 
While in our oversimplified model,  the interfaces between phases  are not explicitly  tracked,  the  kinetic softening  is still present  as it can be shown through the analysis of  radiative damping by  propagating phase fronts  \cite{Truskinovsky2005-tq,gorbushin2022transition}.

\section
{One dimensional continuum model}  

 %\subsection{Continuum model} 

Given  that in a  discrete model   we were able  to  reach the   regime of extended  scale-free  behavior by  adjusting the value of a single parameter $C$,  it is tempting to  argue that   scaling  in martensites is entirely due to an interplay between inertia and dissipation. It is then  natural  to  check if this conjecture survives  in the corresponding continuum model. 
 To this end we  need to perform in \eqref{3} a formal limit  $\delta\rightarrow0$.   The resulting dimensionless partial differential equation is of the form:  
\begin{equation}
\label{contdisp}
C\ddot u  =\partial_{x}   \phi ' (\partial_{x}u )  +    \partial_{xx} \dot u   -G \partial_{xxxx}u .
\end{equation}
Here $u (x,t)$ is the continuum displacement,  $\phi$ is our double well potential  and the only dimensionless  parameters are the ratio of inertia over dissipation $C$ and the measure of nonlocality $G$. The corresponding  total  energy is again 
 $\mathscr E=\mathscr K+\mathscr F,$ 
where   the (dimensionless) kinetic energy is 
\begin{equation}
\mathscr K=\int  \frac{C}{2}\dot u^2 (x,t) dx. 
\end{equation} 
 The  (dimensionless) elastic energy   has the   form:
\begin{equation} \label{contdisp1}
\mathscr F=  \int \left(\phi   (e) +\frac{G}{2} (\partial_{x}e )^2 \right) dx, 
\end{equation}
where we introduced the  continuum strain $e=\partial_{x}u$ which plays  in this case the role  of the classical Landau-Ginzburg  order parameter.  The only difference from  \eqref{3} is that in \eqref{contdisp}, instead of environmental friction,   we used  the (dimensionless)   viscoelastic  Rayleigh dissipative potential  
 \begin{equation}
 \label{rayleigh_term_c4}
 \mathscr R=  \int \sum_i^3 \frac{1}{2}{\dot e ^2 (\bold{r},t)}  d^2x. 
 \end{equation}
 This choice is dictated by the necessity to eventually extend the continuum model from 1D to 3D  where the environmental viscosity would not make much sense. 
 
The one-dimensional continuum  model \eqref{contdisp} has been studied extensively in both statics and dynamics, see for instance \cite{Truskinovsky2008-fn} and the references therein. Here we  used \eqref{contdisp} to perform   numerical experiments  with a   system subjected to  quasi-static cyclic loading.  

Our main  conclusion is that, independently of the value of parameter $C$, the  model \eqref{contdisp} does not produce  either intermittency or scaling. The  reason is a rather  limited ability of such  continuum model   to support pinning of phase boundaries and, more generally, to generate  elastic  metastability which is necessary for the emergence of the recurrent quiescence and the associated intermittency. One can say that  with discreteness lost, the  effective energy landscape becomes  too simple which prevents the emergence of   ( near)  marginality and prevents the system from reaching sufficiently complex non-equilibrium steady state.  

It may sound counter intuitive, but in this respect  the  finite dimensional discrete systm  is much richer than its straightforward infinite dimensional analog.  Indeed, since springs   can transform only after reaching individual thresholds, the effective energy landscape of a discrete system has  an exponentially  large number of local minima and, while this  number diverges in  the continuum limit,  the minima themselves become progressively more  shallow and eventually disappear \cite{puglisi2000mechanics,puglisi2002mechanism,puglisi2002rate,Truskinovsky2004-xp}. In particular,  the continuum model \eqref{contdisp} loses  the capability of  trapping  phase boundaries which    compromises  the  hysteretic nature   of martensitic transformations. 

In the next two sections we study whether the problem can be remedied by  moving from an over schematic 1D continuum model, towards   its  2D and 3D continuum analogs  which can already represent  realistic  martensitic transformations.

\section{Two dimensional continuum model}

The goal of  this  section is to analyze a straightforward  2D analog of \eqref{contdisp}.  To focus exclusively on the effect of higher dimensionality, we  have  chosen to  intentionally simplify the targeted  martensitic transformation by neglecting its  volumetric effect.  This  makes  the energy wells, describing the high symmetry  austenite and  the low symmetry  martensite,  automatically rank one connected and therefore,  kinematically compatible. By  neglecting  this  important source of metastability in the system,  we naturally place  our 2D model  between the automatically compatible 1D model and the generically  incompatible 3D model  \cite{bhattacharya1994restrictions,smyshlyaev1999relation, ball2015incompatible,grabovsky2023class,grabovsky2024rigidity}.

A simplistic, but nevertheless meaningless  example of a martensitic   phase transition in 2D  involves   square and  rectangle phases \cite{Shenoy:1999ri,Curnoe2000-rq,PhysRevB.67.024114}. To describe the corresponding   transformation,  first  turn to dimensional variables and again write the  total  energy in the   form 
 $\mathscr E=\mathscr K+\mathscr F,$ 
where  the  kinetic energy has  the standard form 
\begin{equation} \label{K1}
\mathscr K=\int  \frac{\rho}{2}\dot u^2 (\bold{r},t) d^2x, 
\end{equation} 
where  $\rho$ is the mass density and  $u (\bold{r},t)$ is the displacement vector. The  elastic  energy  must be   specialized to reflect the chosen crystal symmetry and we write it in the form
 \begin{equation}\label{F1}
\mathscr F=\int  \left(\frac{A_1}{2}e_1^2 + \frac{A_3}{2}e_3^2 + \phi(e_2,\tau) + K e_1e_2^2 +\frac{G}{2}|\nabla e_2|^2 \right) d^2x.
\end{equation} 
Note first that the elastic  energy density in \eqref{F1} is presented   as a function of three  components of the linear elastic strain  tensor 
  \begin{align}
    e_{1} &= \left( \frac{\partial u_x}{\partial x} + \frac{\partial u_y}{\partial y} \right),  
    e_{2}  = \left( \frac{\partial u_x}{\partial x} - \frac{\partial u_y}{\partial y} \right),  
    e_{3}  = \frac{1}{2} \left( \frac{\partial u_y}{\partial x} + \frac{\partial u_x}{\partial y} \right).
\end{align} 
As it is clear from  \eqref{F1},  we   assumed that  the strain component $e_2$ is the primary order parameter. The first two harmonic terms  in \eqref{F1}  with the coefficients  $A_1$ and $A_2$ describe the classical physically linear elasticity in a solid with rectangular symmetry.  The cubic term in \eqref{F1}  is needed to  potentially introduce   a  nonzero  volumetric effect  of the phase  transition and therefore the third order elastic modulus   $K$ can be viewed as controlling the   strength of the corresponding  nonlinear shear-dilatation coupling  \cite{budiansky1993mechanics,fadda2002unified}.   However, as we have already mentioned, in our  numerical experiments   we  set  $K=0$ and the corresponding term is mentioned here only because it will reappear in the three-dimensional model.  
We further assumed that the regularizing gradient  energy term in \eqref{F1} , mimicking   NNN interactions in the discrete  model,   includes only the gradient of the  order parameter $e_2$ with the coefficient  $G$ bringing into the   continuum model a finite internal length scale.  To complete the model  we  introduce  viscoelastic  Rayleigh dissipative potential:
 \begin{equation}
 \label{rayleigh_term_c4}
 \mathscr R=  \int \sum_i^3 \frac{\gamma_i }{2}{\dot e_i^2 (\bold{r},t)}  d^2x, 
 \end{equation}
where $\gamma_i$ are the corresponding effective viscosity coefficients.

\begin{figure}[h!]
%\begin{center}
\includegraphics[scale=0.2]{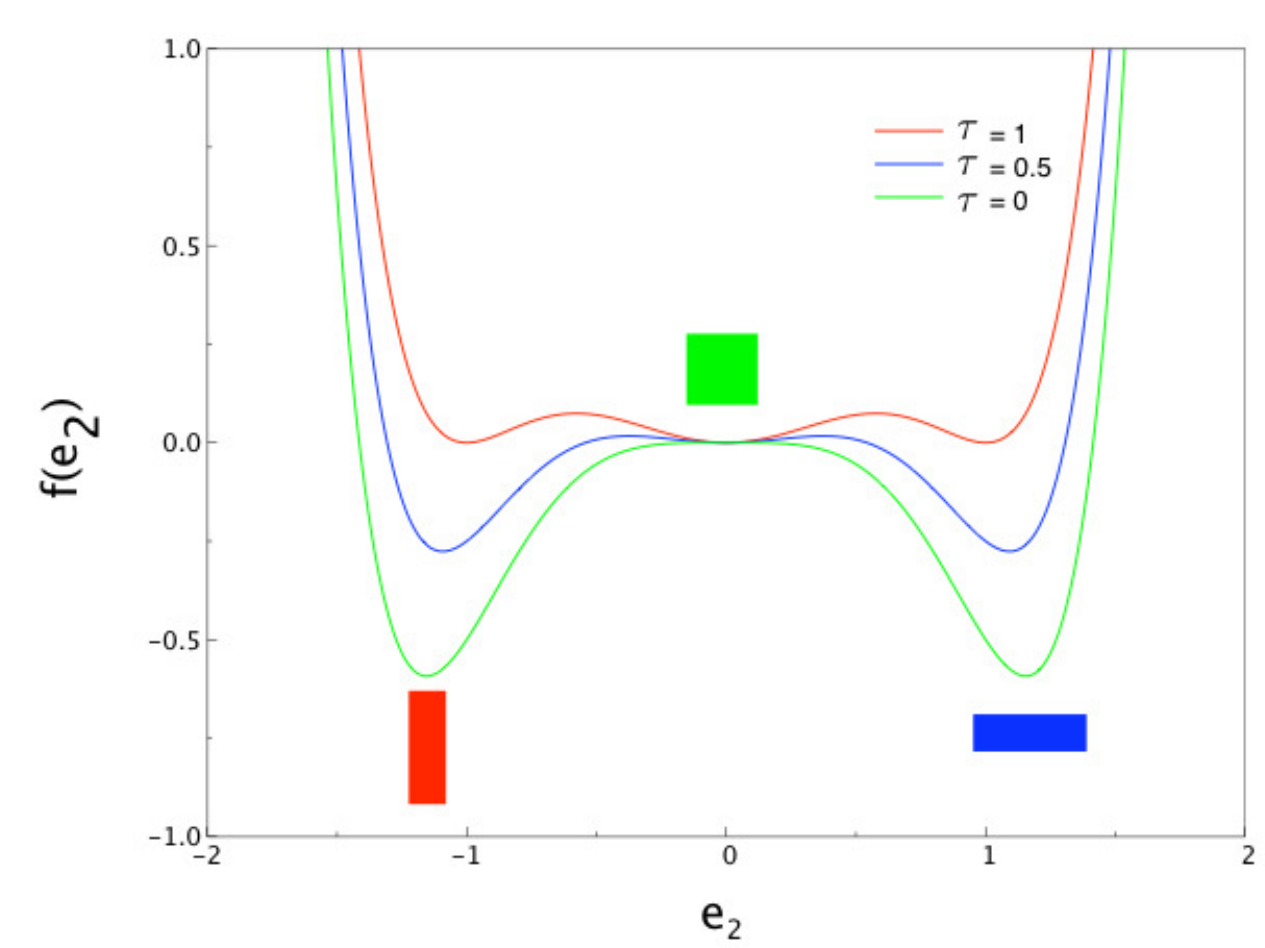} 
\caption{ The elastic energy density $\phi(e_2)$ at different values of the temperature $\tau$.}
\label{func_246_c4}
%\end{center}
\end{figure}

%where   comma   represents partial derivative.
%, for instance  
%$
%    u_{i,j} =  \partial u_i/\partial x_j.
%$ 
The numerical implementation of the equation of motion requires spatial and temporal discretization. If we use as  a spatial scale the grid  size $d_0$, the dimensionless spatial coordinates $\tilde{x}$ and displacements $\tilde{u}$ become
\begin{equation}
\tilde{x} = \frac{x}{d_0}, \quad \tilde{u} = \frac{u}{d_0}. 
\end{equation}
By introducing the characteristic time  scale  $t_0$  we similarly normalize time 
\begin{equation}
\quad \tau = \frac{t}{t_0}.
\end{equation}
We also  introduce  the  energy density scale $f_0$ which allows us to write the remaining dimensionless  parameters of the model in the form:
\begin{equation}
\tilde{\rho} = \frac{\rho d_0^2}{t_0^2 f_0},\,\,
\tilde{G}  = \frac{G}{d_0^2 f_0},\,\,  
\tilde{\gamma}  = \frac{\gamma}{t_0 f_0},\,\,
\tilde{A_i}  = \frac{A_i}{f_0}.
\end{equation}
Finally, the  minimal   Landau-type energy density function $\phi$ capturing  the schematics of square-to-rectangle transition can be written  in the form   \begin{equation}
  \label{eq:lansoc}
 \phi(e_2,\tau)= \frac{\tau}{2} e_2^2 - e_2^4  + \frac{e_2^6}{2}, 
 \end{equation}
where $\tau$ is the dimensionless temperature. In Fig. \ref{func_246_c4} we show that, for instance  at  $\tau=1$, which  is the point of a first-order transition,  the  energy density \eqref{eq:lansoc} exhibits   three equivalent minima describing two variants of the low symmetry  martensite (rectangular phase) phase and one minimum describing the high symmetry austenite (square phase); below this temperature  the austenite stops being the ground state.   

%and the dimensionless Rayleigh energy density is given by
%\begin{equation}
%\tilde{R} = \frac{1}{2}\tilde{\gamma}_i\dot{e}_i^2, \tag{7.41}
%\end{equation}
%with
%
%
%\tilde{F}_L &= \frac{F_L}{f_0}, \tag{7.37} \\
%\text{and} \quad \tilde{R} &= \frac{R}{f_0}. \tag{7.38}
%\end{align}
%

%This dynamical equation may be written in the general form:
%\begin{equation}
%\rho \ddot{u}_i = \frac{\partial}{\partial x_j} \frac{\partial F_L}{\partial u_{i,j}} - \frac{\partial}{\partial x_j} \frac{\partial}{\partial x_k} \frac{\partial F_G}{\partial u_{i,jk}} + \frac{\partial}{\partial x_j} \frac{\partial R}{\partial \dot{u}_{i,j}} \label{dynamical}
%\end{equation}
%where as usual $u_{i,j}$ and $u_{i,jk}$ stand for $\frac{\partial u_i}{\partial x_j}$ and $\frac{\partial^2 u_i}{\partial x_j \partial x_k}$, respectively and $ F_G = \frac{G}{2}|\nabla e_2|^2$. 

In our numerical experiments, the dynamic equations were   discretized   on a rigid grid of size $512\times512$  with  periodic boundary conditions and solved using  a Fourier pseudo-spectral spatial scheme with  the corresponding time marching temporal algorithm detailed in Appendix~\ref{app:discretisation}, see  also \cite{Salman2009-qv}.  Since we study the  2D model only to provide qualitative illustrations of  the effects of inertial dynamics, we used generic values of   dimensionless parameters.  For instance, throughout this study we   assumed that   $ \tilde G=5$, other parameters are specified below. A  physically meaningful calibration of the model  is postponed till  the next section where we consider a fully realistic 3D model of the same basic type.

\begin{figure}[h!]
%\hspace{0.5in}
%\begin{minipage}[t]{30\linewidth}
\subfigure[][]{\includegraphics[scale=0.15]{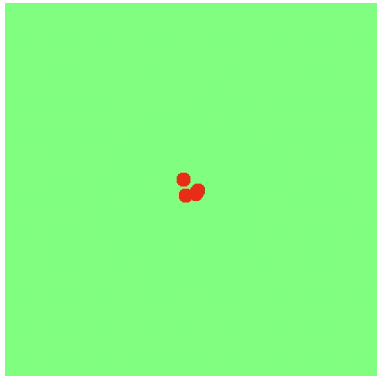}}
\subfigure[][]{\includegraphics[scale=0.15]{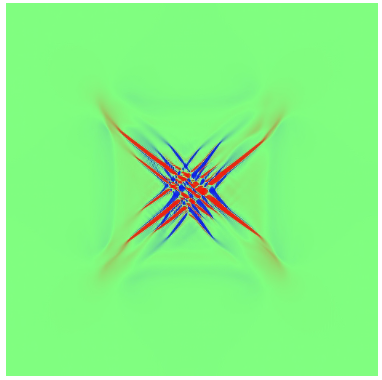}}
\subfigure[][]{\includegraphics[scale=0.15]{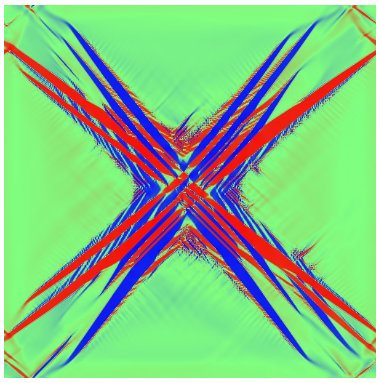}} 
\subfigure[][]{\includegraphics[scale=0.15]{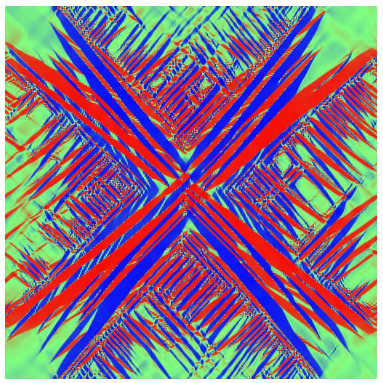}}
\subfigure[][]{\includegraphics[scale=0.15]{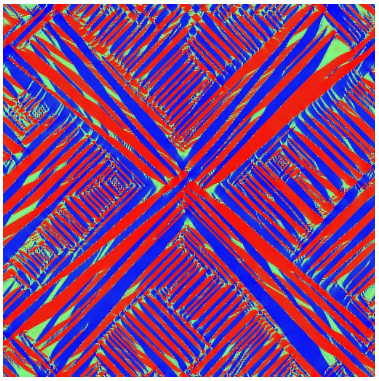}}
\subfigure[][]{\includegraphics[scale=0.15]{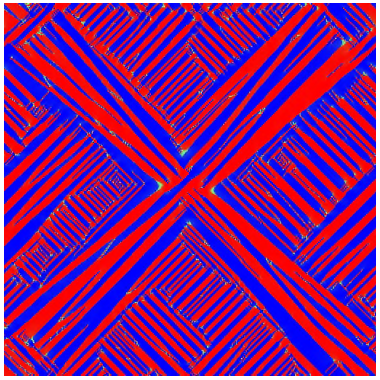}}
%\end{minipage}
\caption{\label{eq:2d_dynam_dissip}  Time evolution of a pre-existent martensitic embryo in an almost underdamped 2D continuum model. Panels (a-f)  illustrate  different  stages of the evolution. Green color represents the austenite, red and blue colors represent the  martensite, see Fig. \ref{func_246_c4} for the  identification of particular variants.} 
\end{figure}

Consider  first    an  almost  \textit{undemped} regime with    $\tilde\rho=1$  and  $\tilde\gamma_2=\tilde\gamma_1=\tilde\gamma_3 =0$.   This choice corresponds to the limit $C\rightarrow\infty$  in the 1D model.   In Fig. \ref{eq:2d_dynam_dissip} we illustrate the results of our  numerical experiments describing dynamic nucleation of a martensite. At time $t=0$  a homogeneous  configuration of  austenite (square) phase was  equilibrated at  $\tau=1$.   To destabilize this  configuration  we  decrease the dimensionless temperature $\tau$  in $10^{-4}$ increments till the transformation begins. 

To break geometrical degeneracy and ensure   controlled nucleation of the martensite we placed a  small   martensitic embryo  in the middle of the square phase.  Specifically, the embryo were installed by initiating the value of the order parameter to 1 in small circular domains in the middle of the computational domain, see red regions in Fig. \ref{eq:2d_dynam_dissip}(a). Our Fig.~\ref{eq:2d_dynam_dissip} (a-f)  illustrate  different  stages of the evolution of the transformation starting from the initial nucleation event. Green color represents the undeformed austenite phase where  $e_2=0$. Red and blue colors correspond to two variants of  the emerging martensite.

%Specializing these parameters  requires the knowledge of experimental data, such as the twin interfacial energy and, the elastic constants of martensite since we are concerned in microstructures of this phase. 

 %add references  detailing the chosen numerical approach 

As we see in Fig.~\ref{eq:2d_dynam_dissip}(a)  the transformation  begins around the infinitesimal embryo in a form of a single  variant of the martensite  effectively chosen    by the type of the inserted  imperfection. Already in Fig.~\ref{eq:2d_dynam_dissip}(b) we see that  the homogeneous  martensitic nucleus  evolves into a complex multiscale two-variant twinned microstructure which grows dynamically  in an apparently self-similar way till it hits the boundaries of the domain, see Fig.~\ref{eq:2d_dynam_dissip}(c). As the transformation process unfolds,  interaction of internally generated elastic waves  produces multiple  new  micro-twinned  patches appearing almost  spontaneously all over the computational domain, see Fig.~\ref{eq:2d_dynam_dissip}(d). As such multi-scale  microstructure formation  process reaches  the boundaries of the domain, austenite phase largely disappears, see Fig.~\ref{eq:2d_dynam_dissip}(e).  Eventually  the microstructure stabilizes by coarsening  and all independently developing  transformed zones  finally   merge into a  single multiscale texture, see  Fig.~\ref{eq:2d_dynam_dissip}(f).  The resulting complex variant mixture  contains  differently oriented but geometrically compatible martensite laminates forming an intricate  hierarchical pattern. In view of the  geometric compatibility of all participating energy wells   the formation of microstructure of  such complexity can be viewed as associated exclusively  with inertial  dynamics.  It implies  generation of elastic  waves by the transformation events,   interaction  of these waves with inhomogeneities,   their reflection   from the boundaries and, most importantly,  their self focusing which triggers  secondary transformation events.

\begin{figure}[h!]
%\hspace{.5in}
%\begin{minipage}[t]{30\linewidth}
\subfigure[][]{\label{sim1a}\includegraphics[scale=0.15]{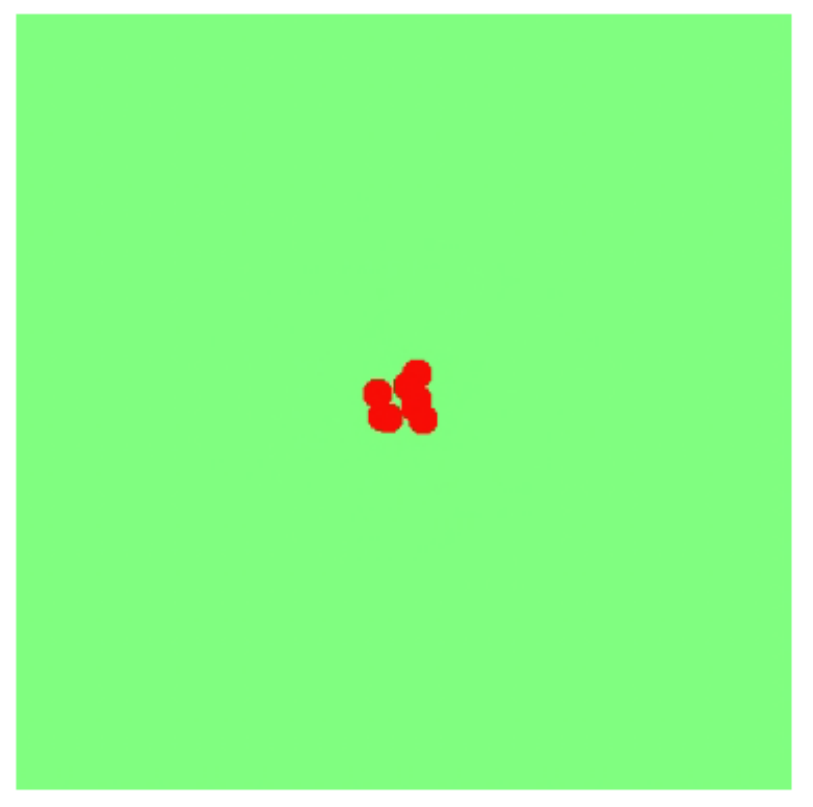}}
%\subfigure[]{\label{sim1b}\includegraphics[scale=0.15]{./c4fig/dyn2.eps}}
\subfigure[]{\label{sim1c}\includegraphics[scale=0.15]{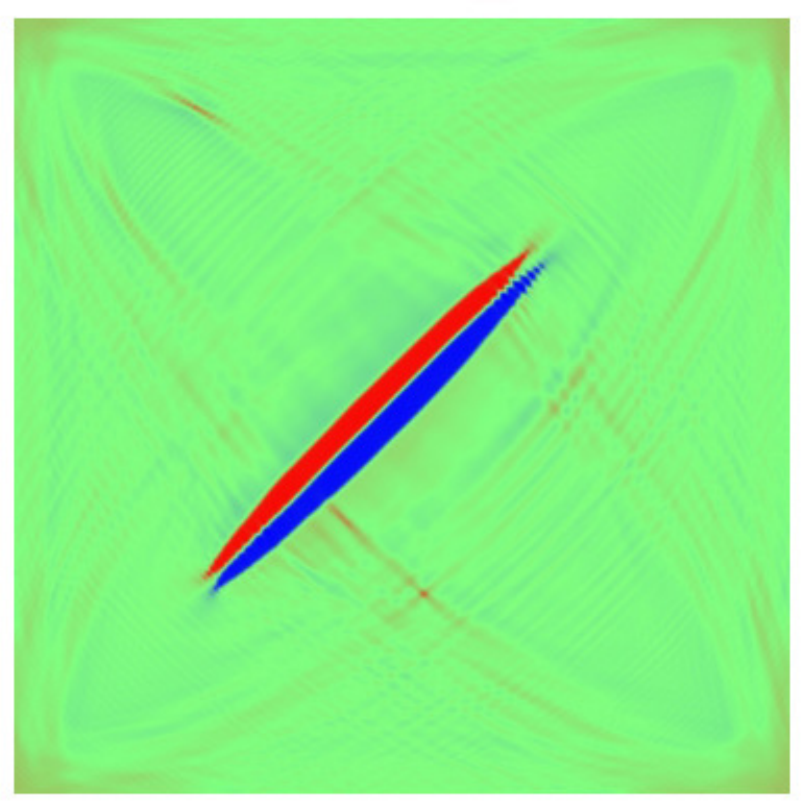}} 
%\subfigure[]{\label{sim1d}\includegraphics[scale=0.15]{./c4fig/dyn4.eps}}
\subfigure[]{\label{sim1e}\includegraphics[scale=0.15]{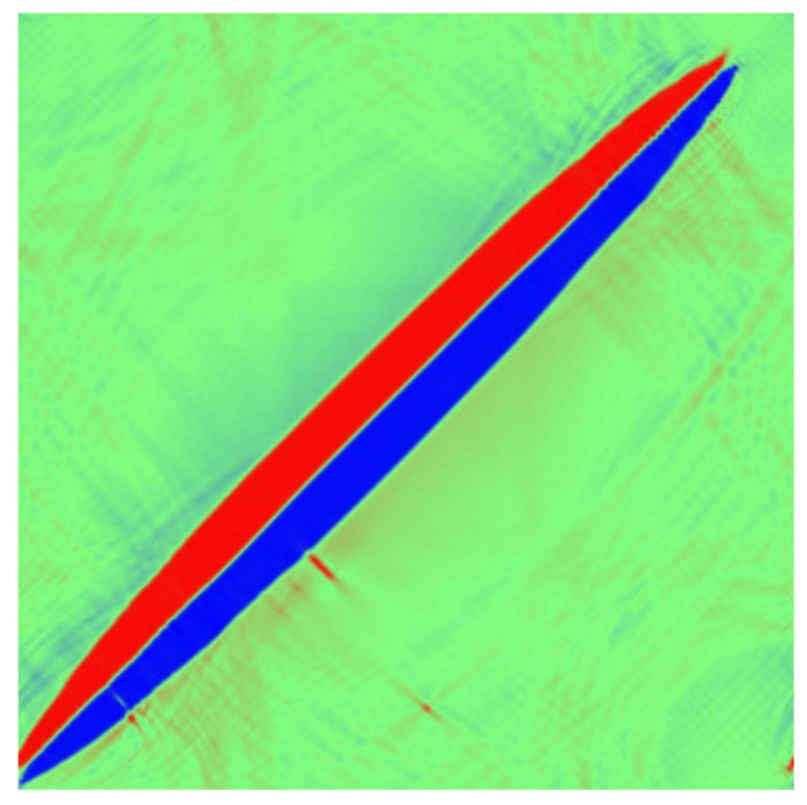}} 
%\subfigure[]{\label{sim1f}\includegraphics[scale=0.15]{./c4fig/dyn6.eps}} 
\subfigure[]{\label{sim1g}\includegraphics[scale=0.15]{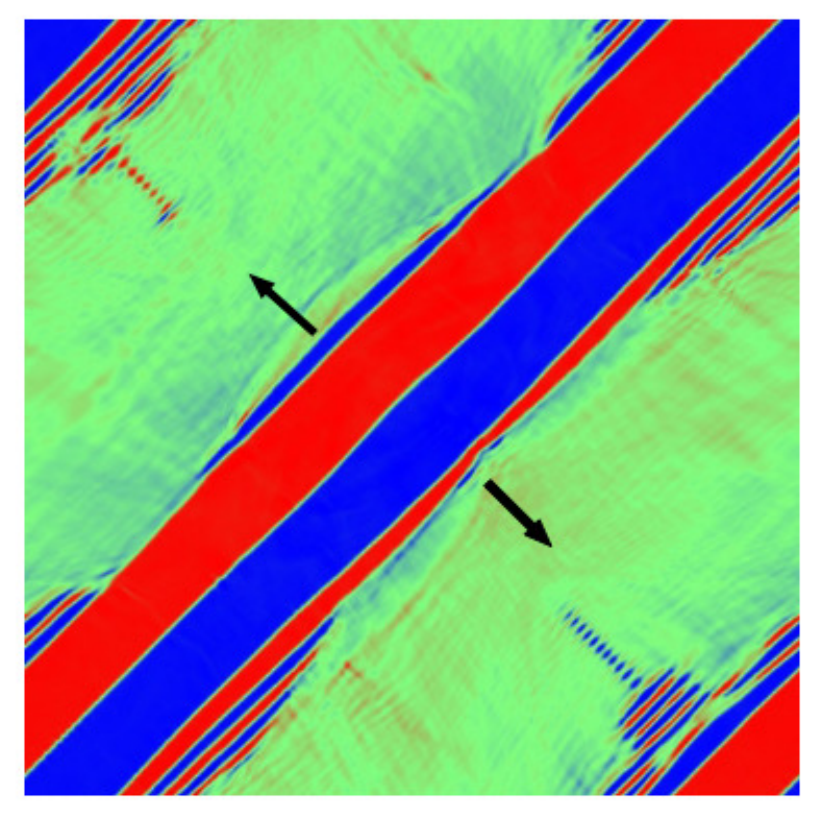}}
\subfigure[]{\label{sim1h}\includegraphics[scale=0.15]{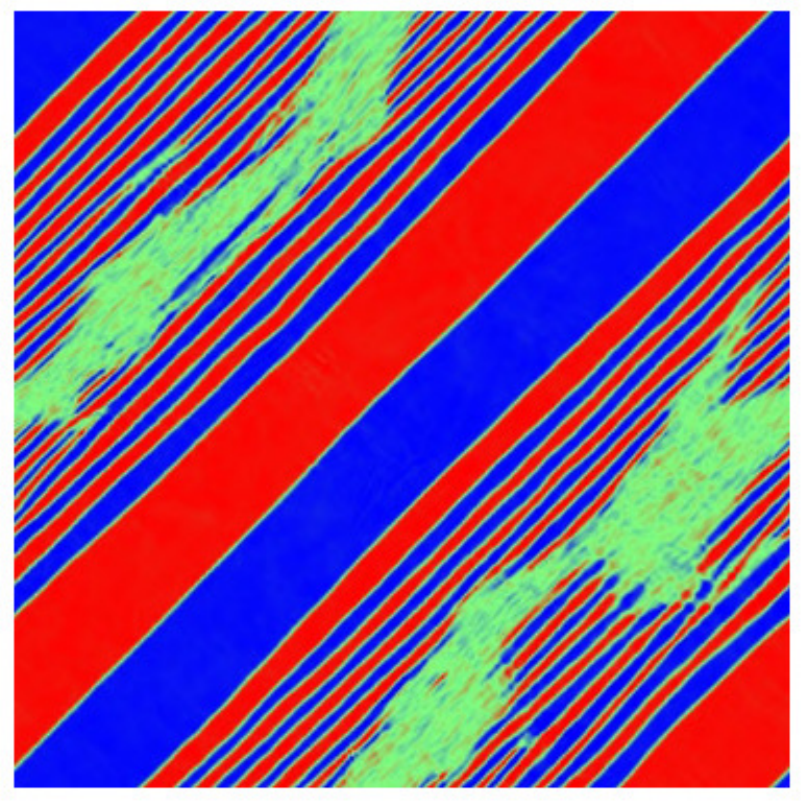}}
\subfigure[]{\label{sim1i}\includegraphics[scale=0.15]{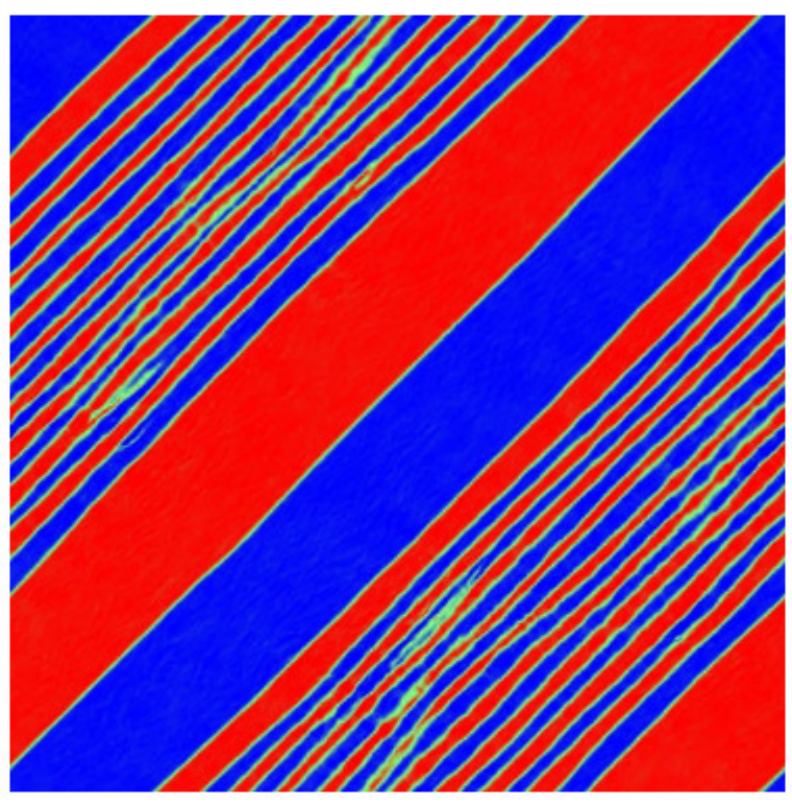}} 
%\end{minipage}
\caption{\label{sim1}   Time evolution of a pre-existent martensitic embryo in an underamped 2D continuum model. Panels (a-f)  illustrate  different  stages of the evolution. Green color represents the austenite, red and blue colors represent the  martensite, see Fig. \ref{func_246_c4} for the  identification of particular variants. }
\end{figure}

 Consider  next    the   \textit{overdamped} regime     with  $\tilde\rho=1$   and  $\gamma_2=1$, $\gamma_1=\gamma_3 =0.5$.
  As we see in Fig. \ref{sim1}(a)  the transformation  begins  again  in a form of a single  variant of the martensite  effectively chosen  by the nature of the  inserted  imperfection. Already in Fig. \ref{sim1}(b) we see that  the  growing martensitic nucleus splits into two variants  and then extends   as a  lenticular domain. The latter  contains two    twinned    martensitic variants forming   fully compatible interfaces between themselves and with  the austenite, see Fig. \ref{sim1}(c).  As this initial band  hits the boundaries of the domain,   the growth process continues along the direction perpendicular to its boundaries    involving thickening of the  two already formed martensitic variants.  The presence of self-generated dynamic activity  compromises  such purely coarsening process producing secondary nucleation events and  generating  sequential appearance  of zipping martensitic variants with  alternating strains, somewhat  similar to what has been predicted in \cite{PhysRevLett.67.3412}. The  resulting  fine poly-twin  pattern  is  regular and the whole pattern shows much less complexity than  the  multi-scale texture observed in the undamped regime.

\begin{figure}[h!] 
%\hspace{0.\begin{center}
\includegraphics[scale=0.02]{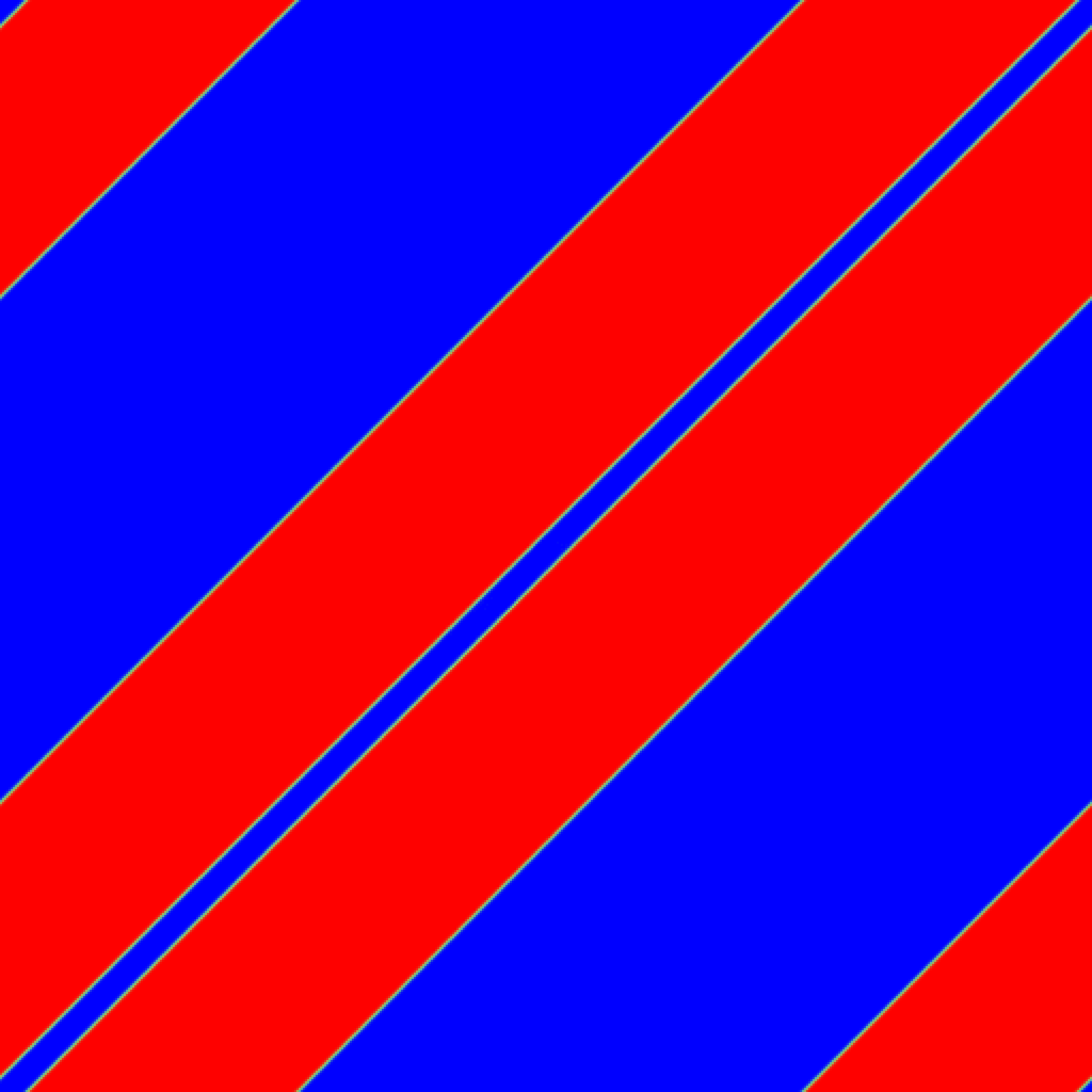}
%\subfigure[][]{\includegraphics[scale=0.3]{./dis6-eps-converted-to.pdf}}
\caption{\label{over}   Post nucleation pattern  in a 2D almost fully damped  continuum model, see  Fig. \ref{func_246_c4} for the  identification of the colors with particular variants of the martensitic phase. }
\end{figure}

%Such  dynamic alternating  can be directly linked to finite propagation time of   elastic waves  and, therefore, it will not be observed 

 Finally, consider the  \textit{damped} regime  with $\tilde\gamma_2=1$, $\tilde\gamma_1=\tilde\gamma_3 =0.5$  and $\tilde\rho\approx0$. This  is equivalent to the assumption that   $C=0$ in Eq. \ref{contdisp}).  In this limit  we obtain time-dependent Ginzburg-Landau model  with viscous damping, see for instance  \cite{PhysRevLett.67.3412,PhysRevB.67.024114}.   In this limit the   breakdown of an  unstable austenite phase  results   a  predictable  emergence of a coarse  highly regular  twin microstructure, see   Fig. \ref{over}. The whole process  can be described as an  elastic  spinodal decomposition with  subsequent coarsening driven by weak interaction of the  twin boundaries with  the possibility of being stuck in one of the kinetically  induced  locking configurations \cite{ball1991dynamics,friesecke1997dynamic}.

The examples presented  in this section show  unambiguously that  suppression  of inertia in this type of problems   blocks the emergence of complexity. Moreover, as our  more exhaustive study have shown,  the  metastability problem detected in the  1D continuum model persists in  the 2D  compatible problem.  Specifically, we have checked   that, independently of the degree of over- or under-damping   the ensuing dynamics does not support  either intermittency or scaling.    Still,  the relative transparency  of the associated  2D  problem   allowed   us   to illustrate  in some detail the  dramatic difference between the  processes of microstructure formation  in  overdamped and underdamped models. In particular,  we could show   that bringing inertia into the model contributes  to  multi-scale nature  of the emerging microstructure through  dynamic creation of  virtual nucleation sites. However, while this continuum model  generates a succession of progressively finer  scales,  we did not see any evidence that it  supports robust  self-organization towards criticality.

%Finally, in the fully damped model,  we observe the standard Ginzburg-Landau kinetics with nucleation of the simplest configuration of phase boundaries compatible with boundary conditions  and their subsequent  quasi-static growth. 

 %Figure \ref{sim1} displays the time evolution of the transformation at different times. 
 
%  First, it is important to note that the Lagrange equations maintain the macroscopic strain to its initial value, see for example (\ref{eq:eqmotlinwps2}).

%\begin{center}
%\begin{figure}
%\subfigure{\includegraphics[scale=0.176]{./image_xyz_10367.pdf}}\hspace{10mm}
%\subfigure{\includegraphics[scale=.24]{./dis6-eps-converted-to.pdf}}
%\caption{\label{time_evolution_ct}\small\textit{Final states (a) Purely dissipative, (b) with inertia.}}
%\end{figure}
%\end{center} 

%

%
%
%\subsubsection{Simulations with dissipation}

%
%

%\subsection{Three dimensional model. }

\section{Three-dimensional continuum  model}

In this section we show that the  desired complexity of the energy landscape,  comparable with what we have seen in our discrete model and,  therefore,  compatible with scaling and extended criticality,  can be   recovered  if we account in our  continuum modeling   for the incompatibility of the energy wells.  Using  3D  instead of 2D framework  is   beneficial in this respect as it allows one to perform a  comparison with  actual physical experiments. It  also considerably increases the set of accessible  data  which ensures the  detection  of scaling regime  with higher statistical certainty.

Specifically, we consider in this section   a  realistic continuum model  of a cubic-to-tetragonal martensitic transition in 3D which, in particular,  takes into account  a nonzero  volumetric  effect of the transformation.  The goal is to  reproduce in our numerical experiments   the result of \cite{Bonnot2008-qg},  where   the authors investigated  the acoustic activity which was measured experimentally as  a single crystal of Fe$_{68.8}$Pd$_{31.2}$  was quasi-statically driven through the  cubic-to-tetragonal phase transition. As we have already mentioned, the analysis of  statistical  distributions of  dissipated  energy   and   avalanche  durations  performed in \cite{Bonnot2008-qg},  revealed  a  power-law statistics and produced particular values of the exponents.

While the  kinetic energy  in the desired 3D  continuum model  can be again taken in the form \eqref{K1},  the elastic  free energy  must be adjusted to capture  the specifics of a particular  martensitic transformation. Specifically, to describe   a generic cubic-to-tetragonal martensitic transition in 3D  we use   the following  expression for the Ginzburg-Landau type energy  \cite{Curnoe2000-rq,PhysRevB.64.064101,PhysRevB.68.224104,Ahluwalia2006-uz} 
\begin{equation}\label{free}
\mathscr F = \int (\phi_L + \phi_G) d^3x
\end{equation}
where the energy densities describing local  and gradient contributions are given by the formulas:
\begin{align}
\phi_L &= \left( A_2(e_2^2 + e_3^2) + A_3(e_4^2 + e_5^2 + e_6^2) + A_4e_3(e_3^2-3e_2^2) \right. \nonumber \\
&\quad \left. + A_6(e_2^2 + e_3^2)^2 + A_1(e_1-K(e_2^2+e_3^2))^2 \right), \label{free_local} 
\end{align}
and 
\begin{equation} \label{free_gradient}.
\phi_G =  \frac{G}{2}(|\nabla e_2|^2 + |\nabla e_3|^2). 
\end{equation}
One can see that  we use here two coupled primary order parameters represented by  two deviatoric   components of the  linear elastic strain tensor  $\bold {e} = (1/2) (\bold \nabla \bold u + \bold \nabla \bold u^T )$  and   describing shear deformations in $\{110\}$-type   and  $<\bar110>$-type  directions:
   \begin{equation}
 e_2   = \frac{1}{\sqrt{2}}(\epsilon_{xx}-\epsilon_{yy}),\,\,\
e_3  = \frac{1}{\sqrt{6}}(\epsilon_{xx}+\epsilon_{yy}-2\epsilon_{zz}).  
\end{equation} 
 The remaining non-order parameter components of the  strain tensor are:  
\begin{equation}
\begin{array}{ll}
&e_1 = \frac{1}{\sqrt{3}}(\epsilon_{xx} + \epsilon_{yy} + \epsilon_{zz}) ,\\
& e_4  = \epsilon_{xy} + \epsilon_{yx}, 
  e_5  = \epsilon_{xz} + \epsilon_{zx},  
 e_6  = \epsilon_{yz} + \epsilon_{zy}.
\end{array}
 \end{equation}
%Here we again adopt a linear  strain displacement relation:  
%\begin{equation}
%\bold {\epsilon} = (1/2) (\bold \nabla \bold u + \bold \nabla \bold u^T ).
% \end{equation}
While the coefficients $A_i$ in  \eqref{free} with $i=1,2,3$  are the classical linear elastic moduli of a tetragonal phase, the coefficients  $A_4$  and $A_6$ describe the coupling between the order parameters which is necessary to destabilize the cubic phase in favor of the tetragonal phase \cite{Curnoe2000-rq,PhysRevB.64.064101,PhysRevB.68.224104,Ahluwalia2006-uz}.  As we have already seen in the 2D case, adding of  the coupling  coefficient $K$ allows one to  capture  shear-induced  volumetric effect of the transformation.  The chosen form of the energy density \eqref{free_local} guarantees that there are in general three compatible   energy wells corresponding to  symmetry related    tetragonal variants of the  martensitic phase and  one  (geometrically incompatible) energy well corresponding to a higher symmetry austenite phase.

\begin{figure}[h!]
%\begin{minipage}[t]{1\linewidth}
%\begin{center}
\includegraphics[scale=0.6]{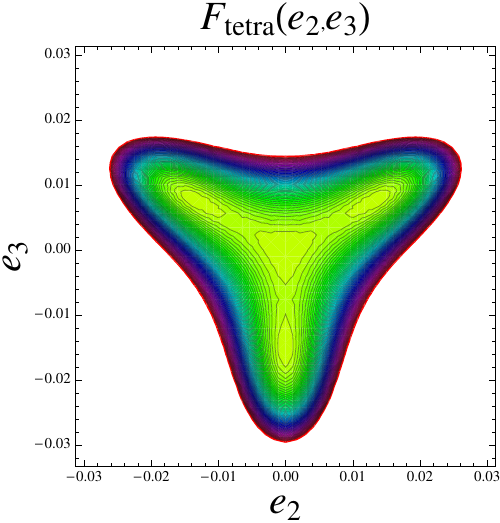}
%\end{center}
%\end{minipage}
\caption{\label{tetra_func_lag}  Typical  energy landscape (contour plot) in the space of primary order parameters adopted in the 3D continuum model. Three minima correspond to  three symmetry related variants of the stable martensite (tetragonal phase).}
\end{figure}

\subsection{Calibration of the model}

Fixing parameters of the  energy  density \eqref{free} requires the knowledge of such experimentally measured  quantities  as the homogeneous transformation strain, the elastic constants and the interfacial energy.  Fortunately, all of these parameters have been  experimentally measured for  FePd alloys  close to the transition temperature \cite{Ahluwalia2006-uz}.  

Specifically, it is known that   $a=3.725$\AA\,  and  $c=3.795$\AA \, are the lattice constants  of the equilibrium martensite (tetragonal phase) at  stress-free state  while  $a_0=3.756$\AA \, is the  lattice constant  of the unstrained austenite (cubic phase). The corresponding  stretch tensors mapping the cubic lattice   to three symmetry related variants of the tetragonal phase   can be then written in the form:  
\begin{eqnarray} \label{Umatrix}
U_1=\left(
\begin{matrix}
c^*       &  0      &  0   \\
0           & a^*   &  0  \\
0           & 0       &  a^*  
\end{matrix}\right),
U_2=\left(
\begin{matrix}
a^*      & 0       &  0  \\
0          & c^*    &  0     \\
0          & 0       &  a^*  
\end{matrix}\right),
U_3=\left(
\begin{matrix}
a^*    & 0       &  0  \\
0        & a^*   &  0  \\
0        & 0       &  c^*  
\end{matrix}\right) ,
\end{eqnarray}
where $c^* = c/a_0 $ and $a^*  = a/a_0 $. In view of these measurements we know, for instance,  that  the equilibrium value  of the nonzero primary order parameter  for, say  variant 3, is $ e_3^{0} =  0.0152$ while the volume change associated with the transformation is  $e_1^{0}= 0.0072$. From  \eqref {Umatrix} one can see  that even without the volumetric effect taken into consideration,   austenitic and martensitic wells are not geometrically compatible.

 The  elastic moduli of FePd  close to the transition temperature have  been also experimentally measured, see the discussion in   \cite{Muto1990-eb, Kartha1995-wj,Ahluwalia2006-uz}. In particular, it was observed that this martensitic transformation is  accompanied by softening of the deviatoric elastic modulus of the austenite $C'^A= (C^A_{11}-C^A_{12})/2$, while the elastic constants $C''^A=(C^A_{11}+C^A_{12}+2C^A_{44})/2$ and $C^A_{44}$ vary only slightly, which is all in accordance with our model.   Since the elastic moduli of the martensitic phase are largely unknown, it is usually assumed  that $C'^M\simeq2C_{44}^M=2C_{44}^A$, were it is implied that  the elastic constants of the martensite are expressed in the undeformed reference state of the austenite.

  If we normalize the elastic moduli in each of the phases  using  the same energy density scale $f_0=20.4$ $GPa$  taken from\cite{Sato1982-mf},  and introduce  the dimensionless constants 
 $$
\tilde A_i = A_i/f_0, \, \, \tilde K =  K/f_0,
 $$
 we can use the experimental data from \cite{Muto1990-eb} to  obtain the following numerical values for the dimensionless parameters of the energy density \eqref{free_local} : 
$
\tilde A_2=1,\,\tilde A_4=-131.57,\,\tilde A_6=4328.2,
\tilde A_1=14\,,\tilde A_3=2\,\text{and}\,\tilde K =31.11.
$
A typical  contour plot of the ensuing elastic potential is shown in Fig. \ref{tetra_func_lag}. 
  
Turning to the gradient term \eqref{free_gradient},  we observe that the adopted expression is the simplest one involving only  primary order parameters $e_2$ and $e_3$. For simplicity, we treated both of these parameters equally   and  introduced a single   coefficient $G$  bringing into the problem an  internal length scale $d_0$. To fix this parameter we  first introduce  the dimensionless   coordinates $\tilde{x} =  x/d_0 $  and   displacements  $\tilde{u} =  u/d_0 $.    We then use   the known dimensional value $G = 3.15\times 10^{-8}J/m$   obtained  from microstructural data  in \cite{Oshima1988-uu}.  and assume arbitrarily that  $G/(d_0^2f_0)=5$ which sets  the value  of the internal length scale  at $d_0=1.81 \rm{nm}$.

To deal with  dissipative dynamics, we  again adopt for  the  Rayleigh potential the  simplest  viscoelastic form:  
\begin{equation}
 \mathscr R=\int \sum_i\frac{\gamma_i }{2}{\dot e_i^2 (\bold{r},t)}  d^2x.
 \end{equation}
  where $\gamma_i$ are the associated generalized viscosity coefficients. They can be nondimensionalized  using  the characteristic time scale $t_0$ of visco-elastic  relaxation which we set  to be  $t_0 \simeq 1$ ps to   ensure  that  in the  numerical experiments the duration of realistic  avalanches  in dimensionless time scale $\tau = t/t_0$ is  of order  1.   We further assume that the non-dimensional viscosity  parameters $\tilde{\gamma}_i =  \gamma_i/(t_0 f_0)$, describing relaxation of the  primary order parameters $\tilde{\gamma}_2$ and $\tilde{\gamma}_3$,  are  both equal to 1. This would mean   that the  corresponding dimensional viscosity coefficients  are of  the order  $\gamma \simeq 21 \times 10^{-3}$ Ns/m$^2$, which is  close to the measured damping parameter  for V$_3$S alloy  also undergoing a cubic-to-tetragonal transition  \cite{Testardi1967-qo}. The remaining viscosity coefficients were chosen to be much smaller: $\tilde{\gamma}_1 = \tilde{\gamma}_4 = \tilde{\gamma}_5 = \tilde{\gamma}_6 = 0.1$.

Finally, we need to fix the degree of underdamping. It is clear  that the scale of inertial effects in our model is characterized by the dimensionless parameter $$\tilde{\rho} = \frac{\rho d_0^2}{t_0^2 f_0},$$ which plays the role analogous to the  parameter $C$ in the discrete model since $\rho$ represents the strength of inertia while the time scale $t_0$ characterizes  the strength of dissipation.  Following \cite{Ahluwalia2006-uz}, we use the estimate $\rho \sim 10^4\,\text{kg}\,\text{m}^{-3}$ which suggests that  the value $\tilde{\rho}  \sim 1$ can be used as representing a typical underdamped regime.

%The dimensionless system of dynamic equation are  given
%\begin{equation}
%\tilde\rho \partial_{\tau\tau}{u}_i = \frac{\partial}{\partial \tilde x_j} \frac{\partial \tilde F_L}{\partial \tilde u_{i,j}} - \frac{\partial}{\partial \tilde x_j} \frac{\partial}{\partial \tilde x_k} \frac{\partial \tilde F_G}{\partial \tilde u_{i,jk}} + \frac{\partial}{\partial \tilde x_j} \frac{\partial \tilde R}{ \partial_{\tau} {\tilde u}_{i,j}} \label{dynamical2}
%\end{equation}

\subsection{Numerical experiments}

Evolution of the system in such  regime was studied numerically by solving the governing equations with a time step $\Delta t = 0.01$ and using periodic boundary conditions, see Appendix~\ref{app:discretisation:continuum} for details. The simulations were carried out using a grid of size of $256\times256\times256$, which corresponds to a domain size of $0.46\rm{\mu m}\times0.46\rm{\mu m}\times0.46\rm{\mu m}$ where we used   the adopted  value of $d_0$. Note that the domain, used in the targeted experimental studies of avalanche-type dynamics, was  much larger, of order of $\rm{mm}$,  which suggests that our experiments are most probably affected by the system size and any quantitative comparison with experiment must incorporate the anticipated  size effect.

%this system size is too small compared to bulk alloys having a size of order of $\rm{mm}$ used in experimental studies of the avalanche-type dynamics.         

In our numerical experiments  we simulated thermally, rather than mechanically,  driven transformation while employing a very small cooling rate.  Specifically,  we effectively decreased  the temperature by reducing the value of $\tilde{A}_2$ using the increments  $10^{-5}$ every 1000 time steps. We started at $t=0$ where 
the  state was  homogeneous and the displacements were absent,  $\tilde u_i(\tilde{\bold r})=0$. To break the degeneracy of such   a state,  we placed a single defect  in the middle of the computational domain  which  created a controlled nucleation site. The implied perturbation  was included into the energy density  through a linear term 
%\begin{equation}
$U_{l}(\tilde{\bold r}) = -s(\tilde{\bold r})(e_2 + e_3),$
%\end{equation}
where $s(\tilde{\bold r})$ is  the external stress field acting on primary order parameters and mimicking a configuration of balanced force couples. One can see that the latter   induce  a locally  tetragonal distortion.   Specifically, following  \cite{Rodney:2001lr}, we assumed that 
\begin{equation} \label{embryo}
s(\tilde{\bold r})= \frac{s_0 e^{-|\tilde{\bold r} - \tilde{\bold r}_0|^2}}{\zeta^2},
\end{equation}
where $\tilde{\bold r}_0$ is the defect coordinate and where we set  $s_0=3$ and $\zeta=4$.

\begin{figure}[h!]
\subfigure[$\bold{t=150}$]{\includegraphics[scale=0.13 ]{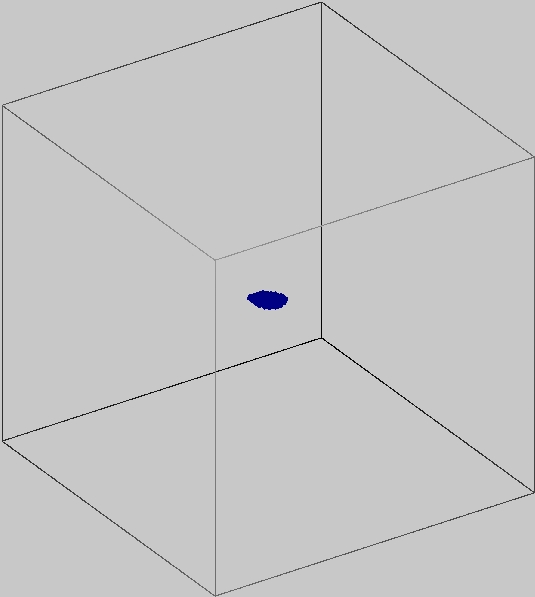}}\hspace{.05in}
\subfigure[$\bold{t=300}$]{\includegraphics[scale=0.13 ]{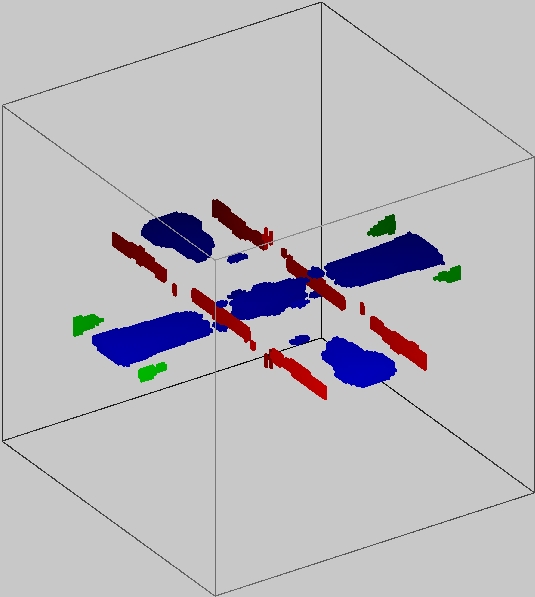}}\hspace{.05in}
\subfigure[$\bold{t=3000}$]{\includegraphics[scale=0.13 ]{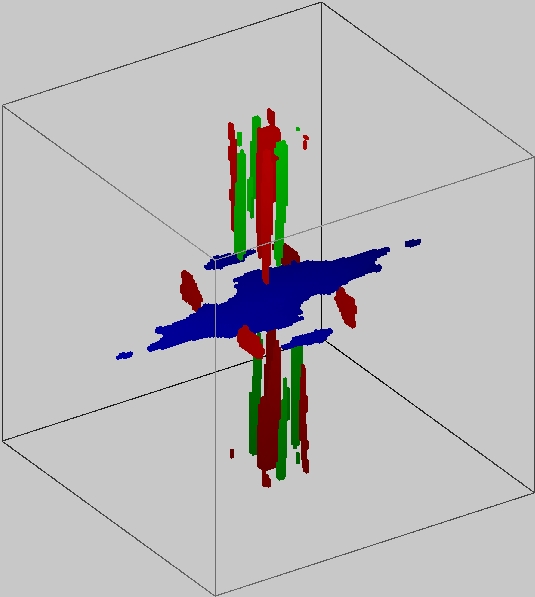}} 
 \subfigure[$\bold{t=4000}$]{\includegraphics[scale=0.13 ]{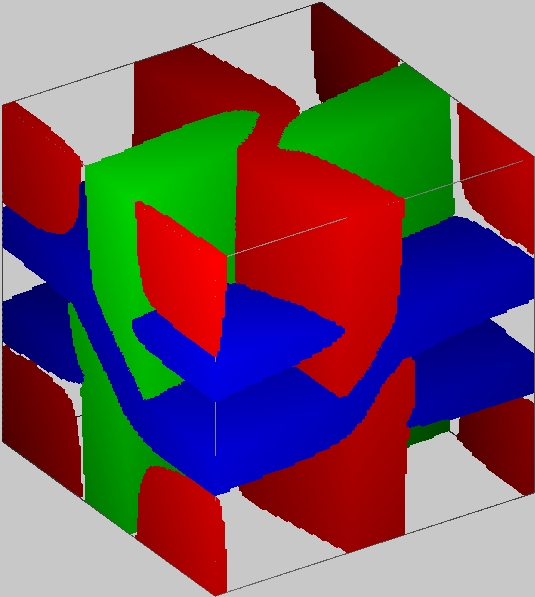}}\hspace{.05in}
%\subfigure[$\bold{t=4500}$]{\includegraphics[scale=0.1 ]{./c4fig/t_90.jpg}}\hspace{.05in}
% \subfigure[$\bold{t=6000}$]{\includegraphics[scale=0.13 ]{./t_120.jpg}} 
%\subfigure[$\bold{t=13000}$]{\includegraphics[scale=0.1 ]{./c4fig/t_260.jpg}}\hspace{.05in}
%\subfigure[$\bold{t=15000}$]{\includegraphics[scale=0.1 ]{./c4fig/t_320.jpg}}\hspace{.05in}
\subfigure[$\bold{t=16000}$]{\includegraphics[scale=0.13 ]{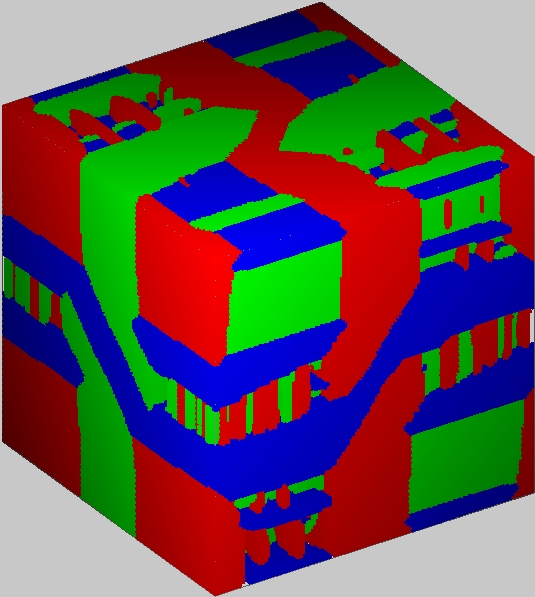}}
\caption{\label{time_evolution_ct}\small Time evolution of the system after a  nucleation event in  the underdamped 3D continuum  system.  Austenite is presented by transparent gray.  Colors represent different martensitic variants.}
\end{figure}

%The gradual decrease in the value of the parameter $\tilde A_2$   changes    the relative depth of the minima of the energy density imitating  homogeneous   cooling  of the system. Therefore, like in our 2D case,  we simulated thermally, rather than mechanically,  driven transformation. This allowed us to avoid additional inhomogeneities and spurious nucleation which are inevitable in the case when the sample is loaded either in soft or hard mechanical loading device. 

The unfolding of the transformation  processes is illustrated in Fig. \ref{time_evolution_ct} where   we show  the evolution of the system from the nucleation moment till the final state is  reached under constant cooling rate.  As we see in Fig. \ref{time_evolution_ct}(a), the homogeneous  austenite   (shown as transparent) transforms first into a single variant of martensite (variant 3 shown in blue color) around the defect which obviously  biases this particular choice. Due to complex interaction of emitted   waves,  other variants start to pop up almost immediately  (variants 1 and 2 shown in green and red colors, respectively), see Fig. \ref{time_evolution_ct}(b).  The  texture of internally twinned martensite laminates continue to grow dynamically while  progressively generating more and more complex   hierarchical pattern, see Fig. \ref{time_evolution_ct}(c,d). 
Finally,  the austenite is completely transformed into the variants of martensite which self-organize to form a multi-scale structure, see Fig. \ref{time_evolution_ct}(e).   Note that each pair of martensitic variants in the ensuing mixture state  is geometrically compatible  and therefore forming  what is known as compound twins  \cite{pitteri2002continuum}.

However,  due to incompatibility of the austenitic and martensitic energy wells,  the  formation of a single variant of martensite inside  an  austenite matrix is hardly possible  since it would have required  a forbiddingly large  energy cost.   Instead, as we have seen, right after the initial instability, we observe  the emergence  of  a complex   combination of  multi-variant  lamellae, incorporating all  symmetry related energy wells  of martensitic phase. The system effectively relies on elastic screening and compensation to minimize globally the effect of the  formation of energetically expensive   boundaries separating  martensite and austenite. The purely elastic drive towards the formation of microstructure is, of course, moderated by the gradient terms in the energy which limit the spatial scale of the emerging microstructure.

%However, because of the regularized nature of the model the corresponding twin width is not infinitesimal. Note also  that  local geometric compatibility conditions are not  respected which creates  at small scale, creates  stray stress fields along  invariant planes. 

%emergence  of twinned microstructures is, in fact,  inevitable since

To show  that the complexity of such microstructure depends crucially on the presence of inertial terms in the governing equations, we also performed n umerical experiments under the assumption of  an overdamped dynamics  with   $\tilde\rho=0.$   In this case  we essentially deal with the classical Ginzburg-Landau model which ends up generating  considerably simpler  equilibrium patterns even though microstructures do form as compatible coexistence of   martensitic variants with the austenite matrix is still  an  issue \cite{Artemev2002-vh,Mamivand2013-cm,Ni2007-ab,Turteltaub2006-ji,PhysRevB.68.224104,PhysRevB.62.5435,Hildebrand2012-is,Idesman2008-ft}.  The overdamped model obviously misses the possibility of self-focusing of elastic waves  and therefore does not  allow for secondary nucleation events  which in the underdamped case contribute in a crucial  way  to the development of multiple length scales. In other words,  without  bringing  inertia into the Ginzburg-Landau model, one completely misses  the effect of \textit{dynamic triggering} of microstructural complexity.

\begin{figure}[h!]
 \subfigure[][]{ \includegraphics[scale=.19]{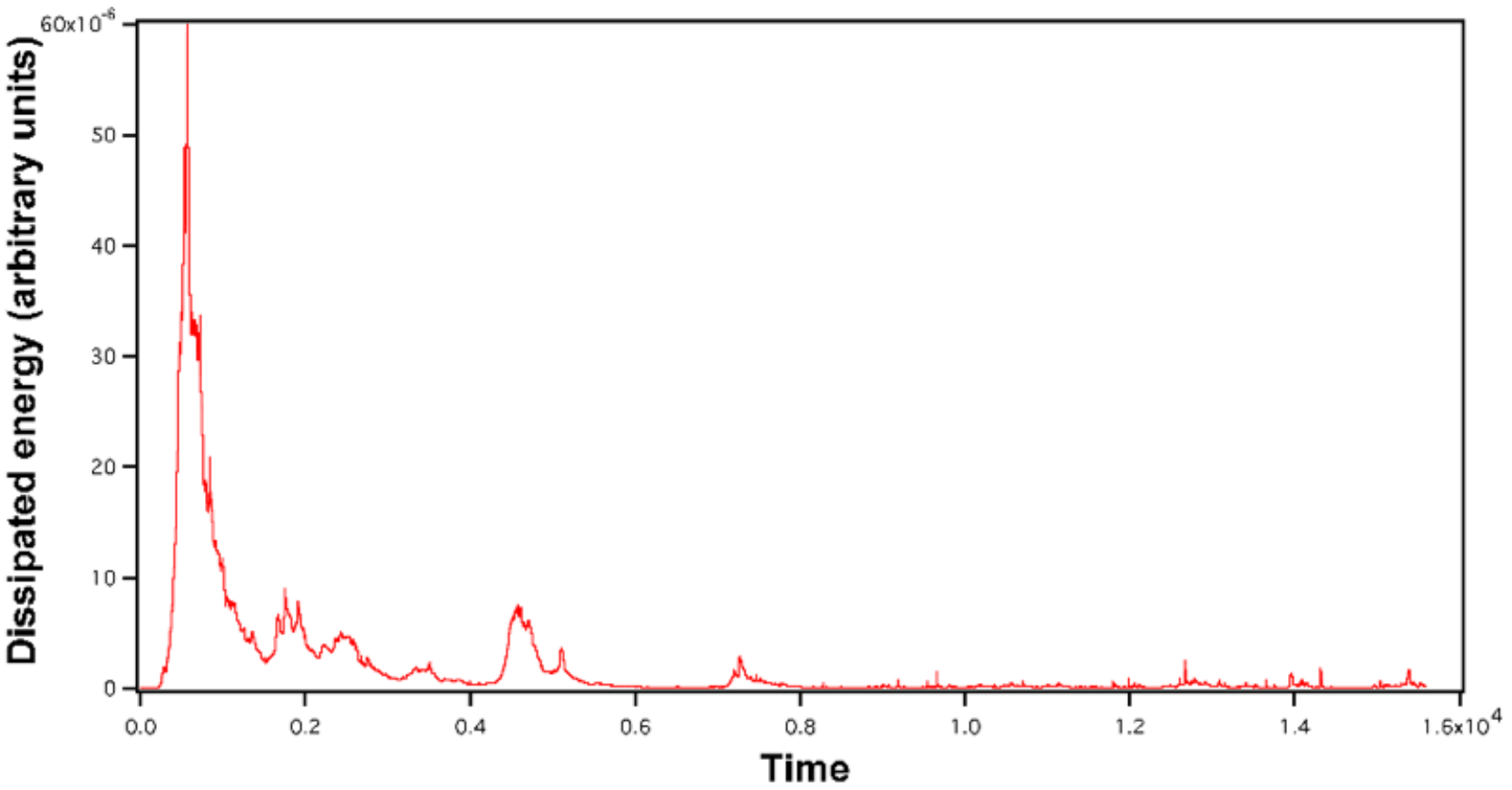}}
 \subfigure[][]{  \includegraphics[scale=0.19]{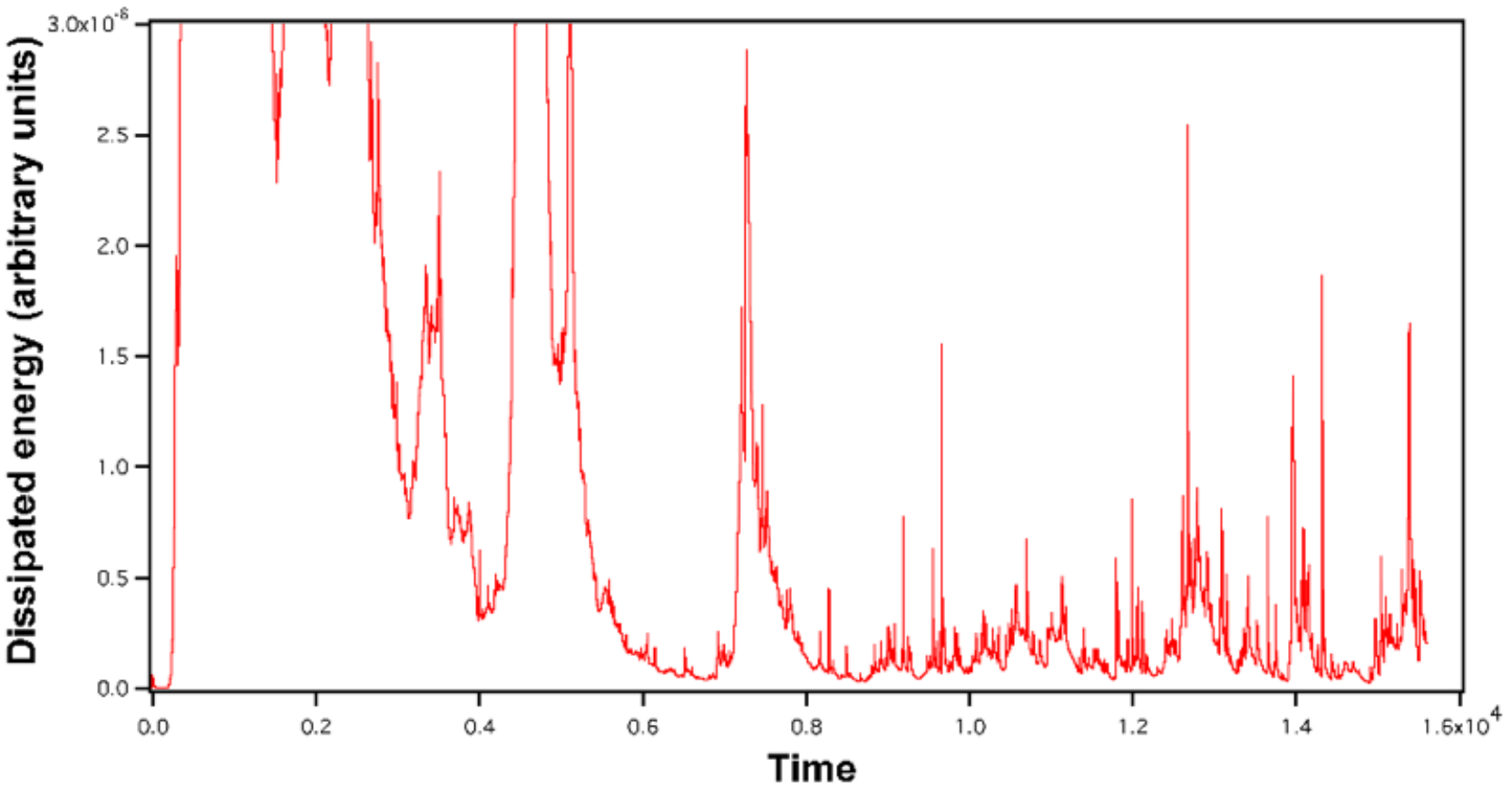}} 
  \subfigure[][]{  \includegraphics[scale=0.19]{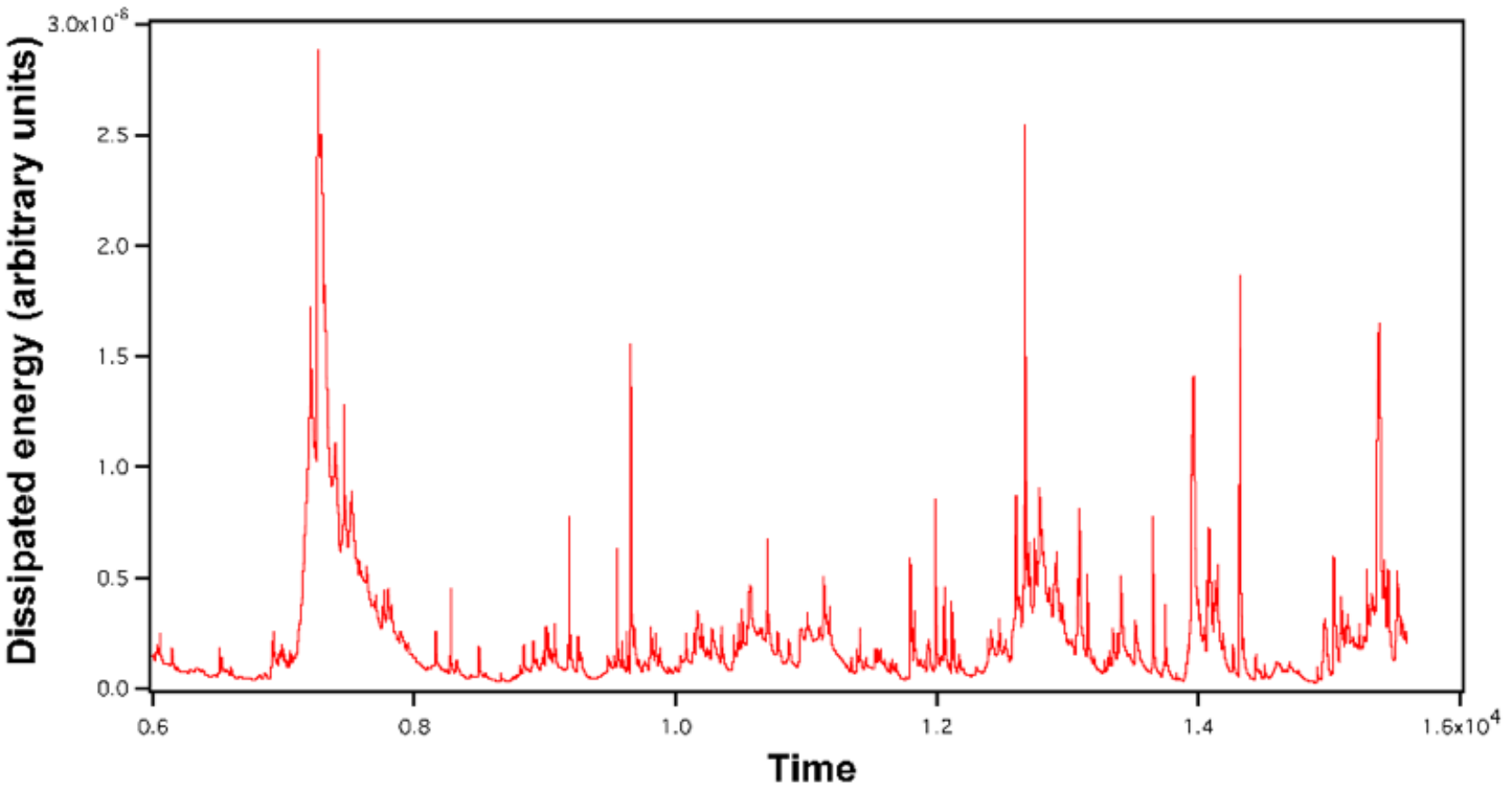}} 
\caption {\small Time evolution of the dissipated energy after the nucleation event in 3D underdamped continuum model. Panels (a-c)  represent the same signal at different  temporal  scales.}
\label{rayleigh_full}
\end{figure}
 
\subsection{Statistics of avalanches}

Consider next  the issues of intermittency and scaling in the  3D model. In Fig. \ref{rayleigh_full} we show  the time evolution of the normalized dissipated energy $ \mathscr R$ during the whole cooling process  beyond the initial nucleation stage.  We use several  levels of magnification in Fig. \ref{rayleigh_full}(a,b,c) to show that the  structure of the time series remains  basically the same. It suggests that the  corresponding  dynamics is   intermittent exhibiting broad distribution of  scales. 

The biggest burst occurs at the very early stage of the phase transformation when the transition is just initiated. As we have seen, at this stage  a complex texture of variously oriented variants of martensite appears  almost instantly inside a significant  part of the volume of the austenite phase, see  Fig. \ref{rayleigh_full} (a).  The system size avalanche, however, is formed  itself  by a large  number of small bursts representing pre- and after-shocks Fig. \ref{rayleigh_full} (b,c).  As a large number of localized  transformation events   occur at almost the same time,  the total dissipated energy is computed  by  integration over the whole sample which makes some  superposition of bursts  inevitable.

\begin{figure}[h!]
%\subfigure[][]{\includegraphics[scale=0.25]{./3D/pt.pdf}}
\includegraphics[scale=0.34]{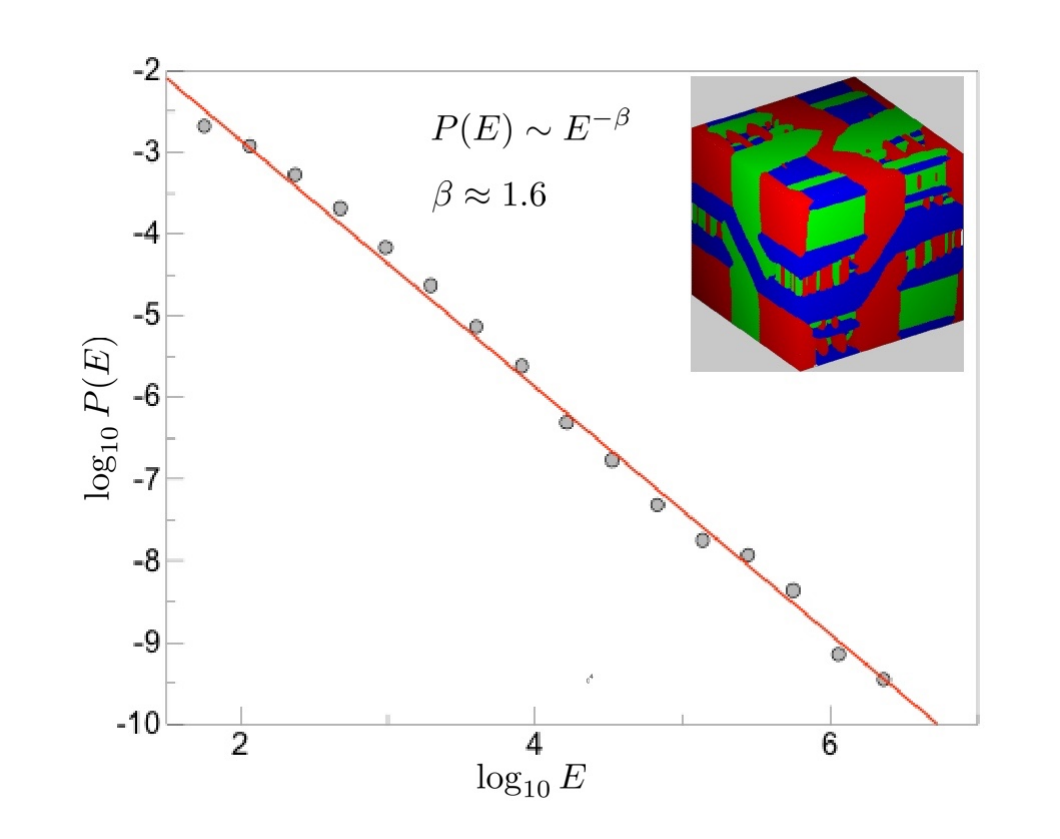}
\caption {\label{acoustic_emission1}Probability density function of avalanche energies in the 3D underdamped continuum model. Inset: final configuration with 3 martensitic variants represented by different colors.}
\end{figure}

%\textbf{ 	Fig. 21: The insets duplicate figures that have been already shown in previous figures. Without further motivation, I would remove such insets.}

The statistical structure of the observed fluctuations is  characterized  using the distribution of our three main observables:  $S$, $E$, and $T$. More specifically, We  interpreted the number of time steps during an avalanche as its duration $T$. The total dissipated energy $E$ is defined again as the discrete sum of incremental values of the energy dissipation over the duration of each avalanche. The avalanche size $S$ is calculated as in 1D using Eqs.~\ref{defs1} and \ref{defs2} within the duration $T$ of an avalanche. As in the 1D model, we used an irrelevant threshold to identify individual  avalanches and ended up processing   approximately 1800 avalanche events. Statistical distributions were constructed using logarithmic binning, with the lower bound $x_{\text{min}}$ determined by minimizing the Kolmogorov-Smirnov distance. Scaling exponents were calculated via  fitting to data above $x_{\text{min}}$, see  \cite{Clauset2009-nv} for details.

%avalanche size distribution. After we fix the simulation parameters via experimental measurements available in the literature, we again characterize the system using our three main observables $S$, $E$, and $T$. 
 
\begin{figure}[h!]
\includegraphics[scale=0.2]{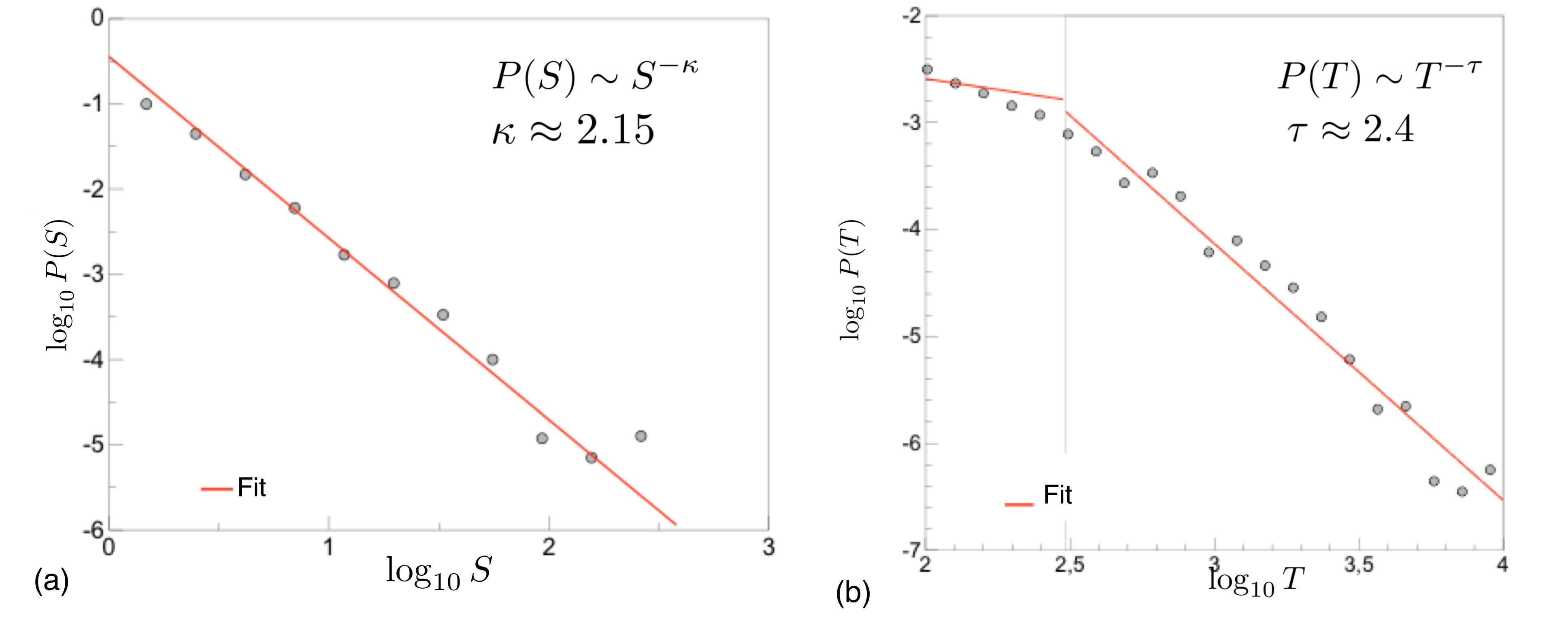}
\caption{\label{ae2}  Statistics of avalanches  in 3D underdamped continuum model: (a)  distribution of avalanche amplitudes; (b) distribution  of avalanche durations.  }
\end{figure} 

\begin{figure}[h!]
 \includegraphics[scale=0.14]{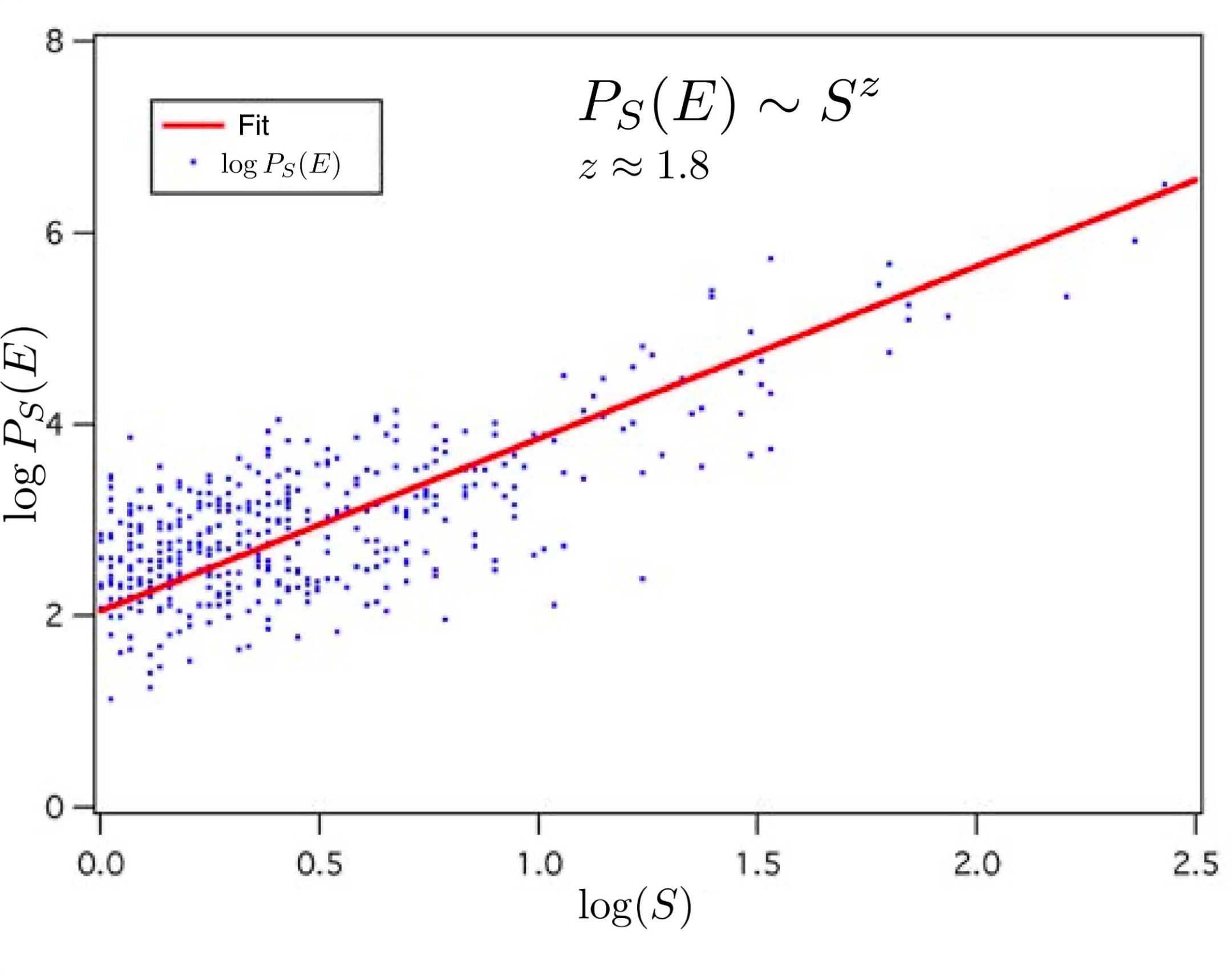}  
\caption{ Correlations between  avalanche energies $E$  and  avalanche amplitudes   in 3D underdamped continuum model.} 
\label{PSvsA}
\end{figure}
By dropping particularly small and large size events, combining  the effects of viscosity, gradient  regularization and  post-processing as well as   the   system size avalanches  at  the beginning of each new loading cycle,   we obtain the  probability density distribution of the dissipated energy  shown in  Fig. \ref{acoustic_emission1}. Over four decades it   is clearly of    power law type
\begin{equation}
P(E)\sim E^{-\beta}
\end{equation}
with exponent   $\beta=1.6$.    The attendant statistical distributions  of  avalanche  durations $T$ and  avalanche amplitudes $S$ are shown in  Fig. \ref{ae2} (a,b). here we show   a   broader range of event sizes with again almost perfect   power-law type ranges characterized by the distributions  
\begin{equation}  
P(T)\sim T^{-\tau},\,\,\,
P(S)\sim S^{-\kappa},
\end{equation}
which   exponents  
$\tau=2.4$ and $\kappa=2.15$. Finally, in Fig. \ref{PSvsA} we show the distributions of avalanche energies  $E$ plotted against the distribution of avalanche amplitudes $S$ which shows a  presence of a correlation of the form 
 \begin{equation}
E\sim S^z
\end{equation}
with exponent  $z=1.8\pm0.3$;  the fact, that $z\neq 2$ is a signature of complexity developing in this system at both temporal and spatial levels.
\begin{table}[h!]
\centering
\begin{tabular}{|c|c|c|c|c|c|c|c|}
\hline
$\kappa^{exp}$ & $\tau^{exp}$ & $\beta ^{exp}$& $z^{exp}$  \\ \hline
$2.26\pm0.1$ & - & $1.64\pm0.1$ & $1.97\pm0.4$ \\  \hline
$\kappa$ & $\tau$& $\beta$ & $z$\\  \hline
 $2.1\pm0.1$ &$2.4\pm0.1$ & $1.6\pm0.05$ & $1.8\pm0.3$  \\  \hline
\end{tabular}
\caption{\label{exp_comp} Comparison of the power law exponents   measured in   experiments \cite{Bonnot2008-qg}  with those  computed  in the 3D underdamped continuum model.}\end{table}

%Note however that   these (untruncated) distributions   show  some signs of supercritical behavior at large event sizes. To confirm the hypothesis of supercriticality (a perceived signature of underdamping)  unambiguously,  our  statistics  needs to be improved through  consideration of much larger systems.

% We reiterate that in  Fig. \ref{acoustic_emission1}, showing an almost perfect power law distribution, we eliminated particularly small events combining  the effects of viscosity, gradient  regularization and  post-processing as well as   particularly large events responsible  for the subcritical exponential finite size cut-off  and   the supercritical effect of   system size avalanches at the beginning of each new cycle.  

%

Given that  we fixed the simulation parameters using  experimental measurements available in the literature for the cubic to tetragonal martensitic transformation in Fe$_{68.8}$Pd$_{31.2}$,   we can   compare  our numerical results with  experimental data  obtained from  the  measurements of  acoustic activity in such crystals subjected to cyclic thermal driving , see  \cite{Bonnot2008-qg} for details.  In experiments, avalanches amplitudes $S$ and avalanche durations $T$ were extracted directly from the  AE signal while the avalanche energies $ E$ were obtained by integrating the square of such  signal.  In our Table \ref{exp_comp}, the numerical values of the exponents found in experiments and simulations are juxtaposed.

One can see that the computed exponents characterizing avalanche amplitudes ($\kappa$) and avalanche  energies ($\beta$)  are in excellent agreement with experiment.  The experimental value of the exponent   characterizing  energy-amplitude correlations ($z$)  is  within the error bars of  our computed  value.  Since  in experiments  apparently no power-law behavior was identified for avalanche durations,  the picture is now completed  by the results of our simulations where small but finite interval of power law behavior was found.

%It should be emphasized, however,  that the difference between experiments and simulations may also reflect  a much larger amount  of data ($\sim 10^4$) obtained and analyzed in experiment, see  \cite{Bonnot2008-qg} for details. Furthermore, the experimental data could be  averaged over thermal loading cycles ($\sim 20$) which also  improved the statistics. Instead, the computational resources available for our  numerical simulations  did not allow us to realize similar cyclic loading in 3D, see though  \cite{Salman2009-qv}.  

To summarize, despite  all   the drastic simplifications made in the modeling of  martensitic transformation in Fe$_{68.8}$Pd$_{31.2}$ crystals,  we seem to have succeeded to  capture surprisingly well the  values of experimentally measured exponents.  This can be seen as a reflection of the scale free (critical) nature of the studied phenomenon  which makes the values of the exponents    insensitive to  the details of the modeling scheme \cite{Carrillo1998-jq,Vives1995-na}.  Given that our model accounts for  neither particular quenched disorder, nor dislocational activity or any other type of  annealed disorder, and knowing that intermittency and scaling are not captured in the corresponding overdamped model, we may conclude  that  the emergence of an extended 'inertial range' in our numerical experiments  is linked  to the underdamped nature of dynamics.

%  We introduced the square root of the maximum value of the energy dissipation  $E$ as the new  measure of the avalanche amplitude. As before, the number of time steps in this interval represented the avalanche duration $T$. 

%.  One can  argue, however, that the complexity of the underlying wave motion may  be also interpreted as the presence of   co-evolving disorder which   self-tunes  the system to criticality.  While all these micro mechanisms can  obviously contribute to  the emergence  of the  scaling regime, more work is needed to identify their individual contributions to the development of   'generic' (extended) criticality.

%On the other hand, as we have seen, in the simulations the avalanches were characterized by the dissipated energy.  However, avalanche energies $E$   can be viewed as the integration of the dissipated energy and therefore can be related to the avalanche energy  measured in the experiments.  

\section{Conclusions}

The goal of this  study was to identify  the origin of the power law acoustic emission (AE)  in martensites subjected to quasistatic loading. Despite numerous proposals regarding the possible  mechanism of observed scaling behavior, the definitive model accounting for self-organization towards  'generic'  criticality in martensites has been missing.  Here, by following some earlier insights,  we explored systematically the  possibility that the implied  extended scaling regime  can be,  at least partially,  interpreted as an  effect of a constructive interplay between inertia and elastic incompatibility.  

Our  interest in the effects of inertia  is rooted in the idea   that   dynamics  should  play an important  role during steady state transformational plasticity   because   the  underlying mechanical system is close to being only marginally stable.  Indeed, it is known that the associated quasi-plastic  yielding   operates  in the part of the energetic  landscape where  the  barriers, separating  generic  metastable states, are extremely small. When inertia is  incorporated into such a model,  randomly generated elastic waves  would not be severely obstructed and scattered by these barriers and can therefore  interact sufficiently freely.  In particular, they can self-focus  creating virtual nucleation sites which can potentially  facilitates  growth of a new phase simultaneously at many scales. In a different perspective, the emerging  wave activity can be   perceived as representing  annealed self-induced disorder that  can self-tune  internally   allowing the system to reach the observed scaling regime. In all these processes the background role of  elastic incompatibility  reduces to  shaping  the energy landscape towards  coexistence of sufficiently rich metastability with  almost  marginality.
 
While the detailed mathematical structure of all the implied  feedback mechanisms  still remains unclear,  in this paper we  provided compelling evidence   that  in a model  accounting for  inertial effects and   at the same time capable of exhibiting both metastability and marginality,  the desired regime of  intermittency and scaling can be  reached.   One can say that  in such  systems  inertia serves as a mechanism of   self-organization  towards an extended  dynamical critical state. 
 
We started with a study of  an elementary example a  1D chain of mass points  connected by   springs.  While the crucial assumption, allowing one to model in this way elastic phase transitions,  was that  the   interactions   between the  nearest neighbors (NN) are  characterized by  a nonconvex  potential,  harmonic next to nearest neighbor (NNN) interactions  were also taken into account. Such an extension of the minimal NN model was  supposed to mimic  the  ferromagnetic interactions  and therefore  the developed   model can be viewed as a soft-spring version of a 1D Random Field Ising Model (RFIM).  The  main   novelty  is  in the focus on  underdamped  dynamics instead of a more conventional overdamped dynamics. We showed   that   in this   setting one can reach the   scale-free  behavior   regime  without quenched disorder  and  without  tuning.  The incorporation of    NNN interactions into a 1D model is as a 'poor man's' attempt to account for strain incompatibility between austenite and martensite energy wells.  In the same spirit, the  assumption of   discreteness, responsible for the abundance of energy wells  in  such model,   can be viewed as a way to  imitate the   complexity of actual energy landscape in transforming martensites.  

 We then  explored the effects of underdamping  in  more  realistic 2D and 3D  continuum models. Here  the  desired  balance between  metastability and marginality  can be attributed  to   elastic incompatibility  which generates long range elastic interactions. While  overdamped   Ginzburg-Landau type  continuum models  do not exhibit either  intermittency  or scaling  in the absence of quenched disorder, we   showed that  the problem can be remedied by the   account  of  inertia. Specifically, using a  physically realistic 3D continuum model,  we were able to show that  the energies, the amplitudes, and the durations of intermittent transformation-induced avalanches  all exhibit the expected   power-law behavior. The computed exponents were found to be in good agreement with those found in experimental studies  which suggests that the model captures adequately the associated inertia-dominated universality class. 
 
 This, of course,  does not completely exclude the relevance of the alternative mechanisms of self-organization towards criticality  in martensites including, for instance,  the one relying on dislocational activity.  In this perspective an interesting    generalization of the present  study  would  be to  take into account the possibility of inertia-induced dislocation nucleation.  This would allow one  to   clarify  the role of plastic activity in  thermoelastic  martensitic transformations. As we have already mentioned, some of the   martensitic transformations exhibit scaling only after extensive cyclic loading and  the role of inertia in the associated self-organization  processes still remains poorly understood. Inertia may be also   playing an important  role in dynamic phase transitions, say  inside shock waves,  which would  be  another class of problems to be addressed using the  tools developed in this paper.  Those are of course severely rate dependent phenomena and the emergence in such problems of additional,  loading dependent,  time scales can be expected to compromise scaling at least  to some extent.  
 
The most immediate extension of the present  work would be to investigate the role of inertia in   martensitic transitions   differing by their  crystallographic symmetry and  the degree of incompatibility of the energy wells. This can help to explain why some of these phase transitions do not  to exhibit scale free  AE, which remains a puzzling question. 

Finally, we mention that the  partition of the energy of wave motion  excited by the transformation  into elastic radiation, measured in AE experiments, and thermal heating,   detected  by infrared cameras, is still not well understood due to    nonlinearity involved in the   energy transfer from long to short waves  and in the process of the  eventual thermalization of lattice scale waves. Studying these phenomena  inside  a single model  can reveal additional relations between inertia and dissipation and contribute to the understanding of how a particular  degree of underdamping, ensuring scaling and criticality,  can be  actually achieved in real systems.

\section{Acknowledgments}
The authors are grateful to Francisco Perez Reche for valuable comments and suggestions. O. U. S. was supported by the grants
ANR-18-CE42-0017-03, ANR-19-CE08-0010-01, ANR-
20-CE91-0010 and  MRTN-CT-2004-505226. A. F. was supported by th grants MRTN-CT-2004-505226.  L. T. was supported by the grants ANR-17-CE08-0047-02, ANR–21-CE08-MESOCRYSP and ERC-H2020-MSCA-RISE-2020-101008140.

\appendix
\section{Spatial and temporal discretisation schemes}
\label{app:discretisation}

Here we provide some technical details behind our numerical implementation of discrete and continuum models.

\subsection{1D discrete problem}

To solve \eqref{3} with periodic boundary conditions we used  discrete Fourier transform. For a system of $N$ nodes   we can write
$$\hat{u}(q_n) = \sum_{j=0}^{N-1} \tilde{u}_j e^{-\mathrm{i}q_n j\delta}, \quad q_n = \frac{2\pi n}{L},$$
where $\mathrm{i} = \sqrt{-1}$. Combining all terms, we arrive at the Fourier space representation:
$$C \delta \frac{d^2\hat{u}(q)}{dt^2} - \frac{d\hat{u}(q)}{dt} = -2\mathrm{i}\sin(q\delta) \frac{\hat{\phi}'}{\delta} +   \frac{4G}{\delta^3}(1 - \cos(q\delta))^2 \hat{u}(q)$$
The nonlinear function $\tilde\phi'$ is first evaluated at each grid point using the current values of $\tilde{u}_i$. Once it is computed   across the entire spatial domain, we  apply the Fourier transform to obtain its spectral representation. 

If we define a mode-dependent operator 
 $L(q) = \frac{4E}{\delta^3}(1 - \cos(q\delta))^2$ 
we can  rewrite our  equation in the form 
$$C \delta \frac{d^2\hat{u}(q)}{dt^2} - \frac{d\hat{u}(q)}{dt} = -2\mathrm{i}\sin(q\delta) \frac{\hat{\phi}'(q)}{\delta} + L(q)\hat{u}(q)$$

To integrate this equation in time  we employ a standard fourth-order Runge-Kutta RK4 algorithm. 

We first define new variables 
\begin{align}
\hat{u}(q) &= y_1\\
\frac{d\hat{u}(q)}{dt} &= y_2
\end{align}
and rewrite our second order equation as a  first-order system 
\begin{align}
\frac{dy_1}{dt} &= y_2\\
\frac{dy_2}{dt} &= \frac{1}{C\delta}\left[y_2 - 2\mathrm{i}\sin(q\delta) \frac{\hat{\phi}'(q)}{\delta} + L(q)y_1\right]
\end{align}
For each time step from $t_n$ to $t_{n+1} = t_n + \Delta t$, the RK4 algorithm proceeds as follows:

1. Compute the nonlinear term $\tilde{\phi}'$ in real space using the current $\tilde{u}_i$ values, then Fourier transform it to obtain $\hat{\phi}'(q)$

2. Execute the four RK4 steps:
  \begin{align}
  k_{1,1} &= y_{2,n}\\
  k_{1,2} &= \frac{1}{C\delta}\left[y_{2,n} - 2\mathrm{i}\sin(q\delta) \frac{\hat{\phi}'(q)}{\delta} + L(q)y_{1,n}\right]\\
  k_{2,1} &= y_{2,n} + \frac{\Delta t}{2}k_{1,2} 
  \end{align}
  To compute  $k_{2,2}$ and perform subsequent steps, we need to update the nonlinear term as follows:
  
  - Transform $y_1 + \frac{\Delta t}{2}k_{1,1}$ back to real space
  
  - Compute the nonlinear term $\tilde{\phi}'$ with these updated values
  
  - Fourier transform it to obtain the update of  $\hat{\phi}'(q)$
%  \begin{align}
%  k_{2,2} &= \frac{1}{C\delta}\left[(y_{2,n} + \frac{\Delta t}{2}k_{1,2}) - 2\mathrm{i}\sin(q\delta) \frac{\hat{\phi}'_{mid1}(q)}{\delta} + L(q)(y_{1,n} + \frac{\Delta t}{2}k_{1,1})\right]\\
%  k_{3,1} &= y_{2,n} + \frac{\Delta t}{2}k_{2,2}\\
%  k_{3,2} &= \frac{1}{C\delta}\left[(y_{2,n} + \frac{\Delta t}{2}k_{2,2}) - 2\mathrm{i}\sin(q\delta) \frac{\hat{\phi}'_{mid2}(q)}{\delta} + L(q)(y_{1,n} + \frac{\Delta t}{2}k_{2,1})\right]\\ 
%  k_{4,1} &= y_{2,n} + \Delta t \, k_{3,2}\\
%  k_{4,2} &= \frac{1}{C\delta}\left[(y_{2,n} + \Delta t \, k_{3,2}) - 2\mathrm{i}\sin(q\delta) \frac{\hat{\phi}'_{mid3}(q)}{\delta} + L(q)(y_{1,n} + \Delta t \, k_{3,1})\right]
%  \end{align}
\begin{align}
k_{2,2} &= \frac{1}{C\delta}\left[(y_{2,n} + \tfrac{\Delta t}{2}k_{1,2}) - 2\mathrm{i}\sin(q\delta) \frac{\hat{\phi}'^{(2)}(q)}{\delta} + L(q)\left(y_{1,n} + \tfrac{\Delta t}{2}k_{1,1}\right)\right]\\
k_{3,1} &= y_{2,n} + \tfrac{\Delta t}{2}k_{2,2}\\
k_{3,2} &= \frac{1}{C\delta}\left[(y_{2,n} + \tfrac{\Delta t}{2}k_{2,2}) - 2\mathrm{i}\sin(q\delta) \frac{\hat{\phi}'^{(3)}(q)}{\delta} + L(q)\left(y_{1,n} + \tfrac{\Delta t}{2}k_{2,1}\right)\right]\\ 
k_{4,1} &= y_{2,n} + \Delta t \, k_{3,2}\\
k_{4,2} &= \frac{1}{C\delta}\left[(y_{2,n} + \Delta t \, k_{3,2}) - 2\mathrm{i}\sin(q\delta) \frac{\hat{\phi}'^{(4)}(q)}{\delta} + L(q)\left(y_{1,n} + \Delta t \, k_{3,1}\right)\right].
  \end{align}

Here, $\hat{\phi}'^{(i)}(q)$ denotes the Fourier transform of the nonlinear force $\phi'$ evaluated at the $i$-th intermediate estimate of the displacement. Specifically, for each stage $i = 2, 3, 4$, we compute:
\begin{itemize}
  \item Transform $y_1^{(i)}$ back to real space via inverse FFT,
  \item Evaluate $\tilde{\phi}'$ pointwise in real space using the updated displacement,
  \item Then apply FFT to obtain $\hat{\phi}'^{(i)}(q)$ in spectral form.
\end{itemize}
This ensures that the nonlinearity is updated consistently at each intermediate Runge-Kutta stage.

3. Update the solution:
  \begin{align}
  y_{1,n+1} &= y_{1,n} + \frac{\Delta t}{6}(k_{1,1} + 2k_{2,1} + 2k_{3,1} + k_{4,1})\\
  y_{2,n+1} &= y_{2,n} + \frac{\Delta t}{6}(k_{1,2} + 2k_{2,2} + 2k_{3,2} + k_{4,2})
  \end{align}

4. Recover the Fourier coefficients:
  \begin{align}
  \hat{u}_{n+1}(q) &= y_{1,n+1}\\
  \left.\frac{d\hat{u}(q)}{dt}\right|_{n+1} &= y_{2,n+1}
  \end{align}

This computational scheme provides fourth-order temporal accuracy and does not introduce numerical dissipation, preserving the spectral properties of our original equation. To implement   non-periodic (fixed) boundary conditions in this setup, we used the ghost points technique which is described in  \cite{Salman2009-qv}.

\subsection{1D continuum model}
\label{app:discretisation:continuum}
To overcome the stiffness of the time dependent problem  we had to use   an implicit-explicit time marching scheme with high temporal accuracy. The more straightforward  explicit methods require less complex structure 
but are not suitable in our  case because they require  small time step $dt$ which does not allow one 
  to reach  steady state   in a reasonable computational time. Below we illustrate  the method  using  the simplest one dimensional framework. Exactly the same type of equation is also solved in our 2D and 3D numerical experiments and the corresponding generalization is straightforward. 
  
In the interest of analytical transparency we present the  analysis  for  the following simplified   quasi-linear differential equation which has the minimal required property of being  second-order in time and fourth-order 
%( \textbf{The parameters here are not non-dimensionalized as in the main text where for instance we do not have $\gamma$. We also use there $\phi$ instead of $f$. All this needs to be harmonized.})
in space
%\begin{equation}
%\rho\ddot{u} = \partial_x f(\partial_x u)-G\partial_{xxxx}  u + \gamma\partial_{xx}   \dot{u}, \label{3.95}
%\end{equation}
\begin{equation}
C\ddot u  =\partial_{x}   \phi ' (\partial_{x}u )  +    \partial_{xx} \dot u   -G \partial_{xxxx}u .
\end{equation}

 Here it is implied that the continuum system   is discretized at  $N$ nodes and that we deal again with periodic boundary conditions.   The nonlinear term $f(\partial_x u)$ is  taken to be the same as in our model  1D equation.  
  At least second order accuracy in time is required \cite{LeVeque:2007ye} and therefore 
we choose second order  approximations  to discretize time  
\begin{equation}
\ddot{u} \approx \frac{2u^{t+1}-5u^t + 4u^{t-1}-u^{t-2}}{dt^2}, \label{3.96}
\end{equation}
\begin{equation}
\dot{u} \approx \frac{-3u^t + 6u^{t-1} + u^{t-2}}{6dt}. \label{3.97}
\end{equation}
The remaining terms are  approximated 
around the time step $t+1$.  Thus, the Taylor expansion
of the nonlinear terms $f^t$ and $f^{t-1}$ gives
\begin{equation}
f^t \approx f^{t+1}-dt f^{t+1}_t + \frac{dt^2 f^{t+1}_{tt}}{2}, \label{3.98}
\end{equation}
\begin{equation}
f^{t-1} \approx f^{t+1}-2dt f^{t+1}_t + 2dt^2 f^{t+1}_{tt}. \label{3.99}
\end{equation}
We can also write  
\begin{equation}
f^{t+1} \approx 2f^t-f^{t-1} + O(dt^2). \label{3.100}
\end{equation}

The  remaining linear terms  are  computed  implicitly at time $(t+1)$. This allows us to 
express   $u^{t+1}_i$ in terms of $u^t_i$, $u^{t-1}_i$ and $u^{t-2}_i$ 
\begin{equation}
\left(\frac{2C}{dt^2}-\frac{2\partial_{xx}}{6dt} + G\partial_{xxxx}\right)u^{t+1}
= \partial_{x}[2f^t_i-f^{t-1}]+ \rho\frac{5u^t-4u^{t-1} + u^{t-2}}{dt^2}-\partial_{xx}\frac{-3u^t_+ 6u^{t-1}+ u^{t-2}}{6dt}.
\label{3.101}
\end{equation}
This equation can be rewritten compactly if we introduce  the
linear operator 
\begin{equation}
H= \left(\frac{2C}{dt^2}-\frac{2\partial_{xx}}{6dt} + G\partial_{xxxx}\right) \label{3.102}
\end{equation}
and   the non linear function $h^t_i$  
\begin{equation}
h^t_i = \partial_x[2f^t-f^{t-1}] + C\frac{5u^t_i-4u^{t-1}_i + u^{t-2}_i}{dt^2}-\partial_{xx}\frac{-3u^t+ 6u^{t-1} + u^{t-2}}{6dt}. \label{3.103}
\end{equation}
Then, we obtain the equation
\begin{equation}
Hu^{t+1} = h^t \label{3.104} 
\end{equation}

Observe  first that dealing with spatial derivatives of  fourth order   is numerically
challenging if they are computed in  real space and therefore we transform  our equation  again  into Fourier space
\begin{equation}
\hat{H}_{q_n}\hat{u}^{t+1}_{q_n} = \hat{h}^t_{q_n}. \label{3.105}
\end{equation}
Here it is implied that  the wave vector $q$ is quantized and takes the following discrete values:
$$q_n = \frac{2\pi n}{L}, \quad n = 0, 1, 2, ..., N-1,$$
where $n$ is the mode index running from $0$ to $N-1$,  $L$ is the   physical size of the system and  $\Delta x = L/N$ is the spacing between nodes. The Fourier representation of the operator 
$$\hat{H}_{q_n} = \frac{2C}{dt^2} + \frac{2 q_n^2}{6dt} + G q_n^4, \quad n = 0, 1, 2, ..., N-1$$
 is particularly suitable for numerical implementations using the Fast Fourier Transform (FFT) algorithm, as the corresponding  operations in Fourier space for a diagonal operator  can be performed efficiently.  In particular   the operator $\hat{H}_{q_n}$   can be easily inverted.  We can then write the solution of the problem in the form
\begin{equation}
\hat{u}^{t+1}_{q_n}  = \hat{H}_{q_n} ^{-1}\hat{h}^t_{q_n}.
\label{eqfinal}
\end{equation}
 The nonlinear terms appearing in the RHS of Eq. \ref{eqfinal} are first calculated in real space and transformed into Fourier space, and their derivative is calculated as $\partial_x[2f^t-f^{t-1}] =\mathcal{F}^{-1}(\mathrm{i}q_n\mathcal{F}([2f^t-f^{t-1}] ))$, where $\mathcal{F}$ and $\mathcal{F}^{-1}$ denote direct and inverse Fourier transforms, respectively. 

%\bibliographystyle{unsrtnat}
%\bibliography{references}

\begin{thebibliography}{172}
\providecommand{\natexlab}[1]{#1}
\providecommand{\url}[1]{\texttt{#1}}
\expandafter\ifx\csname urlstyle\endcsname\relax
  \providecommand{\doi}[1]{doi: #1}\else
  \providecommand{\doi}{doi: \begingroup \urlstyle{rm}\Url}\fi

\bibitem[Planes and Vives(2017)]{Planes2017}
A.~Planes and E.~Vives.
\newblock Avalanche criticality in thermal-driven martensitic transitions: the
  asymmetry of the forward and reverse transitions in shape-memory materials.
\newblock \emph{J. Phys.: Condens. Matter}, 29\penalty0 (33):\penalty0 334001,
  2017.

\bibitem[Porta et~al.(2019)Porta, Cast{\'a}n, Saxena, and Planes]{Porta2019}
M.~Porta, T.~Cast{\'a}n, A.~Saxena, and A.~Planes.
\newblock Influence of the number of orientational domains on avalanche
  criticality in ferroelastic transitions.
\newblock \emph{Phys. Rev. E}, 100\penalty0 (6):\penalty0 062115, 2019.

\bibitem[Planes et~al.(2013)Planes, Ma{\~n}osa, and Vives]{Planes2013-ag}
Antoni Planes, Llu{\'\i}s Ma{\~n}osa, and Eduard Vives.
\newblock Acoustic emission in martensitic transformations.
\newblock \emph{J. Alloys Compd.}, 577:\penalty0 S699--S704, November 2013.

\bibitem[Perez-Reche(2017)]{PerezReche2017}
Francisco~J. Perez-Reche.
\newblock Modelling avalanches in martensites.
\newblock In \emph{Avalanches in functional materials and geophysics}, pages
  99--136. 2017.

\bibitem[Balandraud et~al.(2015)Balandraud, Barrera, Biscari, Gr{\'e}diac, and
  Zanzotto]{Balandraud2015}
X.~Balandraud, N.~Barrera, P.~Biscari, M.~Gr{\'e}diac, and G.~Zanzotto.
\newblock Strain intermittency in shape-memory alloys.
\newblock \emph{Phys. Rev. B}, 91\penalty0 (17):\penalty0 174111, 2015.

\bibitem[Sethna et~al.(2017)Sethna, Bierbaum, Dahmen, Goodrich, Greer, Hayden,
  and Zapperi]{Sethna2017}
J.~P. Sethna, M.~K. Bierbaum, K.~A. Dahmen, C.~P. Goodrich, J.~R. Greer, L.~X.
  Hayden, and S.~Zapperi.
\newblock Deformation of crystals: Connections with statistical physics.
\newblock \emph{Annu. Rev. Mater. Res.}, 47\penalty0 (1):\penalty0 217--246,
  2017.

\bibitem[Alava et~al.(2006)Alava, Nukala, and Zapperi]{Alava2006}
M.~J. Alava, P.~K. Nukala, and S.~Zapperi.
\newblock Statistical models of fracture.
\newblock \emph{Adv. Phys.}, 55\penalty0 (3-4):\penalty0 349--476, 2006.

\bibitem[Alava et~al.(2014)Alava, Laurson, and Zapperi]{Alava2014}
M.~J. Alava, L.~Laurson, and S.~Zapperi.
\newblock Crackling noise in plasticity.
\newblock \emph{Eur. Phys. J. Spec. Top.}, 223\penalty0 (11):\penalty0
  2353--2367, 2014.

\bibitem[Miguel et~al.(2001)Miguel, Vespignani, Zapperi, Weiss, and
  Grasso]{Miguel2001}
M.~C. Miguel, A.~Vespignani, S.~Zapperi, J.~Weiss, and J.~R. Grasso.
\newblock Intermittent dislocation flow in viscoplastic deformation.
\newblock \emph{Nature}, 410:\penalty0 667--671, 2001.

\bibitem[Zaiser(2006)]{Zaiser2006}
M.~Zaiser.
\newblock Scale invariance in plastic flow of crystalline solids.
\newblock \emph{Adv. Phys.}, 55\penalty0 (1-2):\penalty0 185--245, 2006.

\bibitem[Borja~da Rocha and Truskinovsky(2022)]{daRocha2022}
H.~Borja~da Rocha and L.~Truskinovsky.
\newblock Mean field fracture in disordered solids: Statistics of fluctuations.
\newblock \emph{J. Mech. Phys. Solids}, 158:\penalty0 104646, 2022.

\bibitem[Salman and Truskinovsky(2011)]{Salman2011-ij}
Oguz~Umut Salman and Lev Truskinovsky.
\newblock Minimal integer automaton behind crystal plasticity.
\newblock \emph{Phys. Rev. Lett.}, 106\penalty0 (17):\penalty0 175503, April
  2011.

\bibitem[Zhang et~al.(2020)Zhang, Salman, Weiss, and
  Truskinovsky]{Zhang2020-ax}
P~Zhang, O~U Salman, J~Weiss, and L~Truskinovsky.
\newblock Variety of scaling behaviors in nanocrystalline plasticity.
\newblock \emph{Phys Rev E}, 102\penalty0 (2-1):\penalty0 023006, August 2020.

\bibitem[Salman et~al.(2012)Salman, Finel, Delville, and Schryvers]{Salman2012}
O.~U. Salman, A.~Finel, R.~Delville, and D.~Schryvers.
\newblock The role of phase compatibility in martensite.
\newblock \emph{J. Appl. Phys}, 111\penalty0 (10), 2012.

\bibitem[Fisher(1985)]{Fisher1985}
D.~S. Fisher.
\newblock Sliding charge-density waves as a dynamic critical phenomenon.
\newblock \emph{Phys. Rev. B}, 31:\penalty0 1396, 1985.

\bibitem[Rosso et~al.(2022)Rosso, Sethna, and Wyart]{Rosso2022}
A.~Rosso, J.~P. Sethna, and M.~Wyart.
\newblock Avalanches and deformation in glasses and disordered systems.
\newblock In \emph{Emergent Dynamics in Glasses and Disordered Systems:
  Correlations and Avalanches}, Spin Glass Theory and Far Beyond: Replica
  Symmetry Breaking After 40 Years, pages 277--305. 2022.

\bibitem[LeBlanc et~al.(2014)LeBlanc, Uhl, and Dahmen]{LeBlanc2014}
M.~LeBlanc, J.~T. Uhl, and K.~A. Dahmen.
\newblock Avalanches in strained bulk metallic glasses.
\newblock \emph{Phys. Rev. Lett.}, 112:\penalty0 155501, 2014.

\bibitem[Denisov et~al.(2016)Denisov, Lorincz, Uhl, Dahmen, and
  Schall]{Denisov2016}
D.~Denisov, K.~Lorincz, J.~Uhl, K.~Dahmen, and P.~Schall.
\newblock Universality of slip avalanches in flowing granular matter.
\newblock \emph{Nature Communications}, 7:\penalty0 10641, 2016.

\bibitem[Liu et~al.(2016)Liu, Ferrero, Puosi, Barrat, and Martens]{Liu2016}
C.~Liu, E.~E. Ferrero, F.~Puosi, J.-L. Barrat, and K.~Martens.
\newblock Driving rate dependence of avalanche statistics and shapes at the
  yielding transition.
\newblock \emph{Phys. Rev. Lett.}, 116:\penalty0 065501, 2016.

\bibitem[Chan et~al.(2010)Chan, Tsekenis, Dantzig, Dahmen, and
  Goldenfeld]{Chan2010}
P.~Y. Chan, G.~Tsekenis, J.~Dantzig, K.~A. Dahmen, and N.~Goldenfeld.
\newblock Plasticity and dislocation dynamics in a phase field crystal model.
\newblock \emph{Phys. Rev. Lett.}, 105:\penalty0 015502, 2010.

\bibitem[Friedman et~al.(2012)Friedman, Jennings, Tsekenis, Kim, Tao, Uhl,
  Greer, and Dahmen]{Friedman2012}
N.~Friedman, A.~T. Jennings, G.~Tsekenis, J.-Y. Kim, M.~Tao, J.~T. Uhl, J.~R.
  Greer, and K.~A. Dahmen.
\newblock Statistics of dislocation slip avalanches in nanosized single
  crystals show tuned critical behavior.
\newblock \emph{Phys. Rev. Lett.}, 109:\penalty0 095507, 2012.

\bibitem[Bak et~al.(1987)Bak, Tang, and Wiesenfeld]{PhysRevLett.59.381}
Per Bak, Chao Tang, and Kurt Wiesenfeld.
\newblock Self-organized criticality: An explanation of the 1/f noise.
\newblock \emph{Phys. Rev. Lett.}, 59\penalty0 (4):\penalty0 381--384, 1987.

\bibitem[Dickman et~al.(2000)Dickman, Mu\~noz, Vespignani, and
  Zapperi]{Dickman2000}
R.~Dickman, M.~A. Mu\~noz, A.~Vespignani, and S.~Zapperi.
\newblock Paths to self-organized criticality.
\newblock \emph{Brazilian Journal of Physics}, 30:\penalty0 27--41, 2000.

\bibitem[Jensen(1998)]{Jensen1998}
H.~J. Jensen.
\newblock \emph{Self-Organized Criticality: Emergent Complex Behavior in
  Physical and Biological Systems}, volume~10.
\newblock Cambridge University Press, 1998.

\bibitem[Dhar(2006)]{Dhar2006}
D.~Dhar.
\newblock Theoretical studies of self-organized criticality.
\newblock \emph{Physica A}, 369\penalty0 (1):\penalty0 29--70, 2006.

\bibitem[Pruessner(2012)]{Pruessner2012}
Gunnar Pruessner.
\newblock \emph{Self-Organised Criticality: Theory, Models and
  Characterisation}.
\newblock Cambridge University Press, 2012.

\bibitem[Sornette(2006)]{Sornette2006}
D.~Sornette.
\newblock \emph{Critical Phenomena in Natural Sciences: Chaos, Fractals,
  Self-Organization and Disorder: Concepts and Tools}.
\newblock Springer Science and Business Media, 2006.

\bibitem[Gros(2024)]{Gros2024}
C.~Gros.
\newblock Self-organized criticality.
\newblock In \emph{Complex and Adaptive Dynamical Systems: A Comprehensive
  Introduction}, pages 203--239. Springer International Publishing, 2024.

\bibitem[Tadi\'c and Melnik(2021)]{Tadic2021}
B.~Tadi\'c and R.~Melnik.
\newblock Self-organised critical dynamics as a key to fundamental features of
  complexity in physical, biological, and social networks.
\newblock \emph{Dynamics}, 1\penalty0 (2):\penalty0 181--197, 2021.

\bibitem[M{\"u}ller and Wyart(2015)]{Muller2015}
M.~M{\"u}ller and M.~Wyart.
\newblock Marginal stability in structural, spin, and electron glasses.
\newblock \emph{Annu. Rev. Condens. Matter Phys.}, 6:\penalty0 177, 2015.

\bibitem[Shang et~al.(2020)Shang, Guan, and Barrat]{Shang2020}
B.~Shang, P.~Guan, and J.~L. Barrat.
\newblock Elastic avalanches reveal marginal behavior in amorphous solids.
\newblock \emph{Proc. Natl. Acad. Sci.}, 117\penalty0 (1):\penalty0 86--92,
  2020.

\bibitem[Ovaska et~al.(2017)Ovaska, Lehtinen, Alava, Laurson, and
  Zapperi]{Ovaska2017}
M.~Ovaska, A.~Lehtinen, M.~J. Alava, L.~Laurson, and S.~Zapperi.
\newblock Excitation spectra in crystal plasticity.
\newblock \emph{Phys. Rev. Lett.}, 119\penalty0 (26):\penalty0 265501, 2017.

\bibitem[Franz and Spigler(2017)]{Franz2017}
S.~Franz and S.~Spigler.
\newblock Mean-field avalanches in jammed spheres.
\newblock \emph{Phys. Rev. E}, 95\penalty0 (2):\penalty0 022139, 2017.

\bibitem[Nishiyama(2012)]{nishiyama2012martensitic}
Zenji Nishiyama.
\newblock \emph{Martensitic transformation}.
\newblock Elsevier, 2012.

\bibitem[Roitburd(1978)]{roitburd1978martensitic}
AL~Roitburd.
\newblock Martensitic transformation as a typical phase transformation in
  solids.
\newblock In \emph{Solid state physics}, volume~33, pages 317--390. Elsevier,
  1978.

\bibitem[James and Hane(2000)]{james2000martensitic}
Richard~D James and Kevin~F Hane.
\newblock Martensitic transformations and shape-memory materials.
\newblock \emph{Acta Mater}, 48\penalty0 (1):\penalty0 197--222, 2000.

\bibitem[Khachaturyan(2013)]{khachaturyan2013theory}
Armen~G Khachaturyan.
\newblock \emph{Theory of structural transformations in solids}.
\newblock Courier Corporation, 2013.

\bibitem[Bhattacharya(2003)]{Bhattacharya2003}
K.~Bhattacharya.
\newblock \emph{Microstructure of martensite}.
\newblock Oxford University Press, 2003.

\bibitem[P\'erez-Reche et~al.(2004)P\'erez-Reche, Stipcich, Vives, Ma\~nosa,
  Planes, and Morin]{PhysRevB.69.064101}
Francisco-Jos\'e P\'erez-Reche, Marcelo Stipcich, Eduard Vives, Llu\'{\i}s
  Ma\~nosa, Antoni Planes, and Michel Morin.
\newblock Kinetics of martensitic transitions in cu-al-mn under thermal
  cycling: Analysis at multiple length scales.
\newblock \emph{Phys. Rev. B}, 69:\penalty0 064101, Feb 2004.

\bibitem[Chandni et~al.(2009)Chandni, Ghosh, Vijaya, and Mohan]{Chandni2009}
U.~Chandni, A.~Ghosh, H.~S. Vijaya, and S.~Mohan.
\newblock Criticality of tuning in athermal phase transitions.
\newblock \emph{Phys. Rev. Lett.}, 102\penalty0 (2):\penalty0 025701, 2009.

\bibitem[Rosinberg and Vives(2011)]{Rosinberg2011}
M.~L. Rosinberg and E.~Vives.
\newblock Metastability, hysteresis, avalanches, and acoustic emission:
  Martensitic transitions in functional materials.
\newblock In \emph{Disorder and Strain-Induced Complexity in Functional
  Materials}, pages 249--272. Springer Berlin Heidelberg, Berlin, Heidelberg,
  2011.

\bibitem[Vives et~al.(2011)Vives, Soto~Parra, Planes, Ma{\~n}osa, Romero,
  Edwards, and Dixon]{Vives2011}
E.~Vives, D.~E. Soto~Parra, A.~Planes, L.~Ma{\~n}osa, R.~Romero, R.~S. Edwards,
  and S.~Dixon.
\newblock Acoustic emission avalanches in martensitic transitions: new
  perspectives for the problem of source location.
\newblock \emph{Solid State Phenom.}, 172:\penalty0 144--149, 2011.

\bibitem[Gallardo et~al.(2010)Gallardo, Manchado, Romero, Del~Cerro, Salje,
  Planes, and Stipcich]{Gallardo2010}
M.~C. Gallardo, J.~Manchado, F.~J. Romero, J.~Del~Cerro, E.~K. Salje,
  A.~Planes, and M.~Stipcich.
\newblock Avalanche criticality in the martensitic transition of cu 67.64 zn
  16.71 al 15.65 shape-memory alloy: A calorimetric and acoustic emission
  study.
\newblock \emph{Phys. Rev. B}, 81\penalty0 (17):\penalty0 174102, 2010.

\bibitem[Beke et~al.(2019)Beke, Dar{\'o}czi, T{\'o}th, Bolg{\'a}r, Samy, and
  Hud{\'a}k]{Beke2019}
D.~L. Beke, L.~Dar{\'o}czi, L.~Z. T{\'o}th, M.~K. Bolg{\'a}r, N.~M. Samy, and
  A.~Hud{\'a}k.
\newblock Acoustic emissions during structural changes in shape memory alloys.
\newblock \emph{Metals}, 9\penalty0 (1):\penalty0 58, 2019.

\bibitem[Illa et~al.(2015)Illa, Winkelmayer, and Vives]{Illa2015}
X.~Illa, P.~Winkelmayer, and E.~Vives.
\newblock Local strain variability and force fluctuations during the
  martensitic transition under different driving mechanisms.
\newblock \emph{Phys. Rev. B}, 92\penalty0 (18):\penalty0 184107, 2015.

\bibitem[Vives et~al.(2009)Vives, Soto-Parra, Ma\~nosa, Romero, and
  Planes]{PhysRevB.80.180101}
Eduard Vives, Daniel Soto-Parra, Llu\'{\i}s Ma\~nosa, Ricardo Romero, and
  Antoni Planes.
\newblock Driving-induced crossover in the avalanche criticality of martensitic
  transitions.
\newblock \emph{Phys. Rev. B}, 80:\penalty0 180101, Nov 2009.

\bibitem[Torrents et~al.(2017)Torrents, Illa, Vives, and
  Planes]{Torrents2017-kk}
Gen{\'\i}s Torrents, Xavier Illa, Eduard Vives, and Antoni Planes.
\newblock Geometrical model for martensitic phase transitions: Understanding
  criticality and weak universality during microstructure growth.
\newblock \emph{Phys Rev E}, 95\penalty0 (1-1):\penalty0 013001, January 2017.

\bibitem[Beke et~al.(2018)Beke, Bolg{\'a}r, T{\'o}th, and
  Dar{\'o}czi]{Beke2018-xv}
D~L Beke, M~K Bolg{\'a}r, L~Z T{\'o}th, and L~Dar{\'o}czi.
\newblock On the asymmetry of the forward and reverse martensitic
  transformations in shape memory alloys.
\newblock \emph{J. Alloys Compd.}, 741:\penalty0 106--115, April 2018.

\bibitem[Song et~al.(2013)Song, Chen, Dabade, Shield, and James]{Song2013-os}
Yintao Song, Xian Chen, Vivekanand Dabade, Thomas~W Shield, and Richard~D
  James.
\newblock Enhanced reversibility and unusual microstructure of a
  phase-transforming material.
\newblock \emph{Nature}, 502\penalty0 (7469):\penalty0 85--88, October 2013.

\bibitem[Sethna et~al.(2001)Sethna, Dahmen, and Myers]{Sethna2001}
J.~P. Sethna, K.~A. Dahmen, and C.~R. Myers.
\newblock Crackling noise.
\newblock \emph{Nature}, 410:\penalty0 242, 2001.

\bibitem[Sethna et~al.(1993)Sethna, Dahmen, Kartha, Krumhansl, Roberts, and
  Shore]{PhysRevLett.70.3347}
James~P. Sethna, Karin Dahmen, Sivan Kartha, James~A. Krumhansl, Bruce~W.
  Roberts, and Joel~D. Shore.
\newblock Hysteresis and hierarchies: Dynamics of disorder-driven first-order
  phase transformations.
\newblock \emph{Phys. Rev. Lett.}, 70\penalty0 (21):\penalty0 3347--3350, May
  1993.

\bibitem[Perkovi{\'c} et~al.(1995)Perkovi{\'c}, Dahmen, and
  Sethna]{PhysRevLett.75.4528}
Olga Perkovi{\'c}, Karin Dahmen, and James~P. Sethna.
\newblock Avalanches, barkhausen noise, and plain old criticality.
\newblock \emph{Phys. Rev. Lett.}, 75\penalty0 (24):\penalty0 4528--4531, Dec
  1995.

\bibitem[Dahmen and Sethna(1996)]{PhysRevB.53.14872}
Karin Dahmen and James~P. Sethna.
\newblock Hysteresis, avalanches, and disorder-induced critical scaling: A
  renormalization-group approach.
\newblock \emph{Phys. Rev. B}, 53\penalty0 (22):\penalty0 14872--14905, Jun
  1996.

\bibitem[Vives and Planes(1994)]{PhysRevB.50.3839}
Eduard Vives and Antoni Planes.
\newblock Avalanches in a fluctuationless first-order phase transition in a
  random-bond ising model.
\newblock \emph{Phys. Rev. B}, 50\penalty0 (6):\penalty0 3839--3848, Aug 1994.

\bibitem[Tadic(1996)]{PhysRevLett.77.3843}
B~Tadic.
\newblock Nonuniversal scaling behavior of barkhausen noise.
\newblock \emph{Phys. Rev. Lett.}, 77\penalty0 (18):\penalty0 3843--3846,
  October 1996.

\bibitem[Vives and Planes(2001)]{PhysRevB.63.134431}
Eduard Vives and Antoni Planes.
\newblock Hysteresis and avalanches in the random anisotropy ising model.
\newblock \emph{Phys. Rev. B}, 63\penalty0 (13):\penalty0 134431, Mar 2001.

\bibitem[P{\'e}rez-Reche et~al.(2008)P{\'e}rez-Reche, Truskinovsky, and
  Zanzotto]{Perez-Reche2008}
Francisco-Jos{\'e} P{\'e}rez-Reche, Lev Truskinovsky, and Giovanni Zanzotto.
\newblock Driving-induced crossover: From classical criticality to
  self-organized criticality.
\newblock \emph{Phys. Rev. Lett.}, 101\penalty0 (23):\penalty0 27, December
  2008.

\bibitem[Sethna et~al.(2006)Sethna, Dahmen, and Perkovic]{Sethna2006}
J.~P. Sethna, K.~A. Dahmen, and O.~Perkovic.
\newblock \emph{Random-field Ising models of hysteresis}.
\newblock Academic Press, New York, 2006.

\bibitem[P{\'e}rez-Reche et~al.(2007)P{\'e}rez-Reche, Truskinovsky, and
  Zanzotto]{Perez-Reche:2007et}
Francisco-Jos{\'e} P{\'e}rez-Reche, Lev Truskinovsky, and Giovanni Zanzotto.
\newblock Training-induced criticality in martensites.
\newblock \emph{Phys. Rev. Lett.}, 99\penalty0 (7), August 2007.

\bibitem[Perez-Reche et~al.(2016)Perez-Reche, Triguero, Zanzotto, and
  Truskinovsky]{PerezReche2016}
F.~J. Perez-Reche, C.~Triguero, G.~Zanzotto, and L.~Truskinovsky.
\newblock Origin of scale-free intermittency in structural first-order phase
  transitions.
\newblock \emph{Phys. Rev. B}, 94\penalty0 (14):\penalty0 144102, 2016.

\bibitem[Perez-Reche et~al.(2009)Perez-Reche, Truskinovsky, and
  Zanzotto]{PerezReche2009}
F.~J. Perez-Reche, L.~Truskinovsky, and G.~Zanzotto.
\newblock Martensitic transformations: from continuum mechanics to spin models
  and automata.
\newblock \emph{Contin. Mech. Thermodyn.}, 21:\penalty0 17--26, 2009.

\bibitem[Carrillo et~al.(1998)Carrillo, Ma{\~n}osa, Ort{\'\i}n, Planes, and
  Vives]{Carrillo1998-jq}
Llu{\'\i}s Carrillo, Llu{\'\i}s Ma{\~n}osa, Jordi Ort{\'\i}n, Antoni Planes,
  and Eduard Vives.
\newblock Experimental evidence for universality of acoustic emission avalanche
  distributions during structural transitions.
\newblock \emph{Phys. Rev. Lett.}, 81\penalty0 (9):\penalty0 1889--1892, August
  1998.

\bibitem[Bonnot et~al.(2008)Bonnot, Romero, Ma{\~n}osa, Vives, and
  Planes]{Bonnot2008-qg}
Erell Bonnot, Ricardo Romero, Llu{\'\i}s Ma{\~n}osa, Eduard Vives, and Antoni
  Planes.
\newblock Elastocaloric effect associated with the martensitic transition in
  shape-memory alloys.
\newblock \emph{Phys. Rev. Lett.}, 100\penalty0 (12):\penalty0 125901, March
  2008.

\bibitem[Eshelby(1957)]{Eshelby1957}
J.~D. Eshelby.
\newblock The determination of the elastic field of an ellipsoidal inclusion,
  and related problems.
\newblock \emph{Proceedings of the Royal Society of London A}, 241:\penalty0
  376, 1957.

\bibitem[Picard et~al.(2004)Picard, Ajdari, Lequeux, and Bocquet]{Picard2004}
G.~Picard, A.~Ajdari, F.~Lequeux, and L.~Bocquet.
\newblock Elastic consequences of a single plastic event: A step towards the
  microscopic modeling of the flow of yield stress fluids.
\newblock \emph{Europhysics Journal E}, 15:\penalty0 371, 2004.

\bibitem[Rossi et~al.(2023)Rossi, Biroli, Ozawa, and Tarjus]{Rossi2023}
S.~Rossi, G.~Biroli, M.~Ozawa, and G.~Tarjus.
\newblock Far-from-equilibrium criticality in the random-field ising model with
  eshelby interactions.
\newblock \emph{Phys. Rev. B}, 108\penalty0 (22):\penalty0 L220202, 2023.

\bibitem[Clapp(1995)]{clapp1995would}
PC~Clapp.
\newblock How would we recognize a martensitic transformation if it bumped into
  us on a dark \& austy night?
\newblock \emph{Le Journal de Physique IV}, 5\penalty0 (C8):\penalty0 C8--11,
  1995.

\bibitem[Miao and Vlassak(2020)]{miao2020explosive}
Yucong Miao and Joost~J Vlassak.
\newblock Explosive martensitic transformation of supercooled austenite in
  cuzr-based thin-film shape memory alloys.
\newblock \emph{Acta Mater}, 200:\penalty0 162--170, 2020.

\bibitem[Schwabe et~al.(2022)Schwabe, L{\"u}nser, Schmidt, Nielsch, Gaal, and
  F{\"a}hler]{schwabe2022speed}
Stefan Schwabe, Klara L{\"u}nser, Daniel Schmidt, Kornelius Nielsch, Peter
  Gaal, and Sebastian F{\"a}hler.
\newblock What is the speed limit of martensitic transformations?
\newblock \emph{Sci. Technol. Adv. Mater.}, 23\penalty0 (1):\penalty0 633--641,
  2022.

\bibitem[Lin and Pence(1998)]{lin1998pulse}
Jiehliang Lin and Thomas~J Pence.
\newblock Pulse attenuation by kinetically active phase boundary scattering
  during displacive phase transformations.
\newblock \emph{J. Mech. Phys. Solids}, 46\penalty0 (7):\penalty0 1183--1211,
  1998.

\bibitem[Truskinovsky(2002)]{truskinovsky2002nucleation}
Lev Truskinovsky.
\newblock Nucleation and growth in elastodynamics.
\newblock In \emph{Dynamics of Crystal Surfaces and Interfaces}, pages
  185--197. Springer, 2002.

\bibitem[Lookman et~al.(2003)Lookman, Shenoy, Rasmussen, Saxena, and
  Bishop]{PhysRevB.67.024114}
T.~Lookman, S.~R. Shenoy, K.~\O{}. Rasmussen, A.~Saxena, and A.~R. Bishop.
\newblock Ferroelastic dynamics and strain compatibility.
\newblock \emph{Phys. Rev. B}, 67\penalty0 (2):\penalty0 024114, 2003.

\bibitem[Truskinovsky and Vainchtein(2008)]{Truskinovsky2008-fn}
Lev Truskinovsky and Anna Vainchtein.
\newblock Dynamics of martensitic phase boundaries: discreteness, dissipation
  and inertia.
\newblock \emph{Contin. Mech. Thermodyn.}, 20\penalty0 (2):\penalty0 97--122,
  May 2008.

\bibitem[Steinbach and Shchyglo(2011)]{steinbach2011phase}
Ingo Steinbach and Oleg Shchyglo.
\newblock Phase-field modelling of microstructure evolution in solids:
  perspectives and challenges.
\newblock \emph{Curr. Opin. Solid State Mater. Sci.}, 15\penalty0 (3):\penalty0
  87--92, 2011.

\bibitem[Cho et~al.(2012)Cho, Idesman, Levitas, and Park]{cho2012finite}
J-Y Cho, AV~Idesman, VI~Levitas, and Taehyo Park.
\newblock Finite element simulations of dynamics of multivariant martensitic
  phase transitions based on ginzburg--landau theory.
\newblock \emph{International Journal of Solids and Structures}, 49\penalty0
  (14):\penalty0 1973--1992, 2012.

\bibitem[Liu et~al.(2024)Liu, Schneider, Reder, Hoffrogge, and
  Nestler]{liu2024modeling}
Xiaoying Liu, Daniel Schneider, Martin Reder, Paul~W Hoffrogge, and Britta
  Nestler.
\newblock Modeling of martensitic phase transformation accounting for inertia
  effects.
\newblock \emph{Int. J. Mech. Sci}, 278:\penalty0 109443, 2024.

\bibitem[Vives et~al.(1994)Vives, Ort\'in, Ma{\~n}osa, R\`afols,
  P\'erez-Magran\'e, and Planes]{PhysRevLett.72.1694}
Eduard Vives, Jordi Ort\'in, Llu\'is Ma{\~n}osa, Ismael R\`afols, Ramon
  P\'erez-Magran\'e, and Antoni Planes.
\newblock Distributions of avalanches in martensitic transformations.
\newblock \emph{Phys. Rev. Lett.}, 72\penalty0 (11):\penalty0 1694--1697, 1994.

\bibitem[Ma{\~n}osa et~al.(1989)Ma{\~n}osa, Planes, Rouby, Morin, Fleischmann,
  and Macqueron]{Manosa1989-fn}
Ll~Ma{\~n}osa, A~Planes, D~Rouby, M~Morin, P~Fleischmann, and J~L Macqueron.
\newblock Acoustic emission field during thermoelastic martensitic
  transformations.
\newblock \emph{Appl. Phys. Lett.}, 54\penalty0 (25):\penalty0 2574--2576, June
  1989.

\bibitem[Bar{\'o} et~al.(2013)Bar{\'o}, Corral, Illa, Planes, Salje, Schranz,
  Soto-Parra, and Vives]{baro2013statistical}
Jordi Bar{\'o}, {\'A}lvaro Corral, Xavier Illa, Antoni Planes, Ekhard~KH Salje,
  Wilfried Schranz, Daniel~E Soto-Parra, and Eduard Vives.
\newblock Statistical similarity between the compression of a porous material
  and earthquakes.
\newblock \emph{Phys. Rev. Lett.}, 110\penalty0 (8):\penalty0 088702, 2013.

\bibitem[Gomberg and Johnson(2005)]{gomberg2005dynamic}
Joan Gomberg and Paul Johnson.
\newblock Dynamic triggering of earthquakes.
\newblock \emph{Nature}, 437\penalty0 (7060):\penalty0 830--830, 2005.

\bibitem[Johnson and Jia(2005)]{johnson2005nonlinear}
Paul~A Johnson and Xiaoping Jia.
\newblock Nonlinear dynamics, granular media and dynamic earthquake triggering.
\newblock \emph{Nature}, 437\penalty0 (7060):\penalty0 871--874, 2005.

\bibitem[Kocharyan(2021)]{kocharyan2021nucleation}
GG~Kocharyan.
\newblock Nucleation and evolution of sliding in continental fault zones under
  the action of natural and man-made factors: A state-of-the-art review.
\newblock \emph{Izvestiya, Physics of the Solid Earth}, 57:\penalty0 439--473,
  2021.

\bibitem[Khfifi and Loulidi(2008)]{Khfifi2008}
M.~Khfifi and M.~Loulidi.
\newblock Scaling properties of a rice-pile model: Inertia and friction
  effects.
\newblock \emph{Phys. Rev. E}, 78:\penalty0 051117, 2008.

\bibitem[Maimon and Schwarz(2004)]{Maimon2004}
R.~Maimon and J.~M. Schwarz.
\newblock Continuous depinning transition with an unusual hysteresis effect.
\newblock \emph{Phys. Rev. Lett.}, 92:\penalty0 255502, 2004.

\bibitem[Papanikolaou(2016)]{Papanikolaou2016}
S.~Papanikolaou.
\newblock Shearing a glass and the role of pinning delay in models of interface
  depinning.
\newblock \emph{Phys Rev E}, 93:\penalty0 032610, 2016.

\bibitem[Carlson and Langer(1989)]{Carlson1989}
J.~M. Carlson and J.~S. Langer.
\newblock Properties of earthquakes generated by fault dynamics.
\newblock \emph{Phys. Rev. Lett.}, 62:\penalty0 2632, 1989.

\bibitem[Prado and Olami(1992)]{Prado1992}
C.~P.~C. Prado and Z.~Olami.
\newblock Inertia and break of self-organized criticality in sandpile
  cellular-automata models.
\newblock \emph{Phys. Rev. A}, 45:\penalty0 665, 1992.

\bibitem[Held et~al.(1990)Held, Solina, Keane, Haag, Horn, and
  Grinstein]{Held1990}
G.~A. Held, D.~H. Solina, D.~T. Keane, W.~J. Haag, P.~M. Horn, and
  G.~Grinstein.
\newblock Experimental study of critical-mass fluctuations in an evolving
  sandpile.
\newblock \emph{Phys. Rev. Lett.}, 65:\penalty0 1120, 1990.

\bibitem[Jaeger et~al.(1989)Jaeger, Liu, and Nagel]{Jaeger1989}
H.~M. Jaeger, C.-h. Liu, and S.~R. Nagel.
\newblock Relaxation at the angle of repose.
\newblock \emph{Phys. Rev. Lett.}, 62:\penalty0 40, 1989.

\bibitem[Marchetti(2006)]{Marchetti2006-oa}
M~C Marchetti.
\newblock Depinning and plasticity of driven disordered lattices.
\newblock In M~Carmen~Miguel and Miguel Rubi, editors, \emph{Jamming, Yielding,
  and Irreversible Deformation in Condensed Matter}, Lecture Notes in Physics,
  pages 137--157. Springer-Verlag, 2006.

\bibitem[Denisov et~al.(2017)Denisov, Lorincz, Wright, Hufnagel, Nawano, Gu,
  Uhl, Dahmen, and Schall]{Denisov2017}
D.~V. Denisov, K.~A. Lorincz, W.~J. Wright, T.~C. Hufnagel, A.~Nawano, X.~Gu,
  J.~T. Uhl, K.~A. Dahmen, and P.~Schall.
\newblock Universal slip dynamics in metallic glasses and granular matter --
  linking frictional weakening with inertial effects.
\newblock \emph{Scientific Reports}, 7:\penalty0 43376, 2017.

\bibitem[Nicolas et~al.(2016)Nicolas, Barrat, and Rottler]{Nicolas2016}
A.~Nicolas, J.-L. Barrat, and J.~Rottler.
\newblock Effects of inertia on the steady-shear rheology of disordered solids.
\newblock \emph{Phys. Rev. Lett.}, 116:\penalty0 058303, 2016.

\bibitem[Karimi and Barrat(2016)]{Karimi2016}
K.~Karimi and J.~L. Barrat.
\newblock Role of inertia in the rheology of amorphous systems: A
  finite-element-based elastoplastic model.
\newblock \emph{Phys Rev E}, 93\penalty0 (2):\penalty0 022904, 2016.

\bibitem[de~Geus and Wyart(2024)]{deGeus2024}
T.~W. de~Geus and M.~Wyart.
\newblock Short-range depinning in the presence of velocity-weakening.
\newblock \emph{arXiv preprint}, 2024.

\bibitem[Clancy and Corcoran(2005)]{Clancy2005}
I.~Clancy and D.~Corcoran.
\newblock Criticality in the burridge-knopoff model.
\newblock \emph{Phys Rev E}, 71\penalty0 (4):\penalty0 046124, 2005.

\bibitem[Salerno et~al.(2012)Salerno, Maloney, and Robbins]{Salerno2012}
K.~M. Salerno, C.~E. Maloney, and M.~O. Robbins.
\newblock Avalanches in strained amorphous solids: does inertia destroy
  critical behavior?
\newblock \emph{Phys. Rev. Lett.}, 109:\penalty0 105703, 2012.

\bibitem[Salerno and Robbins(2013)]{Salerno2013}
K.~M. Salerno and M.~O. Robbins.
\newblock Effect of inertia on sheared disordered solids: Critical scaling of
  avalanches in two and three dimensions.
\newblock \emph{Phys Rev E}, 88:\penalty0 062206, 2013.

\bibitem[Karimi et~al.(2017)Karimi, Ferrero, and Barrat]{Karimi2017}
K.~Karimi, E.~E. Ferrero, and J.-L. Barrat.
\newblock Inertia and universality of avalanche statistics: the case of slowly
  deformed amorphous solids.
\newblock \emph{Phys Rev E}, 95:\penalty0 013003, 2017.

\bibitem[Truskinovsky(1994)]{truskinovsky1994transition}
Lev Truskinovsky.
\newblock Transition to detonation in dynamic phase changes.
\newblock \emph{Arch. Ration. Mech. Anal.}, 125:\penalty0 375--397, 1994.

\bibitem[Reid and Gooding(1997)]{reid1997pattern}
ACE Reid and RJ~Gooding.
\newblock Pattern formation in a 2d elastic solid.
\newblock \emph{Physica A}, 239\penalty0 (1-3):\penalty0 1--10, 1997.

\bibitem[Bales and Gooding(1991)]{PhysRevLett.67.3412}
G.S. Bales and R.J. Gooding.
\newblock Interfacial dynamics at a first-order phase transition involving
  strain: Dynamical twin formation.
\newblock \emph{Phys. Rev. Lett.}, 67\penalty0 (24):\penalty0 3412--3415, 1991.

\bibitem[Elmer(1994)]{Elmer1994}
F.~J. Elmer.
\newblock Avalanches in the weakly driven frenkel-kontorova model.
\newblock \emph{Phys Rev E}, 50\penalty0 (6):\penalty0 4470, 1994.

\bibitem[Ding et~al.(2013)Ding, Lookman, Zhao, Saxena, Sun, and
  Salje]{ding2013dynamically}
X~Ding, T~Lookman, Z~Zhao, A~Saxena, J~Sun, and EKH Salje.
\newblock Dynamically strained ferroelastics: Statistical behavior in elastic
  and plastic regimes.
\newblock \emph{Phys. Rev. B}, 87\penalty0 (9):\penalty0 094109, 2013.

\bibitem[Ahluwalia and Ananthakrishna(2001)]{Ahluwalia2001-cy}
Rajeev Ahluwalia and G~Ananthakrishna.
\newblock Power-law statistics for avalanches in a martensitic transformation.
\newblock \emph{Phys. Rev. Lett.}, 86\penalty0 (18):\penalty0 4076--4079, April
  2001.

\bibitem[Sreekala et~al.(2004)Sreekala, Ahluwalia, and
  Ananthakrishna]{PhysRevB.70.224105}
S.~Sreekala, Rajeev Ahluwalia, and G.~Ananthakrishna.
\newblock Precursors and power-law statistics of acoustic emission and shape
  memory effect in martensites.
\newblock \emph{Phys. Rev. B}, 70:\penalty0 224105, Dec 2004.

\bibitem[Paul et~al.(2008)Paul, Bhattacharya, Sengupta, and Rao]{paul2008non}
Arya Paul, Jayee Bhattacharya, Surajit Sengupta, and Madan Rao.
\newblock Non-affine deformation in microstructure selection in solids ii:
  Elastoplastic theory for thedynamics of solid state transformations.
\newblock \emph{J. Phys.: Condens. Matter}, 20\penalty0 (36):\penalty0 365211,
  2008.

\bibitem[Ball and James(2015)]{ball2015incompatible}
John~M Ball and RD~James.
\newblock Incompatible sets of gradients and metastability.
\newblock \emph{Arch. Ration. Mech. Anal.}, 218:\penalty0 1363--1416, 2015.

\bibitem[Bouville and Ahluwalia(2007)]{bouville2007effect}
Mathieu Bouville and Rajeev Ahluwalia.
\newblock Effect of lattice-mismatch-induced strains on coupled diffusive and
  displacive phase transformations.
\newblock \emph{Phys. Rev. B}, 75\penalty0 (5):\penalty0 054110, 2007.

\bibitem[Zhang et~al.(2009)Zhang, James, and M{\"u}ller]{zhang2009energy}
Zhiyong Zhang, Richard~D James, and Stefan M{\"u}ller.
\newblock Energy barriers and hysteresis in martensitic phase transformations.
\newblock \emph{Acta Mater}, 57\penalty0 (15):\penalty0 4332--4352, 2009.

\bibitem[Atli et~al.(2013)Atli, Franco, Karaman, Gaydosh, and
  Noebe]{atli2013influence}
KC~Atli, BE~Franco, I~Karaman, D~Gaydosh, and RD~Noebe.
\newblock Influence of crystallographic compatibility on residual strain of
  tini based shape memory alloys during thermo-mechanical cycling.
\newblock \emph{Mater. Sci. Eng. A}, 574:\penalty0 9--16, 2013.

\bibitem[Yang et~al.(2023)Yang, Xu, Zhou, Ding, Sun, Lookman, and
  Xue]{yang2023nonthermoelastic}
Yuanchao Yang, Yangyang Xu, Yumei Zhou, Xiangdong Ding, Jun Sun, Turab Lookman,
  and Dezhen Xue.
\newblock Nonthermoelastic martensitic features in ideal martensites due to
  volume effects.
\newblock \emph{Phys. Rev. B}, 108\penalty0 (2):\penalty0 024102, 2023.

\bibitem[Shenoy et~al.(1999)Shenoy, Lookman, Saxena, and Bishop]{Shenoy:1999ri}
S.~R. Shenoy, T.~Lookman, A.~Saxena, and A.~R. Bishop.
\newblock Martensitic textures: Multiscale consequences of elastic
  compatibility.
\newblock \emph{Phys. Rev. B}, 60\penalty0 (18):\penalty0 R12537--R12541, 1999.

\bibitem[Ahluwalia et~al.(2006)Ahluwalia, Lookman, and
  Saxena]{Ahluwalia2006-uz}
Rajeev Ahluwalia, Turab Lookman, and Avadh Saxena.
\newblock Dynamic strain loading of cubic to tetragonal martensites.
\newblock \emph{Acta Mater}, 54\penalty0 (8):\penalty0 2109--2120, 2006.

\bibitem[Kert{\'e}sz and Kiss(1990)]{kertesz1990noise}
J{\'a}nos Kert{\'e}sz and L{\'a}szl{\'o}B Kiss.
\newblock The noise spectrum in the model of self-organised criticality.
\newblock \emph{Journal of Physics A: Mathematical and General}, 23\penalty0
  (9):\penalty0 L433, 1990.

\bibitem[Kuntz and Sethna(2000)]{PhysRevB.62.11699}
Matthew~C. Kuntz and James~P. Sethna.
\newblock Noise in disordered systems: The power spectrum and dynamic exponents
  in avalanche models.
\newblock \emph{Phys. Rev. B}, 62:\penalty0 11699--11708, Nov 2000.

\bibitem[Laurson et~al.(2005)Laurson, Alava, and Zapperi]{laurson2005power}
Lasse Laurson, Mikko~J Alava, and Stefano Zapperi.
\newblock Power spectra of self-organized critical sandpiles.
\newblock \emph{J. Stat. Mech.}, 2005\penalty0 (11):\penalty0 L11001, 2005.

\bibitem[Laurson and Alava(2006)]{laurson20061}
Lasse Laurson and Mikko~J Alava.
\newblock 1/ f noise and avalanche scaling in plastic deformation.
\newblock \emph{Phys Rev E}, 74\penalty0 (6):\penalty0 066106, 2006.

\bibitem[Nandi et~al.(2022)Nandi, Sarracino, Herrmann, and
  de~Arcangelis]{nandi2022scaling}
Manoj~Kumar Nandi, Alessandro Sarracino, Hans~J Herrmann, and Lucilla
  de~Arcangelis.
\newblock Scaling of avalanche shape and activity power spectrum in neuronal
  networks.
\newblock \emph{Phys. Rev. E}, 106\penalty0 (2):\penalty0 024304, 2022.

\bibitem[Ruseckas and Kaulakys(2014)]{ruseckas2014scaling}
J~Ruseckas and B~Kaulakys.
\newblock Scaling properties of signals as origin of 1/f noise.
\newblock \emph{J. Stat. Mech.}, 2014\penalty0 (6):\penalty0 P06005, 2014.

\bibitem[Kay(1993)]{kay1993fundamentals}
Steven~M Kay.
\newblock \emph{Fundamentals of statistical signal processing: estimation
  theory}.
\newblock Prentice-Hall, Inc., 1993.

\bibitem[Truskinovsky and Vainchtein(2004)]{Truskinovsky2004-xp}
Lev Truskinovsky and Anna Vainchtein.
\newblock The origin of nucleation peak in transformational plasticity.
\newblock \emph{J. Mech. Phys. Solids}, 52\penalty0 (6):\penalty0 1421--1446,
  2004.

\bibitem[Puglisi and Truskinovsky(2005)]{Puglisi2005-lg}
G~Puglisi and L~Truskinovsky.
\newblock Thermodynamics of rate-independent plasticity.
\newblock \emph{J. Mech. Phys. Solids}, 53\penalty0 (3):\penalty0 655--679,
  March 2005.

\bibitem[Mielke and Truskinovsky(2011)]{Mielke2011-ck}
Alexander Mielke and Lev Truskinovsky.
\newblock From discrete visco-elasticity to continuum rate-independent
  plasticity: Rigorous results.
\newblock \emph{Arch. Ration. Mech. Anal.}, 203\penalty0 (2):\penalty0
  577--619, September 2011.

\bibitem[Salman and Truskinovsky(2012)]{Salman2012a}
O.U. Salman and L.~Truskinovsky.
\newblock On the critical nature of plastic flow: One and two dimensional
  models.
\newblock \emph{Int. J. Eng. Sci.}, 59:\penalty0 219--254, 2012.

\bibitem[Truskinovsky and Vainchtein(2005)]{Truskinovsky2005-tq}
Lev Truskinovsky and Anna Vainchtein.
\newblock Kinetics of martensitic phase transitions: Lattice model.
\newblock \emph{SIAM J. Appl. Math.}, 66\penalty0 (2):\penalty0 533--553, 2005.

\bibitem[Efendiev and Truskinovsky(2010)]{Efendiev2010-kq}
Yalchin~R Efendiev and Lev Truskinovsky.
\newblock Thermalization of a driven bi-stable {FPU} chain.
\newblock \emph{Continuum Mech. Thermodyn.}, 22\penalty0 (6-8):\penalty0
  679--698, September 2010.

\bibitem[West and Shlesinger(1989)]{West1989-gr}
Bruce~J West and Michael~F Shlesinger.
\newblock On the ubiquity of 1/f noise.
\newblock \emph{Int. J. Mod. Phys. B}, 03\penalty0 (06):\penalty0 795--819,
  June 1989.

\bibitem[Milotti(2002)]{Milotti2002-pt}
E~Milotti.
\newblock 1/f noise: a pedagogical review.
\newblock \emph{arXiv: Classical Physics}, April 2002.

\bibitem[Polizzi et~al.(2021)Polizzi, P{\'e}rez-Reche, Arneodo, and
  Argoul]{polizzi2021power}
Stefano Polizzi, Francisco-Jos{\'e} P{\'e}rez-Reche, Alain Arneodo, and
  Fran{\c{c}}oise Argoul.
\newblock Power-law and log-normal avalanche size statistics in random growth
  processes.
\newblock \emph{Phys Rev E}, 104\penalty0 (5):\penalty0 L052101, 2021.

\bibitem[Sornette(1994{\natexlab{a}})]{sornette1994sweeping}
Didier Sornette.
\newblock Sweeping of an instability: an alternative to self-organized
  criticality to get powerlaws without parameter tuning.
\newblock \emph{Journal de Physique I}, 4\penalty0 (2):\penalty0 209--221,
  1994{\natexlab{a}}.

\bibitem[Sornette(1994{\natexlab{b}})]{sornette1994power}
Didier Sornette.
\newblock Power laws without parameter tuning: An alternative to self-organized
  criticality.
\newblock \emph{Phys. Rev. Lett.}, 72\penalty0 (14):\penalty0 2306,
  1994{\natexlab{b}}.

\bibitem[Viv{\`e}s et~al.(1995)Viv{\`e}s, Ort{\'\i}n, Ma{\~n}osa, R{\`a}fols,
  and Planes]{vives1995distribution}
Eric Viv{\`e}s, J~Ort{\'\i}n, Ll~Ma{\~n}osa, I~R{\`a}fols, and A~Planes.
\newblock Distribution of acoustic emission avalanches in martensitic
  transformations.
\newblock \emph{Le Journal de Physique IV}, 5\penalty0 (C2):\penalty0 C2--59,
  1995.

\bibitem[Urbach et~al.(1995)Urbach, Madison, and Markert]{Urbach1995-ui}
J~S Urbach, R~C Madison, and J~T Markert.
\newblock Interface depinning, self-organized criticality, and the barkhausen
  effect.
\newblock \emph{Phys. Rev. Lett.}, 75\penalty0 (2):\penalty0 276--279, July
  1995.

\bibitem[Durin et~al.(1995)Durin, Bertotti, and Magni]{Durin1995-cg}
Gianfranco Durin, Giorgio Bertotti, and Alessandro Magni.
\newblock Fractals, scaling and the question of self-organized criticality in
  magnetization processes.
\newblock \emph{Fractals}, 03\penalty0 (02):\penalty0 351--370, June 1995.

\bibitem[Schmelzer and Ulbricht(1987)]{schmelzer1987thermodynamics}
J~Schmelzer and H~Ulbricht.
\newblock Thermodynamics of finite systems and the kinetics of first-order
  phase transitions.
\newblock \emph{J. Colloid Interface Sci.}, 117\penalty0 (2):\penalty0
  325--338, 1987.

\bibitem[Salman et~al.(2019)Salman, Muite, and Finel]{salman2019origin}
Oguz~Umut Salman, Benson Muite, and Alphonse Finel.
\newblock Origin of stabilization of macrotwin boundaries in martensites.
\newblock \emph{Eur. Phys. J. B}, 92:\penalty0 1--9, 2019.

\bibitem[Carlson et~al.(1991)Carlson, Langer, Shaw, and Tang]{Carlson1991-jh}
J~M Carlson, J~S Langer, B~E Shaw, and C~Tang.
\newblock Intrinsic properties of a burridge-knopoff model of an earthquake
  fault.
\newblock \emph{Phys. Rev. A}, 44\penalty0 (2):\penalty0 884--897, July 1991.

\bibitem[Schmittbuhl et~al.(1996)Schmittbuhl, Vilotte, and
  Roux]{https://doi.org/10.1029/96JB00653}
Jean Schmittbuhl, Jean-Pierre Vilotte, and St{\'e}phane Roux.
\newblock Velocity weakening friction: A renormalization approach.
\newblock \emph{J. Geophys. Res.}, 101\penalty0 (B6):\penalty0 13911--13917,
  1996.

\bibitem[Elmer(1996)]{Elmer1996-mm}
F~J Elmer.
\newblock Is self-organized criticality possible in dry friction?
\newblock In \emph{Physics of Sliding Friction}, pages 433--447. Springer
  Netherlands, Dordrecht, 1996.

\bibitem[Gorbushin et~al.(2022)Gorbushin, Vainchtein, and
  Truskinovsky]{gorbushin2022transition}
N~Gorbushin, A~Vainchtein, and L~Truskinovsky.
\newblock Transition fronts and their universality classes.
\newblock \emph{Phys Rev E}, 106\penalty0 (2):\penalty0 024210, 2022.

\bibitem[Puglisi and Truskinovsky(2000)]{puglisi2000mechanics}
G~Puglisi and Lev Truskinovsky.
\newblock Mechanics of a discrete chain with bi-stable elements.
\newblock \emph{J. Mech. Phys. Solids}, 48\penalty0 (1):\penalty0 1--27, 2000.

\bibitem[Puglisi and Truskinovsky(2002{\natexlab{a}})]{puglisi2002mechanism}
G~Puglisi and L~Truskinovsky.
\newblock A mechanism of transformational plasticity.
\newblock \emph{Contin. Mech. Thermodyn.}, 14:\penalty0 437--457,
  2002{\natexlab{a}}.

\bibitem[Puglisi and Truskinovsky(2002{\natexlab{b}})]{puglisi2002rate}
G~Puglisi and L~Truskinovsky.
\newblock Rate independent hysteresis in a bi-stable chain.
\newblock \emph{J. Mech. Phys. Solids}, 50\penalty0 (2):\penalty0 165--187,
  2002{\natexlab{b}}.

\bibitem[Bhattacharya et~al.(1994)Bhattacharya, Firoozye, James, and
  Kohn]{bhattacharya1994restrictions}
Kaushik Bhattacharya, Nikan~B Firoozye, Richard~D James, and Robert~V Kohn.
\newblock Restrictions on microstructure.
\newblock \emph{Proc. R. Soc. Edinb. A}, 124\penalty0 (5):\penalty0 843--878,
  1994.

\bibitem[Smyshlyaev and Willis(1999)]{smyshlyaev1999relation}
VP~Smyshlyaev and JR~Willis.
\newblock On the relation of a three-well energy.
\newblock \emph{Proc. R. Soc. Lond. A}, 455\penalty0 (1983):\penalty0 779--814,
  1999.

\bibitem[Grabovsky and Truskinovsky(2023)]{grabovsky2023class}
Yury Grabovsky and Lev Truskinovsky.
\newblock A class of nonlinear elasticity problems with no local but many
  global minimizers.
\newblock \emph{J. Elast.}, 154\penalty0 (1):\penalty0 147--171, 2023.

\bibitem[Grabovsky and Truskinovsky(2024)]{grabovsky2024rigidity}
Yury Grabovsky and Lev Truskinovsky.
\newblock Rigidity-induced critical points.
\newblock \emph{Phys Rev E}, 110\penalty0 (6):\penalty0 064114, 2024.

\bibitem[Curnoe and Jacobs(2000)]{Curnoe2000-rq}
S~H Curnoe and A~E Jacobs.
\newblock Twin wall of proper cubic-tetragonal ferroelastics.
\newblock \emph{Phys. Rev. B: Condens. Matter Mater. Phys.}, 62\penalty0
  (18):\penalty0 R11925--R11928, 2000.

\bibitem[Budiansky and Truskinovsky(1993)]{budiansky1993mechanics}
Bernard Budiansky and Lev Truskinovsky.
\newblock On the mechanics of stress-induced phase transformation in zirconia.
\newblock \emph{J. Mech. Phys. Solids}, 41\penalty0 (9):\penalty0 1445--1459,
  1993.

\bibitem[Fadda et~al.(2002)Fadda, Truskinovsky, and Zanzotto]{fadda2002unified}
Giuseppe Fadda, Lev Truskinovsky, and Giovanni Zanzotto.
\newblock Unified {L}andau description of the tetragonal, orthorhombic, and
  monoclinic phases of zirconia.
\newblock \emph{Phys. Rev. B}, 66\penalty0 (17):\penalty0 174107, 2002.

\bibitem[Salman(2009)]{Salman2009-qv}
O~U Salman.
\newblock \emph{Modeling of spatio-temporal dynamics and patterning mechanisms
  of martensites by phase-field and Lagrangian methods}.
\newblock PhD thesis, Universit{\'e} Pierre et Marie Curie, 2009.

\bibitem[Ball et~al.(1991)Ball, Holmes, James, Pego, and
  Swart]{ball1991dynamics}
JM~Ball, Philip~J Holmes, RD~James, RL~Pego, and PJ~Swart.
\newblock On the dynamics of fine structure.
\newblock \emph{J. Nonlinear Sci.}, 1:\penalty0 17--70, 1991.

\bibitem[Friesecke and McLeod(1997)]{friesecke1997dynamic}
G~Friesecke and JB~McLeod.
\newblock Dynamic stability of non--minimizing phase mixtures.
\newblock \emph{Proc. R. Soc. Lond. A}, 453\penalty0 (1966):\penalty0
  2427--2436, 1997.

\bibitem[Curnoe and Jacobs(2001)]{PhysRevB.64.064101}
S.~H. Curnoe and A.~E. Jacobs.
\newblock Time evolution of tetragonal-orthorhombic ferroelastics.
\newblock \emph{Phys. Rev. B}, 64\penalty0 (6):\penalty0 064101, 2001.

\bibitem[Jacobs et~al.(2003)Jacobs, Curnoe, and Desai]{PhysRevB.68.224104}
A.~E. Jacobs, S.~H. Curnoe, and R.~C. Desai.
\newblock Simulations of cubic-tetragonal ferroelastics.
\newblock \emph{Phys. Rev. B}, 68\penalty0 (22):\penalty0 224104, 2003.

\bibitem[Muto et~al.(1990)Muto, Oshima, and Fujita]{Muto1990-eb}
S~Muto, R~Oshima, and F~E Fujita.
\newblock Elastic softening and elastic strain energy consideration in the
  fcc---fct transformation of fe-pd alloys.
\newblock \emph{Acta Metall. Mater.}, 38\penalty0 (4):\penalty0 685--694, 1990.

\bibitem[Kartha et~al.(1995)Kartha, Krumhansl, Sethna, and
  Wickham]{Kartha1995-wj}
S~Kartha, J~A Krumhansl, J~P Sethna, and L~K Wickham.
\newblock Disorder-driven pretransitional tweed pattern in martensitic
  transformations.
\newblock \emph{Phys. Rev. B}, 52\penalty0 (2):\penalty0 803--822, July 1995.

\bibitem[Sato et~al.(1982)Sato, Grier, Shapiro, and Miyajima]{Sato1982-mf}
M~Sato, B~H Grier, S~M Shapiro, and H~Miyajima.
\newblock Effect of magnetic ordering on the lattice dynamics of {FCC}
  {Fe1}-{xPdx}.
\newblock \emph{J. Phys.}, 12\penalty0 (10):\penalty0 2117--2129, October 1982.

\bibitem[Oshima et~al.(1988)Oshima, Sugiyama, and Fujita]{Oshima1988-uu}
R~Oshima, M~Sugiyama, and F~E Fujita.
\newblock Tweed structures associated with fcc-fct transformations in fe-pd
  alloys.
\newblock \emph{Metall. Trans.}, 19\penalty0 (4):\penalty0 803--810, April
  1988.

\bibitem[Testardi and Bateman(1967)]{Testardi1967-qo}
L~R Testardi and T~B Bateman.
\newblock Lattice instability of high-transition-temperature superconductors.
  {II}. single-{CrystalV3Si} results.
\newblock \emph{Phys. Rev.}, 154\penalty0 (2):\penalty0 402--410, February
  1967.

\bibitem[Rodney et~al.(2001)Rodney, Bouar, and Finel]{Rodney:2001lr}
D.~Rodney, Y.~Le Bouar, and A.~Finel.
\newblock Phase field methods and dislocations.
\newblock \emph{Acta Mater}, 51:\penalty0 17--30, 2001.

\bibitem[Pitteri and Zanzotto(2002)]{pitteri2002continuum}
Mario Pitteri and Giovanni Zanzotto.
\newblock \emph{Continuum models for phase transitions and twinning in
  crystals}.
\newblock Chapman and Hall/CRC, 2002.

\bibitem[Artemev et~al.(2002)Artemev, Jin, and Khachaturyan]{Artemev2002-vh}
A~Artemev, Yongmei Jin, and A~G Khachaturyan.
\newblock Three-dimensional phase field model and simulation of cubic →
  tetragonal martensitic transformation in polycrystals.
\newblock \emph{Philos. Mag. A}, 82\penalty0 (6):\penalty0 1249--1270, April
  2002.

\bibitem[Mamivand et~al.(2013)Mamivand, Zaeem, and El~Kadiri]{Mamivand2013-cm}
Mahmood Mamivand, Mohsen~Asle Zaeem, and Haitham El~Kadiri.
\newblock A review on phase field modeling of martensitic phase transformation.
\newblock \emph{Comput. Mater. Sci.}, 77:\penalty0 304--311, September 2013.

\bibitem[Ni et~al.(2007)Ni, Jin, and Khachaturyan]{Ni2007-ab}
Y~Ni, Y~M Jin, and A~G Khachaturyan.
\newblock The transformation sequences in the cubic→tetragonal decomposition.
\newblock \emph{Acta Mater}, 55\penalty0 (14):\penalty0 4903--4914, August
  2007.

\bibitem[Turteltaub and Suiker(2006)]{Turteltaub2006-ji}
S~Turteltaub and A~S~J Suiker.
\newblock A multiscale thermomechanical model for cubic to tetragonal
  martensitic phase transformations.
\newblock \emph{Int. J. Solids Struct.}, 43\penalty0 (14-15):\penalty0
  4509--4545, July 2006.

\bibitem[Ichitsubo et~al.(2000)Ichitsubo, Tanaka, Koiwa, and
  Yamazaki]{PhysRevB.62.5435}
Tetsu Ichitsubo, Katsushi Tanaka, Masahiro Koiwa, and Yoshihiro Yamazaki.
\newblock Kinetics of cubic to tetragonal transformation under external field
  by the time-dependent ginzburg-landau approach.
\newblock \emph{Phys. Rev. B}, 62:\penalty0 5435--5441, Sep 2000.

\bibitem[Hildebrand and Miehe(2012)]{Hildebrand2012-is}
F~E Hildebrand and C~Miehe.
\newblock A phase field model for the formation and evolution of martensitic
  laminate microstructure at finite strains.
\newblock \emph{Philos. Mag. (Abingdon)}, 92\penalty0 (34):\penalty0
  4250--4290, December 2012.

\bibitem[Idesman et~al.(2008)Idesman, Cho, and Levitas]{Idesman2008-ft}
A~Idesman, J~Cho, and V~Levitas.
\newblock Finite element modeling of dynamics of martensitic phase transitions.
\newblock \emph{Appl. Phys. Lett.}, 93:\penalty0 043102, July 2008.

\bibitem[Clauset et~al.(2009)Clauset, Shalizi, and Newman]{Clauset2009-nv}
Aaron Clauset, Cosma~Rohilla Shalizi, and M~E~J Newman.
\newblock Power-law distributions in empirical data.
\newblock \emph{SIAM Rev. Soc. Ind. Appl. Math.}, 51\penalty0 (4):\penalty0
  661--703, November 2009.

\bibitem[Vives et~al.(1995)Vives, Goicoechea, Ort{\'\i}n, and
  Planes]{Vives1995-na}
E~Vives, J~Goicoechea, J~Ort{\'\i}n, and A~Planes.
\newblock Universality in models for disorder-induced phase transitions.
\newblock \emph{Phys. Rev. E Stat. Phys. Plasmas Fluids Relat. Interdiscip.
  Topics}, 52\penalty0 (1):\penalty0 R5--R8, July 1995.

\bibitem[LeVeque(2007)]{LeVeque:2007ye}
Randall~J. LeVeque.
\newblock \emph{Finite Difference Methods for Ordinary and Partial Differential
  Equations, Steady State and Time Dependent Problems}.
\newblock SIAM, 2007.
\newblock ISBN 978-0-898716-29-0.

\end{thebibliography}
% Produces the bibliography via BibTeX.
%
%
%
\end{document}